\documentclass[1p]{elsarticle}

\usepackage{lineno,hyperref}
\modulolinenumbers[5]
\usepackage{amsmath,amssymb}
\usepackage{graphicx}
\usepackage{float}
\usepackage{color}
\usepackage{caption}
\usepackage{subcaption}
\journal{Journal of Computational Physics}

\usepackage{subcaption}

\graphicspath{{/}} 

\renewcommand{\vec}[1]{\ensuremath\boldsymbol{#1}}

\renewcommand{\Re}{\ensuremath \textnormal{Re}}
\newcommand{\Ca}{\ensuremath \textnormal{Ca}}
\newcommand{\We}{\ensuremath \textnormal{We}}
\newcommand{\Fr}{\ensuremath \textnormal{Fr}}
\newcommand{\sgn}{\ensuremath \textnormal{sgn}}

\biboptions{sort&compress}

\begin{document}

\begin{frontmatter}

\title{Modeling Droplets with Slippery Interfaces}

\author{Afsoun Rahnama Falavarjani, David Salac\fnref{myfootnote}}
\address{Department of Mechanical and Aerospace Engineering, University at Buffalo}
\fntext[myfootnote]{Corresponding author.}

\begin{abstract}
  Many multiphase fluid systems, such as those involving immiscible polymers or liquid-liquid systems with surfactants, have shown a breakdown of the no-slip condition at the material interface. This results in systems where the tangential velocity of the inner and outer fluid can differ, with the jump in velocity dependent not only the material properties of the interface but also the stresses applied by the surrounding fluid. In this work a numerical model is presented which is capable of investigating general multiphase fluid systems involving interfacial slip in both two- and three-dimensions. To make the system computationally feasible, a hybrid Navier-Stokes projection method is used, whereby the viscosity, density, and pressure are assumed to be continuous across the interface while the velocity field can experience a jump, which is handled via the Immersed Interface Method. The numerical model is compared to experimental results involving polymer-polymer mixtures and computational results for droplets in extensional flows, showing excellent agreement with both. It is then used to explore the influence of interfacial slip in a number of common multiphase fluid systems, including the shearing of a planar interface, droplet and filament relaxation, and droplets in shear flow, both unbounded and wall-bound.

\end{abstract}


\end{frontmatter}

\linenumbers

\section{Introduction}\label{sec:intro}

A very common assumption is that a liquid-liquid interface obeys a no-slip condition whereby the velocity of the inner and outer fluid match in both the normal and tangential directions~\cite{Batchelor1967}. While this condition reproduces many macroscopic phenomena, it can break down when the immiscible liquids are poorly mixed, if molecular forces result in a molecular depletion at the interface, or if the length scales of interest are small~\cite{Poesio2017}. For example, it has been observed blends of immiscible polymers can have uncharacteristically low viscosity, sometimes less than the viscosity of each individual polymer~\cite{Utracki1982,Utracki1983,Rauwendaal1988,Lin1979,Han1972}. This has led to recent investigations of the slip in systems along a single interface~\cite{LeeP.2009,Zartman2011,Zhao2002}. Other systems, such as water and oil~\cite{Scarratt2020} have also demonstrated a breakdown of the no-slip condition.

At the micro- and nano-scale the breakdown of the no-slip becomes extremely important. Due to the ratio of the interfacial area compared to the volume being much larger than at the macroscopic level, small changes in conditions at the interface can have large consequences. Microfluidic systems with uses ranging from separation of particles and the mixing of reagents to vesicle fabrication~\cite{lu2015,DiCarlo2007,Zhu2013,Ortseifen2020} have been proposed, and the efficiency of such systems will depend on how much slip, if any, occurs. With this in mind it has been proposed that the use of hydrophobic beads~\cite{Ehlinger2013} or surfactants~\cite{Das2018,Das2017a,Das2017,Ramachandran2012a,Ramachandran2012} can be used to enhance slip at liquid-liquid interfaces for such small systems.

Investigations of slip at liquid-solid interfaces are much more common than those at liquid-liquid interfaces. Beginning with Navier, where the velocity at a solid boundary is related to the gradient of the velocity field~\cite{Navier1823}, slip at liquid-solid interfaces has been shown to occur experimentally~\cite{Baudry2001,Migler1993,Kumar2022,Chen2019} and modeled using a number of techniques~\cite{Miksis1994,Thompson1997,Wang2020,Zhang2022}. The influence of liquid-solid slip on moving contact lines~\cite{Kirkinis2013}, flow rates through microfilters~\cite{Jensen2014}, and propulsion efficiency of solid micro-swimmers~\cite{Guo2021} has also been performed. The interested reader can refer to Refs.~\cite{Lauga2007,Sochi2011,Wang2021}.

There have been several prior modeling efforts for fluid-fluid systems involving interfacial slip. These prior works generally fall into one of three categories. The first use perturbation methods for spherical or nearly-spherical droplets. Sharanya et al~\cite{Sharanya2018} and Mandal et al~\cite{Mandal2015} investigated the cross-migration speed of buoyant droplets in Couette and Poiseuille flow, respectively, and determined that the cross-migration velocity of a droplet increases with increasing slip. Feng et al determined that the presence of interfacial slip can significantly reduce the drag force of a droplet in small Reynolds number flows~\cite{Feng2012}. The effect of interfacial slip on viscoelastic drop deformation~\cite{Das2017a} and suspension rheology~\cite{Ramachandran2012a,Das2018} have also been investigated using perturbation methods.

The second type of modeling effort is based on the boundary integral method. In particular Ramachandran et al utilized this technique to investigate the time-evolution of highly deformed droplets subjected to extensional flows~\cite{Ramachandran2012}. Subsequently the equations arising in this situation were also analyzed~\cite{Ramachandran2012b}.
The third class of numerical investigations utilize molecular dynamics to probe not only slip at interfaces but also possible causes. Hu et al investigated the dependence of the interfacial boundary condition as a function of surfactant concentration~\cite{Hu2010}. It was determined that whether a systems has interfacial slip depends on the concentration of surfactants, with the interfacial slip-length being shear-rate dependent. Poesio utilized molecular dynamics to investigate water-carbon tetrachloride, water-octane, and heptane-ethylene glycol systems, finding that slip can occur in all three~\cite{Poesio2017}.

While recent work has demonstrated direct numerical simulations of two-phase bubbly jets with a slip velocity, the shape of the bubbles was not taken into account~\cite{Seo2022}. To our knowledge there is no two- or three-dimensional numerical model of liquid-liquid interfaces with slip for arbitrary shapes and finite Reynolds numbers, which we address in this work. The rest of this paper is organized as follows. In Section~\ref{sec:math}, the governing equations for the fluid flow and the slip condition are presented. In Section~\ref{sec:method}, the numerical implementation is given, which is followed by sample two- and three-dimensional numerical experiments in Section~\ref{sec:res}. The model will be compared to published prior experimental and computational works and used to explore the influence of slip on a number of multiphase systems. A brief conclusion is then followed in Section~\ref{sec:conc}.

\section{Governing equations}\label{sec:math}
In this work we consider two Newtonian, immiscible fluids separated by an interface. Both fluids can have different fluid properties such as density and viscosity. To describe the dynamics of this system it is necessary to consider the Navier-Stokes equations to describe the fluid flows and the boundary conditions on the interface. The Navier-Stokes equations are coupled with the interfacial stress, which includes forces opposing the interfacial slip. In this section the governing equations, material approximations, and non-dimensionalization are presented.

\begin{figure}[H]
  \centering
  \includegraphics[width=0.4\textwidth]{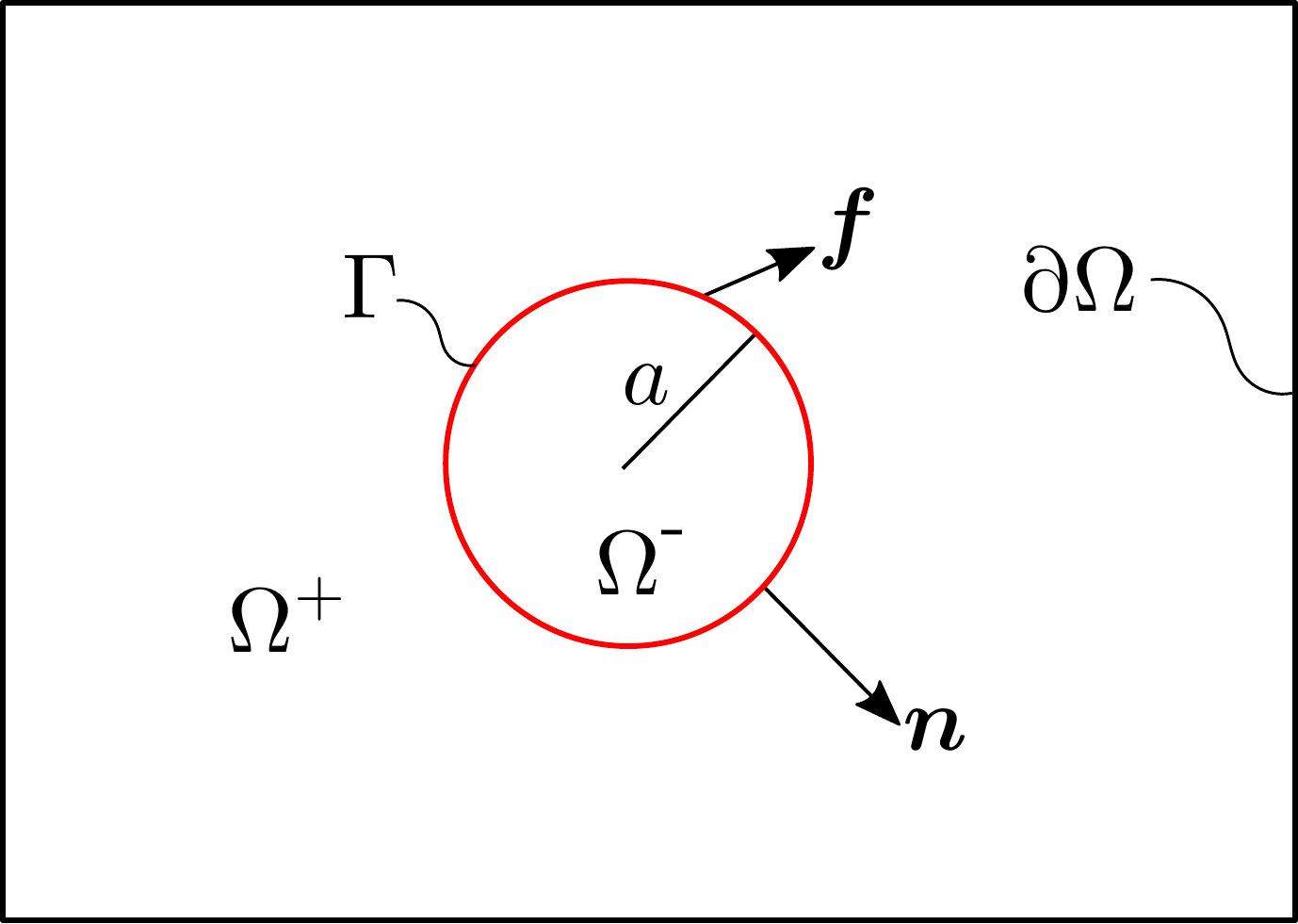}
  \caption{Schematic of a droplet of radius $a$ in a rectangular domain with the exterior boundary of $\partial \Omega$. The interface is denoted as $\Gamma$ and the fluids inside and outside the droplet are marked as $\Omega^-$ and $\Omega^+$, respectively.}
  \label{fig:schem}
\end{figure}

\subsection{Fluid flow equations}\label{sec:NSE}
Assuming that the inner and outer fluids are Newtonian and incompressible, the fluid can be described by the Navier-Stokes Equations
\begin{align}
  \rho^{\pm}\frac{D\vec{u}^{\pm}}{Dt} &= -\nabla p^{\pm} + \nabla \cdot(\mu^{\pm}(\nabla \vec{u}^{\pm} + \nabla^T \vec{u}^{\pm})) + \vec{b}^{\pm}\;\textnormal{in}\;\Omega^{\pm}\label{eq:NS},\\
  \nabla \cdot\vec{u}^{\pm} &= 0\;\textnormal{in}\;\Omega^{\pm},\label{eq:NSdiv}
\end{align}
where $\Omega^-$ and $\Omega^+$ represent the inner and outer domains shown in Fig.~\ref{fig:schem}, Here $D/Dt$ is the total (material) derivative, $\rho$ and $\mu$ are the density and viscosity of the fluid, $\vec{u}$ and $p$ are the velocity and pressure fields, and $\vec{b}$ are any body force term which may be present, such as gravity.

\begin{figure}[H]
  \centering
  \begin{subfigure}[b]{0.48\textwidth}
    \centering
    \includegraphics[height=4cm]{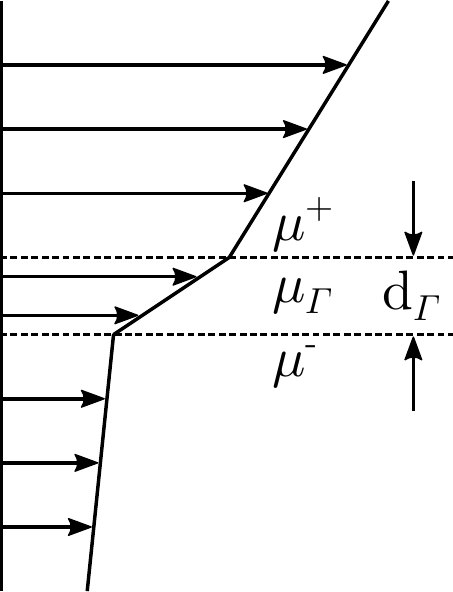}
    \caption{The three-fluid model indicating the interfacial layer of finite thickness.}
    \label{fig:ThreeFluidModelA}
  \end{subfigure}
  \hfill
  \begin{subfigure}[b]{0.48\textwidth}
    \centering
    \includegraphics[height=4cm]{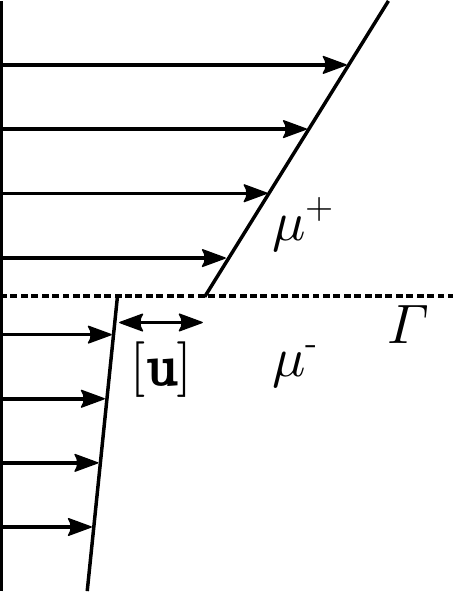}
    \caption{Approximation of the three-fluid model at the length scales of interest.}
    \label{fig:ThreeFluidModelB}
  \end{subfigure}
  \caption{The three-fluid model and it's modeling approximation.}
  \label{fig:ThreeFluidModel}
\end{figure}

The interface is assumed to follow the three-fluid model for multiphase fluid systems, Fig.~\ref{fig:ThreeFluidModelA}. In this model the interface between the inner and outer fluid is assumed to have a finite thickness $d_{\Gamma}$, which is much smaller than the overall system's length scale. In this interfacial layer the viscosity, $\mu_{\Gamma}$, differs from the viscosities of the surrounding fluid and is usually much smaller than the bulk viscosities which allows for a rapid change of velocity (i.e. a large shear rate). From a modeling perspective this interfacial layer can be thought of as an interface of zero-thickness, Fig.~\ref{fig:ThreeFluidModelB}, which allows for a jump in the bulk velocities,
\begin{align}
  [\vec{u}] &= \vec{u}^+ - \vec{u}^- = \vec{j}\;\textnormal{on}\;\Gamma,\label{eq:NSjump}
\end{align}
where $\vec{j}$ is a function to be derived. If the velocity is continuous (i.e. $d_{\Gamma}=0$) then $\vec{j}=\vec{0}$. This work assumes the absence of interfacial permeability
and therefore the normal component of the velocity will be continuous across the interface,
\begin{equation}\label{eq:normalVelocityJump}
  [\vec{u}]\cdot\vec{n} = 0 \; \textnormal{on} \; \Gamma,
\end{equation}
where $\vec{n}$ is the outward-facing unit normal. This results in $\vec{j}$ only being a function of tangential velocities.

In addition to the velocity jump condition, the inner and outer velocities are coupled to forces on either side of the interface, $\vec{f}_{\Gamma}^{\pm}$,
\begin{align}
  \vec{t}_{hd}^\pm = \vec{f}_{\Gamma}^\pm \; \textnormal{on} \; \Gamma.
\end{align}
The traction forces applied to each fluid are defined as $\vec{t}_{hd}^\pm = \vec{T}_{hd}^{\pm} \cdot \vec{n}^\pm$ where the bulk hydrodynamic stress tensor is given by $\vec{T}_{hd}^{\pm} = -p^{\pm}\vec{I}+\mu^{\pm}(\nabla \vec{u}^{\pm} + \nabla^T \vec{u}^{\pm})$, $\vec{I}$ is the identity matrix, and $\vec{n}=\vec{n}^+=-\vec{n}^-$ are the normals which point into each fluid.

In this work we consider two interfacial forces: a simple tension and a friction (dissipative) force due to a difference in tangential velocities such that $\vec{f}_\Gamma=\vec{f}_\sigma + \vec{f}_s$ where
\begin{align}
  \vec{f}_{\sigma}^\pm &= \pm\tfrac{1}{2}\sigma \kappa\vec{n}^\pm,\\
  \vec{f}_{s}^\pm &= \pm b(\vec{u}_s^+ - \vec{u}_s^-) = \pm b[\vec{u}_s].
\end{align}
Here $\kappa=\nabla \cdot \vec{n}$ is twice the mean curvature and $\sigma$ the equilibrium tension. The assumption here is that the standard surface tension
force, $\sigma\kappa\vec{n}$, is split evenly between the inner and outer fluid. The friction force is proportional to the jump in the tangential velocities given by
$\vec{u}_s^\pm=\vec{P}\cdot\vec{u}^\pm$, where $\vec{P}=\vec{I}-\vec{n}\otimes\vec{n}$ is the surface projection operator, $b$ is the friction coefficient with units of Pa\;s/m, with this force having an equal magnitude but opposite sign between the inner and outer fluid. This is essentially a Navier-Slip Condition where the friction coefficient is given by $b=\mu_{\Gamma}/d_{\Gamma}$~\cite{Leal2007,Ramachandran2012}. Note that in much of the literature the inverse of this coefficient is used and called a slip coefficient.

These expressions can now be brought into a single expression,
\begin{align}
  \vec{t}_{hd}^{\pm} = \tfrac{1}{2}\sigma\kappa\vec{n} \pm b[\vec{u}_s], \label{eq:stressBalancePM}
\end{align}
where the fact that $\vec{n}^+=-\vec{n}^-$ has been taken into account.
Adding together $\vec{t}_{hd}^+$ and $\vec{t}_{hd}^-$ gives us
\begin{eqnarray}
  (\vec{t}_{hd}^{+} + \vec{t}_{hd}^{-}) = \sigma\kappa\vec{n} + b[\vec{u}_s] - b[\vec{u}_s] = \sigma\kappa\vec{n},
\end{eqnarray}
as $\vec{t}_{hd}^+ + \vec{t}_{hd}^- = \vec{T}_{hd}^+\cdot\vec{n}^+ + \vec{T}_{hd}^-\cdot\vec{n}^- = \vec{T}_{hd}^+\cdot\vec{n} - \vec{T}_{hd}^-\cdot\vec{n} = \left[\vec{T}_{hd}\right]\cdot\vec{n}$. From this we recover the standard stress-balance expression for simple-tension flow,
\begin{align}
  \left[\vec{T}_{hd}\right]\cdot\vec{n} = \sigma\kappa\vec{n}.
\end{align}
On the other hand, by subtracting the two components of Eq.~(\ref{eq:stressBalancePM}) we get
\begin{eqnarray}\label{eq:tp-tm}
  \vec{t}_{hd}^+ - \vec{t}_{hd}^- = \tfrac{1}{2}\sigma\kappa\vec{n} - \tfrac{1}{2}\sigma\kappa\vec{n} + b[\vec{u}_s] + b[\vec{u}_s] = 2b[\vec{u}_s].
\end{eqnarray}
Taking the inner-product of Eq.~\eqref{eq:tp-tm} and the unit normal results in
\begin{align}
  \left(\vec{t}_{hd}^+ - \vec{t}_{hd}^-\right)\cdot\vec{n} = 2b[\vec{u}_s]\cdot\vec{n}=0
\end{align}
as $[\vec{u}_s]$ only captures the tangential component of the velocity.
We can determine the tangential components of Eq.~\eqref{eq:tp-tm} by using the projection operator $\vec{P}$:
\begin{align}
  \vec{P}\cdot\left(\vec{t}_{hd}^+ - \vec{t}_{hd}^-\right) = 2b\vec{P}\cdot[\vec{u}_s] = 2b[\vec{u}_s] \label{eq:tp-tm,tau}
\end{align}
as the projection operator is idempotent, resulting in $\vec{P}\cdot[\vec{u}_s] = \left[\vec{P}\cdot\vec{u}_s\right] = \left[\vec{P}\cdot\vec{P}\cdot\vec{u}\right]=\left[\vec{P}\cdot\vec{u}\right]=[\vec{u}_s]$.
Expanding Eq.~\eqref{eq:tp-tm,tau} results in
\begin{align}
  2b[\vec{u}_s] &= \vec{P}\cdot\left(\vec{t}_{hd}^+ - \vec{t}_{hd}^-\right) \nonumber\\
                &= \vec{P}\cdot\left(\vec{T}_{hd}^+\cdot\vec{n}^+ - \vec{T}_{hd}^-\cdot\vec{n}^-\right) \nonumber\\
                &= \vec{P}\cdot\left(\vec{T}_{hd}^+\cdot\vec{n} + \vec{T}_{hd}^-\cdot\vec{n}\right) \nonumber\\
                &= \vec{P}\cdot\left(\left(-p^{+}\vec{I}+\mu^{+}\left(\nabla \vec{u}^{+} + \nabla^T \vec{u}^{+}\right)\right)\cdot \vec{n} + \left(-p^{-}\vec{I}+\mu^{-}\left(\nabla \vec{u}^{-} + \nabla^T \vec{u}^{-}\right)\right)\cdot \vec{n} \right) \nonumber \\
                &= \vec{P}\cdot\left(-(p^{+}+p^{-})\vec{n}+\mu^{+}(\nabla \vec{u}^{+} + \nabla^T \vec{u}^{+})\cdot \vec{n} + \mu^{-}(\nabla \vec{u}^{-} + \nabla^T \vec{u}^{-})\cdot \vec{n} \right) \nonumber \\
                &= \vec{P}\cdot\left(\mu^{+}(\nabla \vec{u}^{+} + \nabla^T \vec{u}^{+}) + \mu^{-}(\nabla \vec{u}^{-} + \nabla^T \vec{u}^{-})\right)\cdot \vec{n} \label{eq:tangentialBalance}
\end{align}
due to $\vec{P}\cdot\vec{n}=0$. From this we can say that the function $\vec{j}$ is the right-hand-side of Eq.~\eqref{eq:tangentialBalance} divided by $2b$.

In summary the full fluid flow equations can be written as
\begin{align}
  \rho^{\pm}\frac{D\vec{u}^{\pm}}{Dt} &= -\nabla p^{\pm} + \nabla \cdot(\mu^{\pm}(\nabla \vec{u}^{\pm} + \nabla^T \vec{u}^{\pm})) + \vec{b}^{\pm}\;\textnormal{in}\;\Omega^{\pm},\label{eq:summary1} \\
  \nabla \cdot\vec{u}^{\pm} &= 0\;\textnormal{in}\;\Omega^{\pm},\label{eq:summary2}\\
  \left[\vec{T}_{hd}\right]\cdot\vec{n} &= \sigma\kappa\vec{n}\;\textnormal{on}\;\Gamma,\label{eq:summary3}\\
  [\vec{u}]\cdot\vec{n} &= 0 \;\textnormal{on}\;\Gamma,\label{eq:summary4}\\
  [\vec{u}_s] &= \dfrac{1}{2b}\vec{P}\cdot\left(\mu^{+}(\nabla \vec{u}^{+} + \nabla^T \vec{u}^{+}) + \mu^{-}(\nabla \vec{u}^{-} + \nabla^T \vec{u}^{-})\right)\cdot \vec{n}\;\textnormal{on}\;\Gamma\label{eq:summary5}
\end{align}

\subsection{Assumptions and non-dimensionalization}\label{sec:nonD}

Even in the absence of a jump in the tangential velocity field, solving multiphase Navier-Stokes systems with varying density, viscosity, and pressure is a challenge. Methods to model such systems
by explicitly accounting for the jumps in material properties and pressure have been developed, but can be cumbersome, see examples including the Immersed Interface Method~\cite{Li2003,Russell2003,Tan2008} and the Ghost Fluid Method~\cite{Fedkiw1999,Gibou2007}.
Another method, the Continuum Surface Force Method, assumes continuity of the fluid field across the interface and that material properties, such as the density and viscosity, and the pressure are continuous across the interface~\cite{Chang1996}. Additionally, this method transforms the singular surface force, Eq.~\eqref{eq:summary3}, into equivalent body-force terms localized around the interface~\cite{Chang1996}.

In this work we take a hybrid approach. We assume that the density, viscosity, and pressure are all continuous across the interface while the velocity is discontinuous. Let the interface be given as the zero of a level-set function $\phi(\vec{x})$: $\Gamma=\left\{\vec{x}:\phi(\vec{x})=0\right\}$ with $\Omega^-=\left\{\vec{x}:\phi(\vec{x})<0\right\}$ and $\Omega^+=\left\{\vec{x}:\phi(\vec{x})>0\right\}$. The density and viscosity can
now be written as
\begin{align}
  \rho_{\epsilon}(\phi) &= \rho^- + (\rho^+ - \rho^-)H_{\epsilon}(\phi)\;\textnormal{and}\\
  \mu_{\epsilon}(\phi) &= \mu^- + (\mu^+ - \mu^-)H_{\epsilon}(\phi)
\end{align}
where $H_{\epsilon}(\phi)$ is a smoothed Heaviside function such that $H_{\epsilon}(\phi<-\epsilon)=0$, $H(\phi>\epsilon)=1$ with a smooth transition
between 0 and 1 for $-\epsilon\leq\phi\leq\epsilon$.

To non-dimensionalize the governing equations a few characteristic parameters need to be defined. Given a characteristic length $a$ and either one of characteristic time $t_0$ or characteristic velocity $u_0$, the other one can be obtained using $u_0 = a/t_0$. The density and viscosity are normalized by the values of the outer domain,
\begin{eqnarray}
\hat{\rho}_{\epsilon}(\phi) &=& \lambda + (1 - \lambda) H_{\epsilon}(\phi)\;\textnormal{and}\\
\hat{\mu}_{\epsilon}(\phi) &=& \eta + (1 - \eta) H_{\epsilon}(\phi),
\end{eqnarray}
where $\lambda = \rho^-/\rho^+$ and $\eta = \mu^-/\mu^+$ are the density and viscosity ratios, respectively. Henceforth, $\hat{\rho}_{\epsilon}(\phi)$ and $\hat{\mu}_{\epsilon}(\phi)$ will be written as $\hat{\rho}$ and $\hat{\mu}$ for simplicity.

In the presence of uniform surface tension and gravitational forces, non-dimensionalizing the fluid equations will introduce two non-dimensional parameters, the Weber number denoted as $\We$ and the Froude number denoted as $\Fr$, along with the Reynolds number, $\Re$. These three parameters are defined as follows
\begin{equation}
\Re = \frac{\rho^+ a u_0}{\mu^+}, \quad \We = \frac{\rho^+ a {u_0}^2}{\sigma} , \quad \Fr =  \frac{{u_0}^2}{a g_0}
\end{equation}
where $g_0$ is the strength of gravity~\cite{Salac2016}. Therefore, using the Continuum Surface Force method to account for singular forces, the dimensionless Navier-Stokes equations will be
\begin{equation}
\hat{\rho}\frac{D\hat{\vec{u}}^\pm}{Dt} = -\hat{\nabla} \hat{p} +\frac{1}{\Re} \hat{\nabla} \cdot(\hat{\mu}(\hat{\nabla} \hat{\vec{u}}^\pm + \hat{\nabla}^T \hat{\vec{u}}^\pm)) - \frac{1}{\We}\delta_{\epsilon}(\phi)\hat{\kappa} \hat{\nabla} \phi + \frac{1}{\Fr}(\hat{\rho}-1)\bar{\vec{g}},
\end{equation}
where $\bar{\vec{g}}$ is the direction of gravity, all non-dimensional quantities are denoted with a $\hat{\;}$, and $\delta_{\epsilon}(\phi)=dH_{\epsilon}(\phi)/d\phi$ is the smoothed Dirac delta function. More information regarding the single-fluid formulation can be found in literature~\cite{Kolahdouz2015a,Chang1996,Salac2016} while the particular discretizations for the Dirac delta and Heaviside functions used in this work can be found in References~\cite{Towers2008,Towers2009,Towers2009a}.

The jump in the tangential velocity, Eq.~\eqref{eq:summary5}, can also be normalized using the characteristic parameters mentioned before as
\begin{equation}\label{eq:tangVelJumpSimp}
\hat{\vec{u}}_{s}^+ - \hat{\vec{u}}_{s}^- = \frac{1}{2\beta}\vec{P}\left(\hat{\nabla} \hat{\vec{u}}^{+} + \hat{\nabla}^T \hat{\vec{u}}^{+} + \eta\left(\hat{\nabla} \hat{\vec{u}}^{-} + \hat{\nabla}^T \hat{\vec{u}}^{-}\right) \right) \cdot \vec{n}
\end{equation}
In this relation $\beta$ is the non-dimensional friction parameter given by
\begin{equation}
\beta = (b a)/\mu^+.
\end{equation}
The hat notation in the non-dimensional equations is dropped henceforth for simplicity.

\section{Numerical methods}\label{sec:method}

In this work a modified collocated finite-difference projection method is implemented to solve for the velocity and pressure sequentially, which has previously been used to model multiphase fluid systems with no-slip interfacial conditions~\cite{Salac2016}. First, a semi-implicit, semi-Lagrangian update is performed to obtain a tentative velocity field, $\vec{u}^{\pm,*}$,
\begin{equation}\label{eq:tentativeVel}
  \rho\frac{\vec{u}^{\pm,*} - \vec{u}^{\pm,n}_{d}}{\Delta t} = -\nabla p^n + \frac{1}{\Re}\nabla\cdot\left(\mu(\nabla\vec{u}^{\pm,*} + \nabla^T\vec{u}^{\pm,n})\right) + \vec{f}^n,
\end{equation}
where dependence on the level-set function has been suppressed for clarity, the superscript $n$ refers to the quantities at time step $t^n$, and $\vec{f}^n$ are any interfacial and body force calculated using the position of the interface at time $t^n$. The departure velocity, $\vec{u}^{\pm,n}_{d}$, can be obtained by determining the velocity at time $t^n$ at the departure location $\vec{x}_d=\vec{x}_i - \Delta t\vec{u}^n_i$ where $\vec{x}_i$ and $\vec{u}^n_i$ are the position and velocity at a grid-location, respectively. Higher-order methods can be easily obtained by storing multiple prior velocity fields, see Ref.~\cite{Kolahdouz2015a}. Unlike a standard Continuum Surface Force-based projection method, the jump in the tangential velocity introduces added complexity. Specifically, in addition to Eq.~\eqref{eq:tentativeVel} we also need to simultaneously determine the jump in the velocity field given by
\begin{equation}
      [\vec{u}_s] = \frac{1}{2\beta}\vec{P}\left(\nabla \vec{u}^{+,*} + \nabla^T \vec{u}^{+,*} + \eta\left(\nabla \vec{u}^{-,*} + \nabla^T \vec{u}^{-,*}\right)\right)\cdot \vec{n}.
\end{equation}
By rearranging these two equations we obtain a pair of coupled differential equations:
\begin{subequations}
  \begin{align}
  &\frac{\rho}{\Delta t}\vec{u}^{\pm,*} - \frac{1}{\Re}\nabla \cdot (\mu \nabla \vec{u}^{\pm,*}) = \frac{\rho}{\Delta t}\vec{u}_d^{\pm,n} -\nabla p^n + \frac{1}{\Re}\nabla \cdot (\mu \nabla^T \vec{u}^{\pm,n}) + \vec{f}^n  \quad & \textnormal{in} \quad \Omega,
  \label{eq:slipRearrangedGammaA}
   \\
  &[\vec{u}_s] -\frac{1}{2\beta}\vec{P}\left(\nabla \vec{u}^{+,*} + \nabla^T \vec{u}^{+,*} + \eta\left(\nabla \vec{u}^{-,*} + \nabla^T \vec{u}^{-,*}\right)\right)\cdot \vec{n} = \vec{0}  \quad & \textnormal{on} \quad \Gamma. \label{eq:slipRearrangedGammaB}
  \end{align}
\end{subequations}

\begin{figure}[H]
  \centering
  \includegraphics[width=0.4\textwidth]{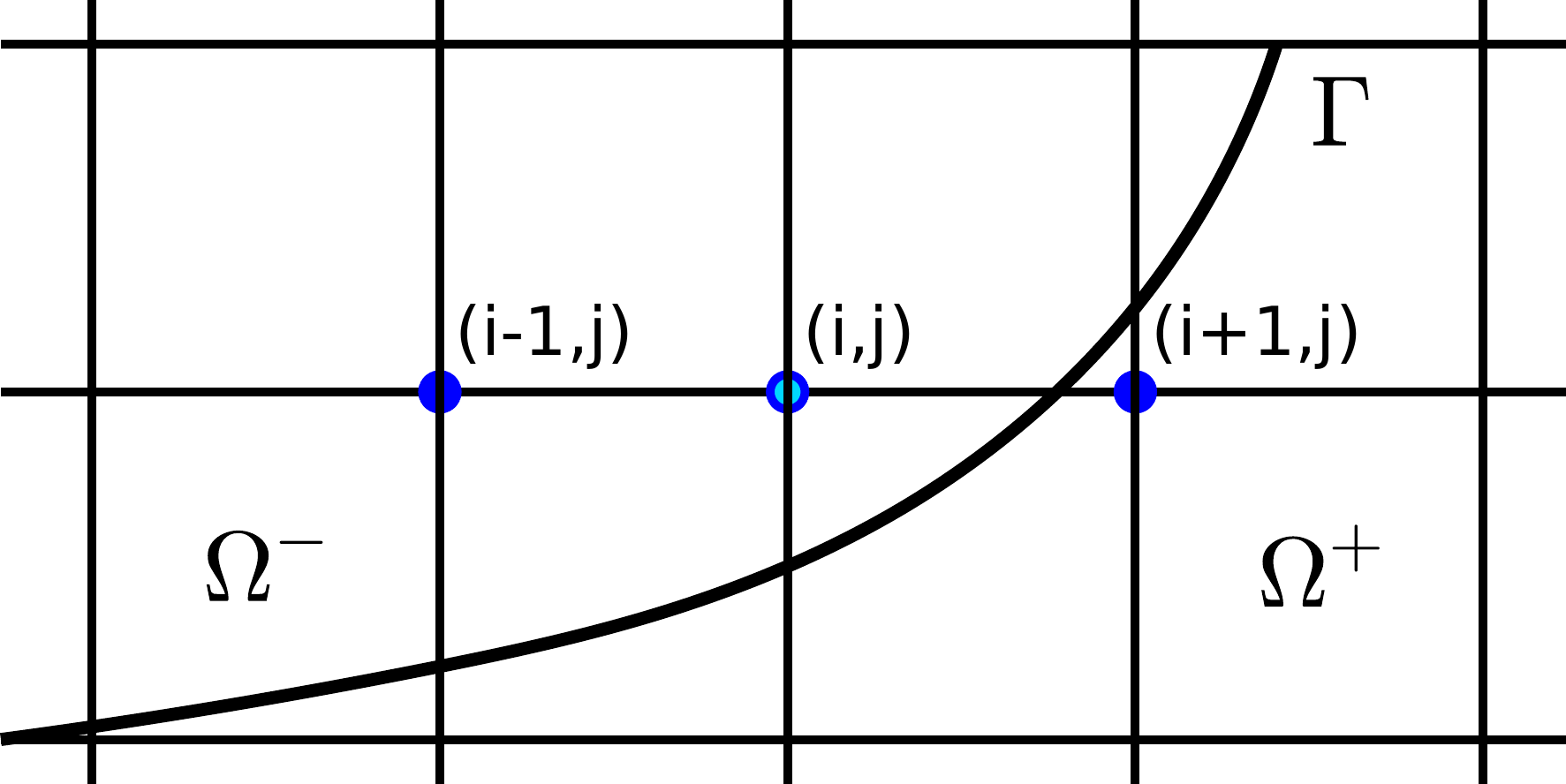}
  \caption{Sample region where the Immersed Interface Method must be used when discretizing the governing equations.}
  \label{fig:iimSimple}
\end{figure}

Discretization of the equations near the interface is handled via an Immersed Interface Method (IIM). Full details of the method are provided in \ref{app:IIM} and briefly provided here for clarity. Consider the discretization of the $x$-derivative of a discontinuous function $f$ where the interface exists within the differentiation stencil, Fig.~\ref{fig:iimSimple}. As the point of interest is located in $\Omega^-$, a proper discretization of the second-derivative would be
\begin{equation}
  \dfrac{\partial^2 f^-}{\partial x^2}\approx \dfrac{f^-_{i+1,j} - 2 f^-_{i,j} + f^-_{i-1,j}}{h^2},
\end{equation}
where $h$ is the grid spacing. The issue is that the value at location $(i+1,j)$ is in $\Omega^+$, meaning the value
$f^-_{i+1,j}$ is not directly available. Instead, assume that the jump in the function is extended from the interface such that we have $[f]_{i+1,j} = f^+_{i+1,j} -f^-_{i+1,j}$. Solving for $f^-_{i+1,j}$ and replacing it in the discretization results in the corrected equation
\begin{equation}
  \dfrac{\partial^2 f^-}{\partial x^2}\approx \dfrac{f^+_{i+1,j} - 2 f^-_{i,j} + f^-_{i-1,j}}{h^2}-\dfrac{[f]_{i+1,j}}{h^2},
  \label{eq:simpleIIM}
\end{equation}
which will be a second-order accurate approximation to the derivative. Additional details regarding the method can be found in References~\cite{Russell2003,Tan2008,Leveque1994}.

Define a function $C_r(L)$ which returns the IIM corrections needed to evaluate a linear operator $L$. Using Eq.~\eqref{eq:simpleIIM} as an example $C_r(\partial_{xx})=-[f]_{i+1,j}/(h^2)$. Note that if the interface does not cross a linear stencil then $C_r(L)=0$.

Now consider the discretization of Eq.~\eqref{eq:slipRearrangedGammaA}. The version accounting for the discontinuous velocity field is written as
\begin{align}
  \frac{\rho}{\Delta t}\vec{u}^* - \frac{1}{\Re}\nabla \cdot (\mu \nabla \vec{u}^*) - \frac{1}{\Re}C_r\left(\nabla\cdot(\mu\nabla\vec{u}^*)\right) = & \frac{\rho}{\Delta t}\vec{u}_d^n -\nabla p^n + \frac{1}{\Re}\nabla \cdot (\mu \nabla^T \vec{u}^n) + \vec{f}^n \nonumber \\
  &+ \frac{\rho}{\Delta t} C_r\left(\vec{u}_d^n\right) + \frac{1}{\Re}C_r\left(\nabla \cdot (\mu \nabla^T \vec{u}^n)\right).
  \label{eq:slipRearrangedGammaA_Corrected}
\end{align}
It is important to note that the corrections on the right-hand side of Eq.~\eqref{eq:slipRearrangedGammaA_Corrected} are known while the correction on the left-hand side is unknown as it is a function of $[\vec{u}_s]$. Additionally, it is not necessary to explicitly account for which particular velocity field (inner or outer) as that is handled via the correction function.

Next, turn to Eq.~\eqref{eq:slipRearrangedGammaB}. Here a slight modification of the correction function nomenclature is introduced. Let $C_r^+(L)$ represent the corrections necessary to evaluate a function using values from $\Omega^+$ and $C_r^-(L)$ represents the corrections necessary to evaluate a function using values from $\Omega^-$. It is now possible to state that $\nabla\vec{u}^{+,*}=\nabla\vec{u}^* + C_r^+\left(\nabla\vec{u}^*\right)$, with a similar expression for the other gradient terms. If all nodes necessary to evaluate $\nabla\vec{u}^*$ are in $\Omega^+$ then $C_r^+(\nabla\vec{u}^*)=0$. If any of the nodes are in $\Omega^-$ then $C_r^+(\nabla\vec{u}^*)\neq0$ will contain the corrections necessary to evaluate the operator. We can thus write the jump at the interface as
\begin{align}
  &-\frac{1+\eta}{2\beta}\vec{P}\left(\nabla {\vec{u}^*} + \nabla^T {\vec{u}^*}\right)\cdot\vec{n} \nonumber\\
  &\quad+ [\vec{u}_s] - \frac{1}{2\beta}\vec{P}\left(C_r^+\left(\nabla\vec{u}^*\right) + C_r^+\left(\nabla^T\vec{u}^*\right) + \eta\left(C_r^-\left(\nabla\vec{u}^*\right) + C_r^-\left(\nabla^T\vec{u}^*\right)\right)\right)\cdot\vec{n} = \vec{0},
  \label{eq:slipRearrangedGammaB_Corrected}
\end{align}
where the terms depending on the velocity (the first term) and the jump in the velocity (second and third terms) have been grouped together.

Before continuing a word needs to be said about the correction functions. It was previously stated that the jumps at the interface must be extended to grid points so that the corrections can be applied. In this work we extend the jumps in the normal direction: $[\vec{u}]_{gp} = [\vec{u}]_\Gamma + d [\vec{u}_n]_\Gamma + \tfrac{1}{2} d^2 [\vec{u}_{nn}]_\Gamma$, where $d$ is the signed distance from the grid point to the interface (equal to $\phi$ if the level set is a signed-distance function) and $[\vec{u}]_\Gamma$, $[\vec{u}_n]_\Gamma$. and $[\vec{u}_{nn}]_\Gamma$ are the velocity jump along with the first- and second-normal derivatives at the closest point on the interface to the grid point~\cite{LAI200899}. The exact forms for these are provided in  \ref{app:jump} and \ref{app:jumpAtGrid}. For the purpose of this discussion it is sufficient to state that these jumps involve linear and non-linear velocity contributions and external forces such as the surface tension. To avoid the need to solve a non-linear system, the discrete form of Eqs.~\eqref{eq:slipRearrangedGammaA_Corrected} and \eqref{eq:slipRearrangedGammaB_Corrected} are written as
\begin{equation}\label{eq:MatVec}
  \begin{bmatrix}
      \vec{H} & \vec{C}_u \\
      \vec{E} & \vec{F}
  \end{bmatrix}
  \begin{bmatrix}
      \vec{u}^* \\
      \vec{q}\\
  \end{bmatrix}
  =
  \begin{bmatrix}
      \vec{r}_u  \\
      \vec{r}_s  \\
  \end{bmatrix}
\end{equation}
where $\vec{q}=[\vec{u}_s]$ is the velocity jump at the interface for the updated iteration and
\begin{eqnarray}
  \vec{H}\vec{u}^*   & = & \frac{\rho}{\Delta t}\vec{u}^* - \frac{1}{\Re}\nabla \cdot (\mu \nabla \vec{u}^*) \textnormal{ and,} \\
  \vec{E}\vec{u}^*   & = & -\frac{1+\eta}{\beta}\vec{P}\Big((\nabla {{\vec{u}}^*} + \nabla^T {\vec{u}^*})\cdot \vec{n}\Big).
\end{eqnarray}
The contributions to the corrections that only depend on the velocity jump are contained in $\vec{C}_u\vec{q}$ and $\vec{F}\vec{q}$, noting that $\vec{F}\vec{q}$ includes the additional $[\vec{u}_s]$ term seen in Eq.~\eqref{eq:slipRearrangedGammaB_Corrected}. Any contributions to the corrections that depend on the velocity or forces are evaluated at the previous time step and stored in $\vec{r}_u$ and $\vec{r}_s$ as required. Full information regarding this split is located in \ref{app:jumpAtGrid}.

Once the tentative velocity field and velocity jump are obtained, the updated velocity field is obtained using the method outlined in Ref.~\cite{Salac2016}. The updated velocity
field is given by
\begin{equation}
  \frac{\vec{u}^{\pm,n+1} - \vec{u}^{*}}{\Delta t} =  - \nabla \psi. \label{eq:projStep}
\end{equation}
where $\psi=\tilde{\psi}+\psi_0$ represents the corrections needed for the pressure. These corrections are split into a spatially varying portion,
$\tilde{\psi}$, and a constant portion, $\psi_0$. Conceptually, $\tilde{\psi}$ enforces the divergence-free condition, $\nabla\cdot\vec{u}^{\pm}=0$,
while $\psi_0$ corrects for any total volume errors. The only difference between the method used here and the prior work is that the computed velocity jumps, $[\vec{u}_s]$,
are used to correct the divergence-free condition calculation near the interface. Once $\psi$ is determined, the updated pressure is given by $p^{n+1}=p^n + \psi$.

In this work the PETSc library~\cite{Balay1997,Balay2022,Balay2022a} is used for solution of the linear systems. The block-matrix in Eq.~\ref{eq:MatVec} is not explicitly formed and stored in memory. The $\vec{H}$ matrix is formed using standard second-order finite difference discretization while the $\vec{F}$ matrix is formed through the use of stencil composition~\cite{Mishra2022}, with the full block-matrix implemented via a matrix-free method. When numerically solving Eq.~\eqref{eq:MatVec} a preconditioner is formed by inverting the upper-triangular portion of the block matrix:
\begin{equation}
  \begin{bmatrix}
      \vec{H} & \vec{C}_u \\
      \vec{0} & \vec{F}
  \end{bmatrix}^{-1},
\end{equation}
where due to the lower dimensionality of the matrix $\vec{F}$, it is possible to use a LU-decomposition solver when solving $\vec{F}\vec{x}=\vec{y}$, while the solution to $\vec{H}\vec{x}=\vec{y}$ and Eq.~\eqref{eq:MatVec} are handled via the GMRES algorithm~\cite{Saad1986}. Note that the number of iterations necessary to solve Eq.~\eqref{eq:MatVec} heavily depends on the value of $\beta$. In particular, the preconditioner shown here works well for $\beta\gtrapprox 0.1$ except for the simplest of cases: a flat, non-moving interface. Alternative preconditioners are currently being investigated to remove this dependence on $\beta$.

Time evolution occurs using a split scheme. After the updated velocity field is obtained using the method described here the level set is advanced. To better describe the interface a gradient-augmented level set is used, whereby not only is the level set function tracked but so are is the gradient of the level set. This allows for the evaluation of interpolating polynomials with compact stencils and has shown to increase the overall accuracy of geometric quantities~\cite{Nave2010,Seibold2012}. In particular this work uses a semi-implicit version which provides additional stability compared to explicit schemes~\cite{Kolahdouz2013,Velmurugan2016}.

\section{Numerical results}\label{sec:res}
In this section, several numerical experiments on the influence of interfacial slip on the dynamics of various multi-phase systems are presented, both for validation and demonstration purposes.
For the first part of this section, a brief convergence study using a droplet in shear flow is presented. This is followed by comparing the numerical results to experimental results of high-density polymer melts~\cite{Lam2003} and theoretical results of droplets in extensional flow~\cite{Ramachandran2012}. The influence of slip on the dynamics of two immiscible fluids in Couette flow, on droplet relaxation, on droplets in shear flow, and on wall-bounded shear flow is then explored. If any parameter is not explicitly mentioned in a section then it can be assumed to be ignored.

\subsection{Convergence study}\label{sec:res_converg}
To explore the convergence of this numerical model we examine the deformation of an initially spherical droplet with initial radius $r=1$ placed in a simple shear flow with a shear rate of $\dot{\gamma}=1$ and friction coefficient $\beta=1.0$. For this example the parameters used are matched viscosity and density, $\eta=1$ and $\lambda=1$, $\Re=1.0$, and $\We=1.0$. The two-dimensional computational domain is of size $[-4, 4]\times[-2, 2]$ and the time-step is given by $\Delta t=0.4 h$, where $h$ is the grid-spacing.

\begin{figure}[H]
  \centering
  \includegraphics[width=0.5\textwidth]{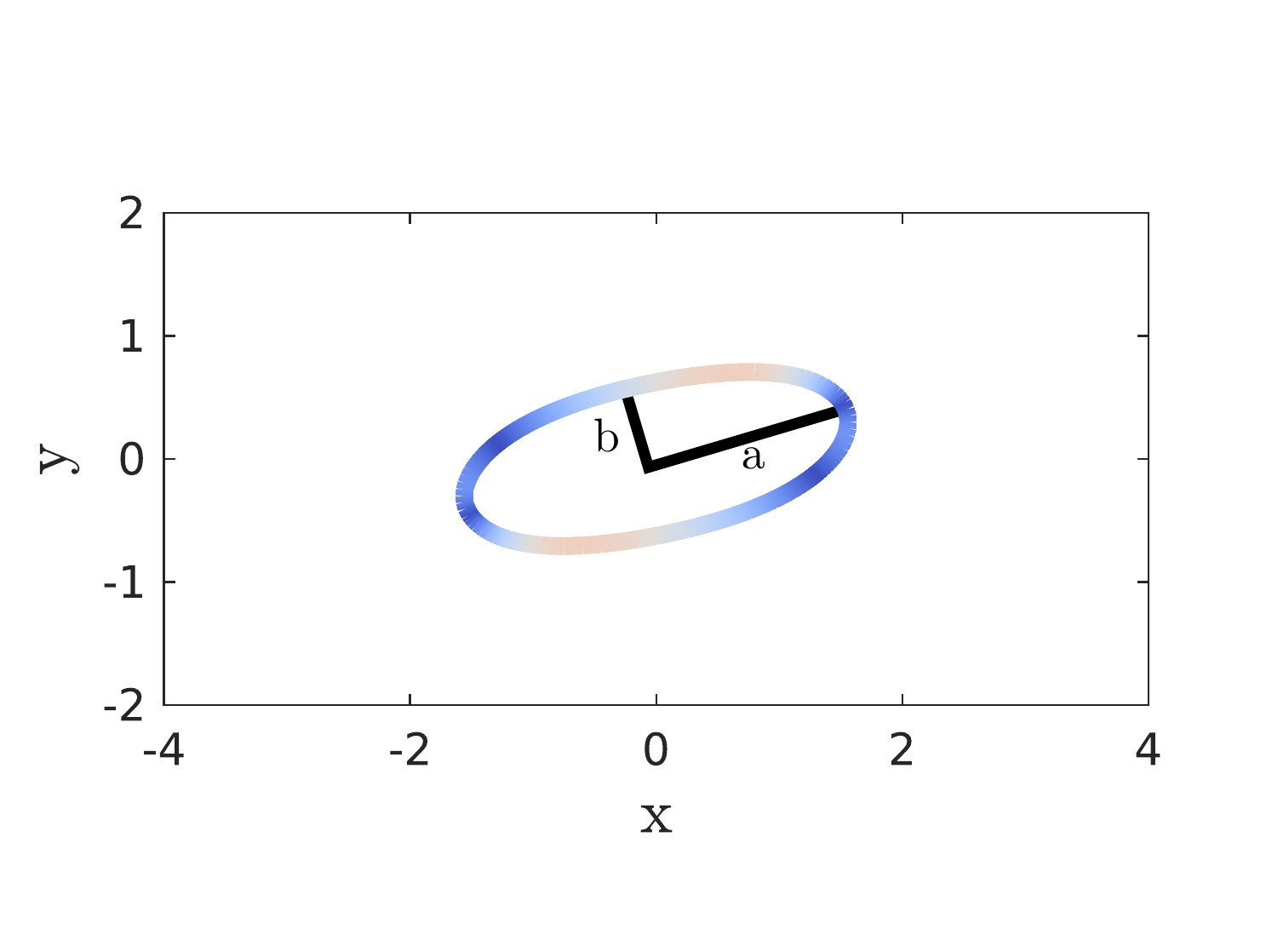}
  \caption{Equilibrium deformation of an initially spherical droplet of unit radius in shear flow, with $\dot{\gamma}=1$ and $\beta=1.0$ and a grid-size of $161\times 81$ with $\Delta t = 0.4h$ at a time of 50. The colors indicate regions of high-slip (red) and low-slip (blue). The major and minor axes used to define the deformation parameter are also shown.}
  \label{fig:convergIllust}
\end{figure}
\begin{figure}[H]
  \centering
  \begin{subfigure}[b]{0.48\textwidth}
    \includegraphics[width=\textwidth]{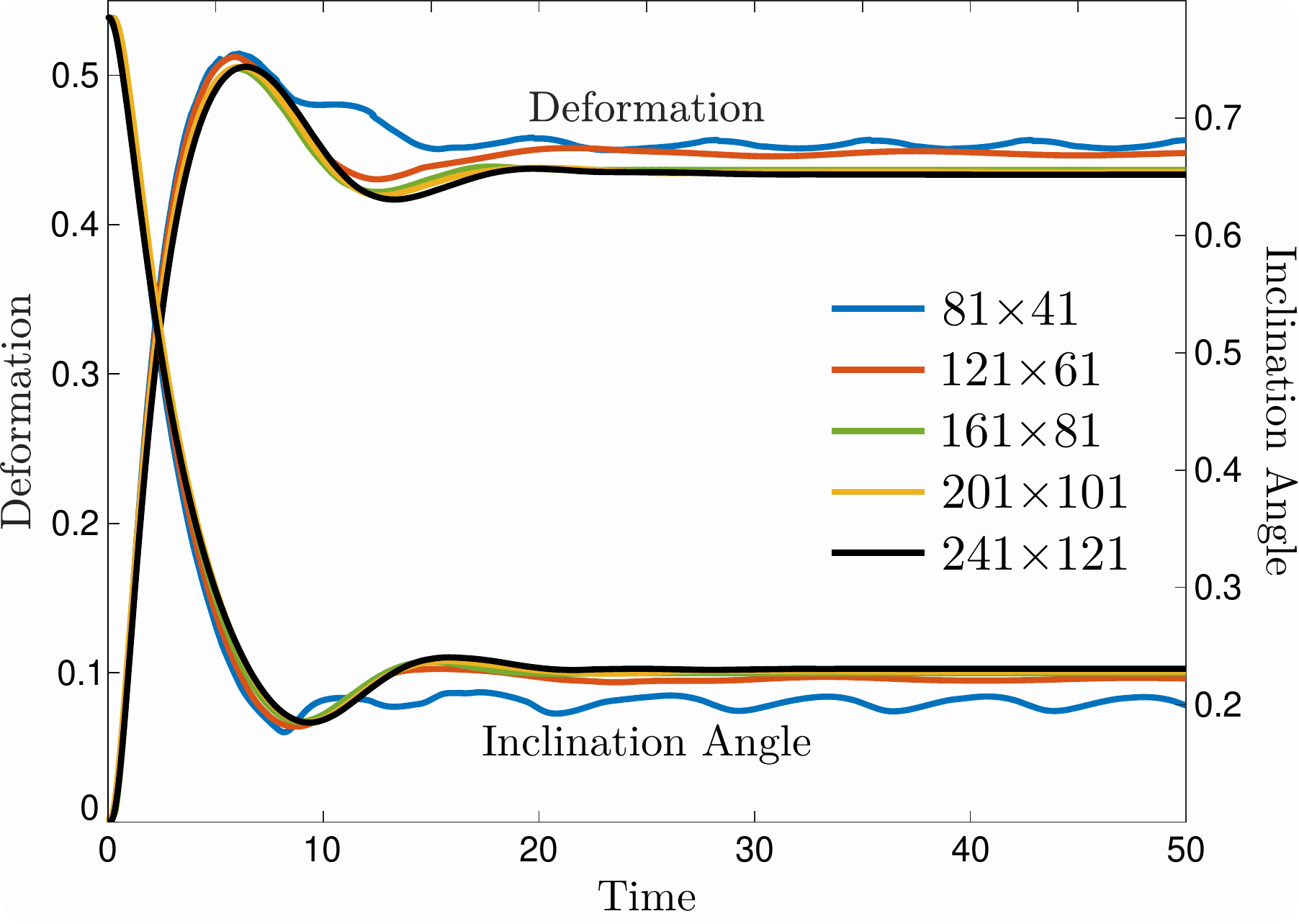}
    \caption{Deformation parameter and inclination angle.}
  \end{subfigure}
  \hfill
  \begin{subfigure}[b]{0.48\textwidth}
    \includegraphics[width=\textwidth]{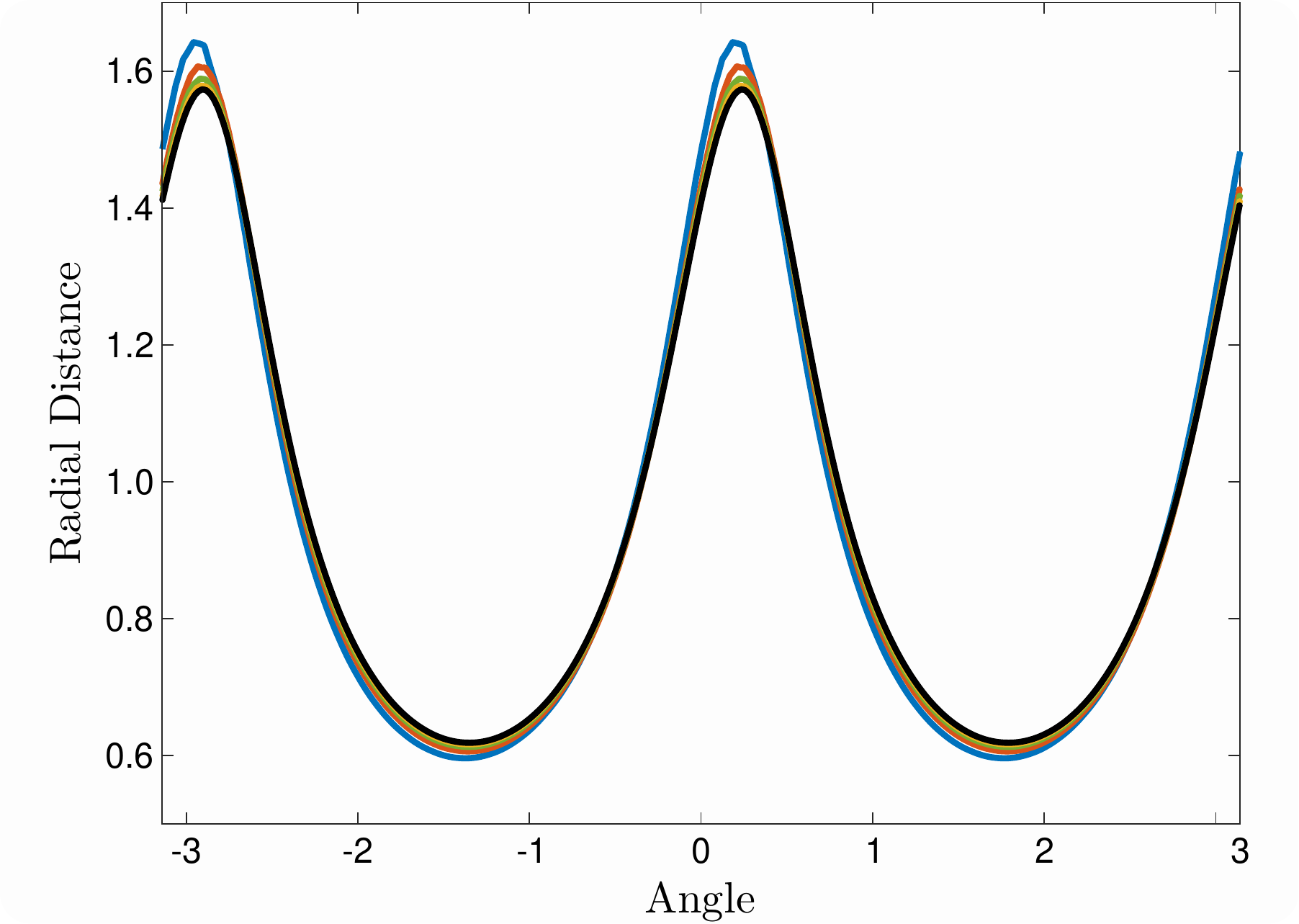}
    \caption{Radial distance from droplet center.}
  \end{subfigure}\\
  \smallskip
  \begin{subfigure}[b]{0.48\textwidth}
    \includegraphics[width=\textwidth]{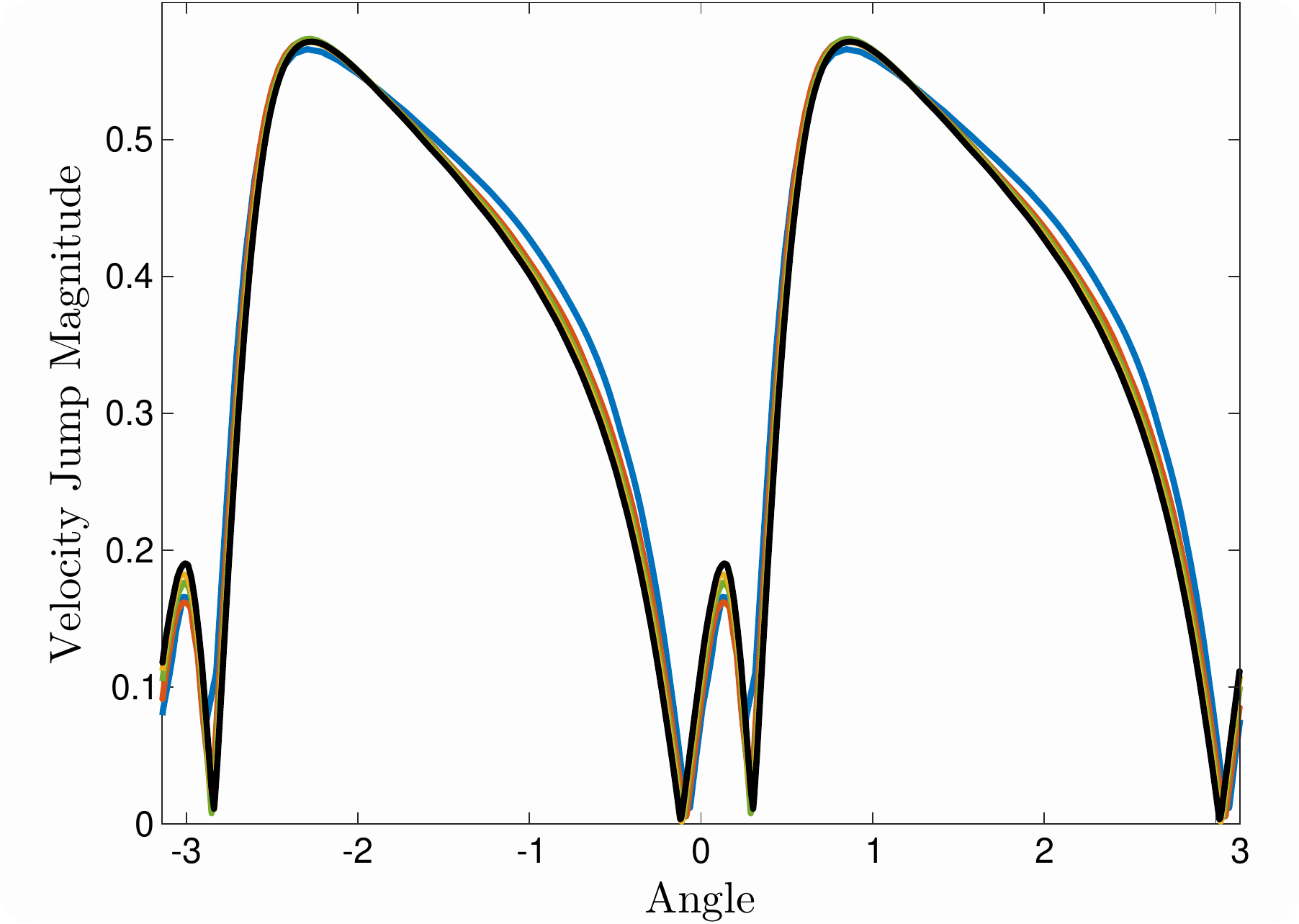}
    \caption{Magnitude of the jump in the velocity.}
  \end{subfigure}
  \caption{Convergence tests using five different grid sizes, where the ratio of the time step to the grid size is fixed and equal to $\Delta t/h = 0.4$. Figures (b) and (c) follow the same legend as (a). The angle is defined with respect to the positive $x$-direction. There are very little variations for grids larger than $161\times 81$.}
  \label{fig:convergence}
\end{figure}

First, a sample equilibrium result for a grid of $161\times 81$ is shown in Fig.~\ref{fig:convergIllust}. Due the the applied shear flow the initially spherical droplet elongates and aligns with the flow. The interfacial regions at the top and bottom of the droplet exhibit the most amount of slip, as shown by the colored interface.

We now consider the dynamics of the Taylor deformation parameter~\cite{Taylor1934} given by $D=\frac{a-b}{a+b}$, where $a$ and $b$ are the eigenvalues of the droplet's inertia/gyration tensor about it's center of mass~\cite{Messlinger2009, Laadhari2014, Salac2012}, along with the inclination angle defined as the angle of the major axis with respect to the positive $x$-axis, the radial distance between the droplet center and the interface, and the resulting magnitude of the jump in the velocity. The results can be seen in Fig.~\ref{fig:convergence} for grids ranging from $81\times 41$ to $241\times 121$. As there is very little variation for grids larger than $161\times 81$, it is decided to use grid spacings of $h\approx\mathcal{O}(0.05)$ or smaller and time steps of $\Delta t=0.4h$ or smaller for all future simulations.

\subsection{Polymer-polymer interfaces}\label{sec:res_PPI}

\begin{figure}[H]
  \centering
  \includegraphics[width=0.5\textwidth]{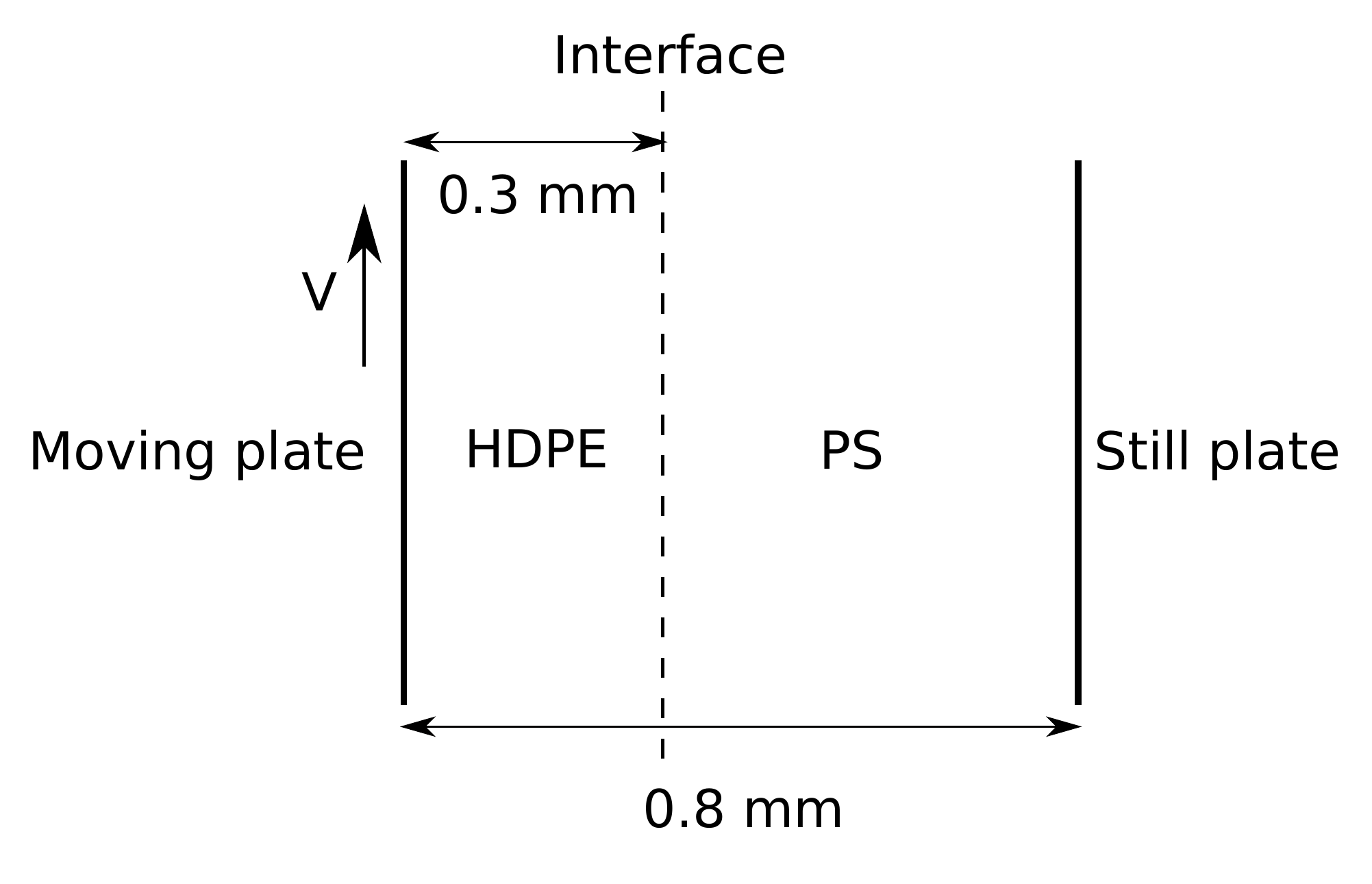}
  \caption{Sketch of the HDPE/PS experiment setup by Lam et al~\cite{Lam2003}.}
  \label{fig:PPIsketch}
\end{figure}

Next, we compare the numerical model to the experimental results presented in Fig. 7 of Lam et al~\cite{Lam2003}. In this experiment the interfacial slip between two high density polymer-melts, polyethylene (HDPE) and polystyrene (PS), is observed while subjected to a steady shear at an elevated temperature of 180$^\circ$C. A schematic of this experimental set up is shown in Fig.~\ref{fig:PPIsketch}, showing the interface at $0.3$ mm while the gap distance between the two plates is $0.8$ mm and the applied shear rate is $0.5$ s$^{-1}$. Material properties were normalized with respect to the PS properties, resulting in a viscosity ratio of $\eta=0.07761$ and density ratio of $\lambda=0.78125$.

\begin{figure}[H]
  \centering
  \includegraphics[width=0.8\textwidth]{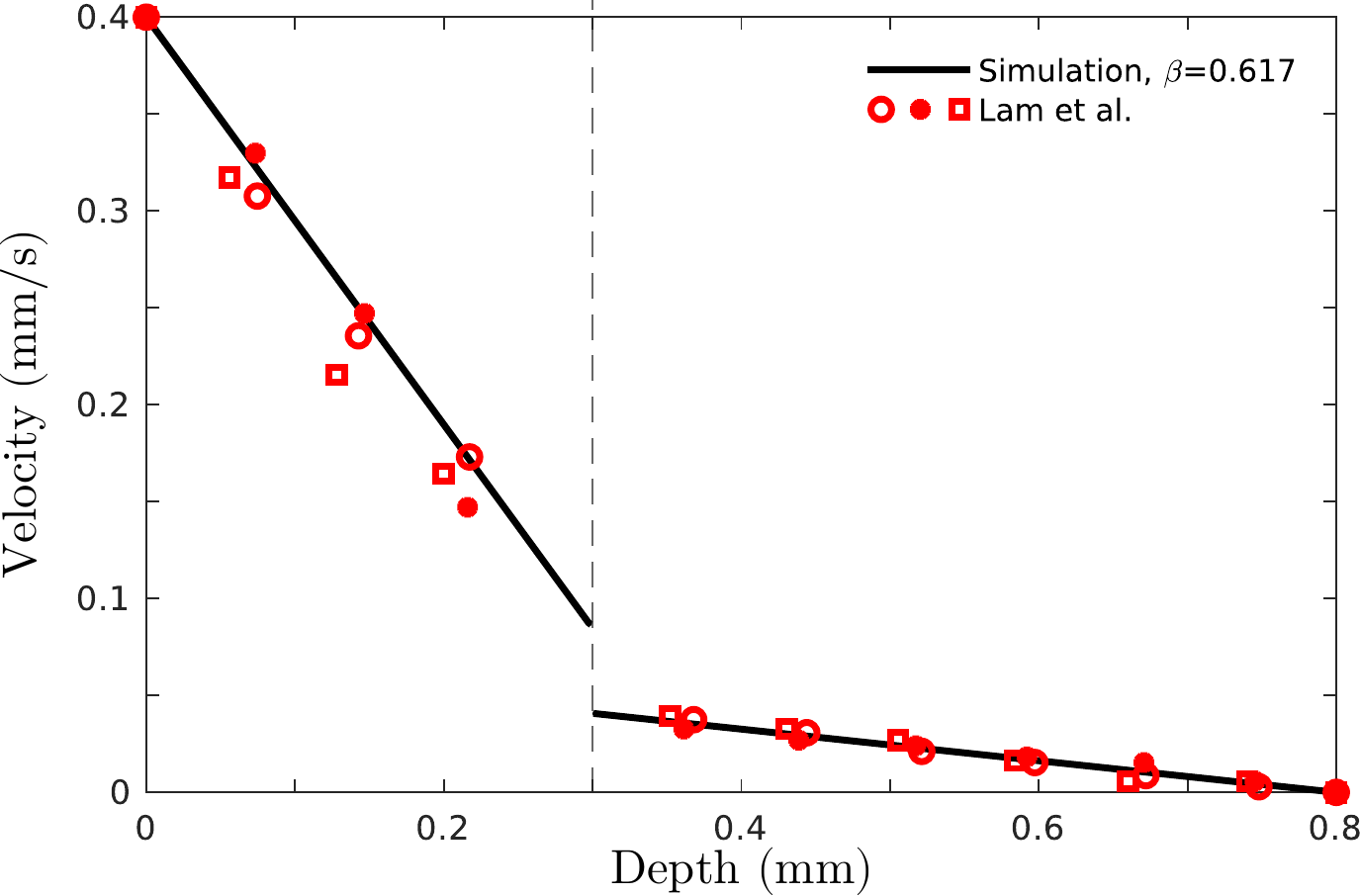}
  \caption{Comparison of the velocity profiles of HDPE/PS in Couette flow with experiments by Lam et al~\cite{Lam2003}, where the shear rate and the friction coefficient are $\dot{\gamma}=0.5\ s^{-1}$ and $\beta=0.617$, respectively. The computational results are shown as the solid line while the symbols represent the three published experiments. }
  \label{fig:PPI_C2}
\end{figure}

To compare to experimentally determined results we utilize the relationship $b=\mu_{\Gamma}/d_{\Gamma}$, where $\mu_{\Gamma}$ is the interfacial viscosity and $d_{\Gamma}$ is the interfacial thickness~\cite{Leal2007,Ramachandran2012}. Note that in Ref.~\cite{Ramachandran2012} they use the inverse of this, calling it the slip coefficient. The interfacial thickness between a HDPE and PS mixture has been calculated to be approximately 21\AA~\cite{Jiang2005, Jiang2003}. The interfacial viscosity can be approximated by requiring that the tangential shear stress to be the same across all three layers. Therefore $\mu_{\Gamma}\dot{\gamma}_{\Gamma} = \mu_T\dot{\gamma}_T$, where $\dot{\gamma}_{\Gamma}$ is the shear rate in the interfacial layer while $\mu_T$ and $\dot{\gamma}_T$ are the overall (effective) viscosity and shear rate, respectively. The interfacial shear rate is provided by using the experimentally determined slip velocity, 0.036 mm/s and the interfacial thickness, resulting in $\dot{\gamma}_{\Gamma}\approx1.71\times 10^{4}$ s$^{-1}$. Using a viscosity of $\mu_T=1967$ Pa\;s~\cite{Lam2003} and applied shear rate of $\dot{\gamma}_T=0.5$ s$^{-1}$ this results in an interfacial viscosity of $\mu_{\Gamma}=5.74\times 10^{-2}$ Pa\;s. Finally, the friction coefficient is determined to be $b\approx 2.7\times 10^{7}$ Pa\;s/m. Using a characteristic length of $a=0.25$ mm and the viscosity of PS, estimated to be 11000 Pa\;s at the given temperature and shear rate, this results in a dimensionless friction coefficient approximately equal to $\beta=0.617$. Using this friction coefficient, along with a Reynolds number of $\Re=10^{-4}$ with a non-dimensional 2D domain of $[-1.2, 2]\times[-1.2,2]$, a grid size of $512 \times 512$, and a time step of $\Delta t=10^{-3}$ the system is modeled until equilibrium. The result, after mapping the simulation results back to dimensional quantities show excellent agreement with the experimental results, Fig.~\ref{fig:PPI_C2}. The jump at the interface from the simulation is determined to be 0.04 mm/s, very close to the experimentally approximated result.

\subsection{Two immiscible fluids in Couette flow}\label{sec:LLI}
In this section we explore how the presence of interfacial slip affects the dynamics of two liquids in simple Couette flow.
In all of these simulations, the 2D domain is of the size $[-2,2]^2$ with a $64\times 64$ grid and a time step of $\Delta t = 0.01$. Wall boundary conditions are held in the $y$-direction while periodicity is assumed in the $x-$ direction. The interface is located at the center of the domain at $y=0$, the Reynolds number is $\Re = 0.5$, the applied non-dimensional shear rate is $\dot{\gamma}=0.5$, and both fluids have matched viscosity and density, $\eta=1$ and $\lambda=1$.

\begin{figure}[H]
  \centering
  \begin{subfigure}[t]{0.495\textwidth}
    \centering
    \includegraphics[width=\textwidth]{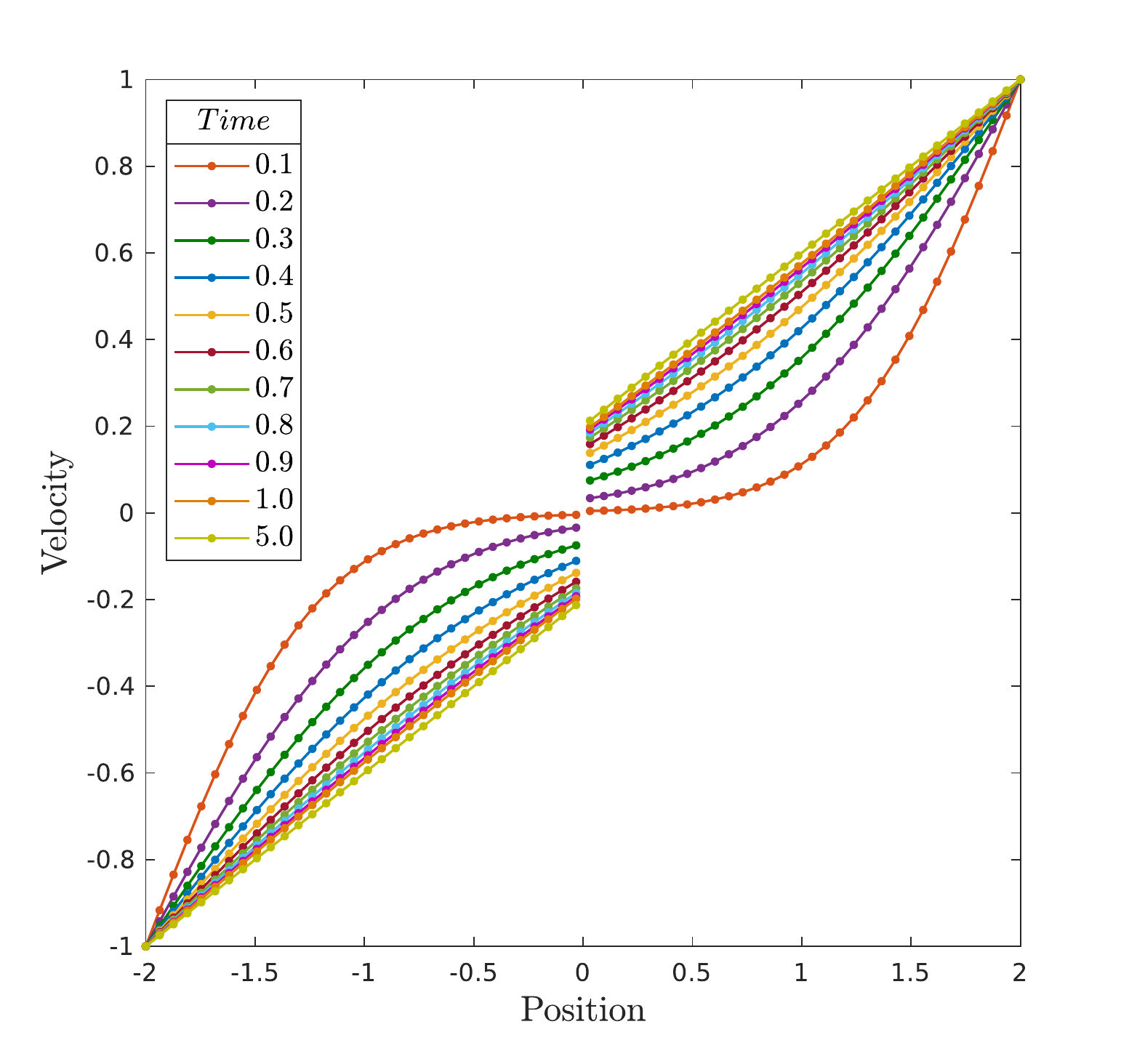}
    \caption{The velocity profile along the $x=0$ line over time.}
  \end{subfigure}
  \hfill
  \begin{subfigure}[t]{0.495\textwidth}
    \centering
    \includegraphics[width=\textwidth]{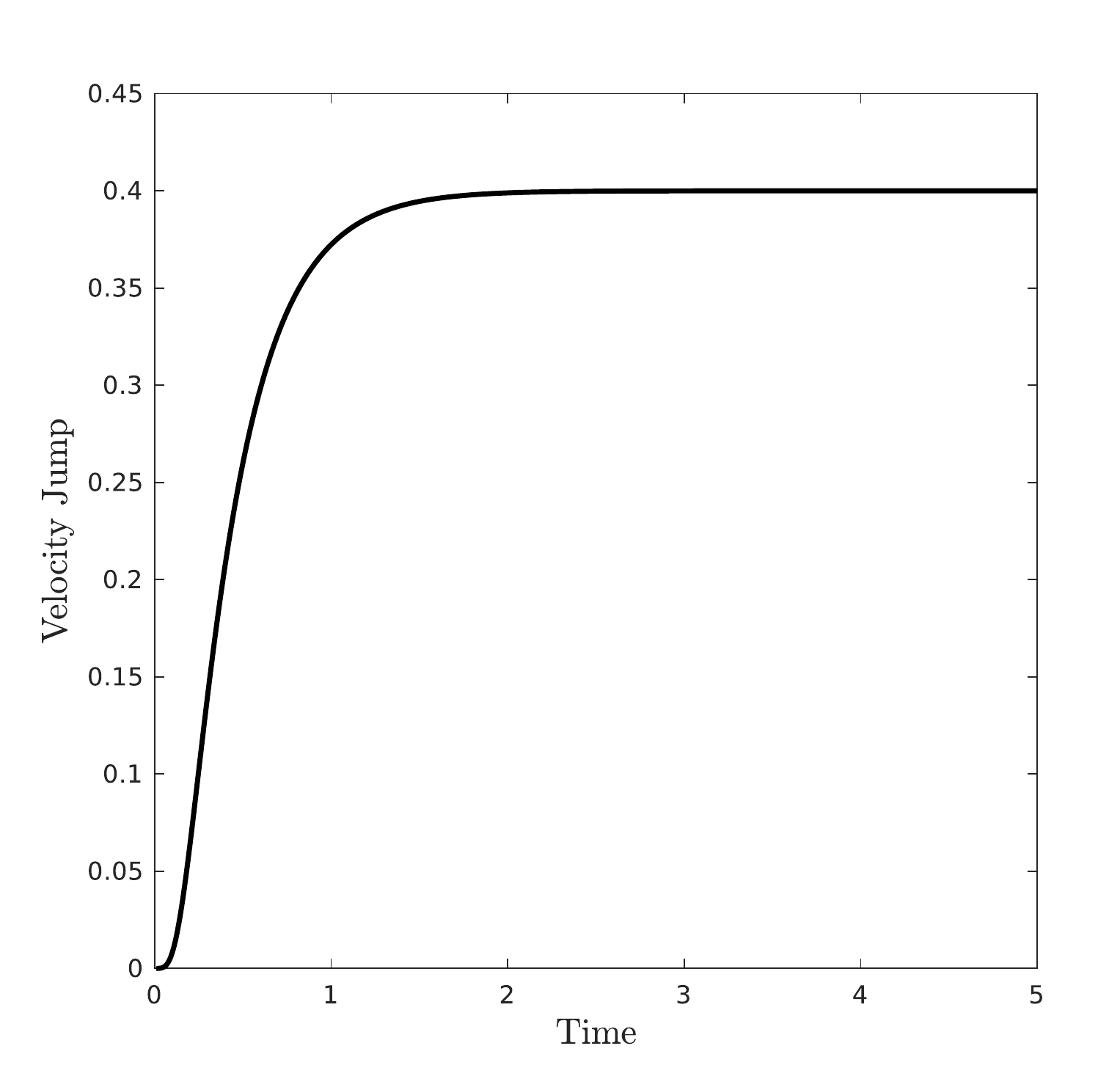}
    \caption{The jump magnitude as a function of time.}
  \end{subfigure}
  \caption{Evolution of the velocity field in time in presence of slip at a liquid-liquid interface with $\beta=1.0$.}
  \label{fig:u_time}
\end{figure}

\begin{figure}[H]
  \centering
  \begin{subfigure}[t]{0.495\textwidth}
    \centering
    \includegraphics[width=\textwidth]{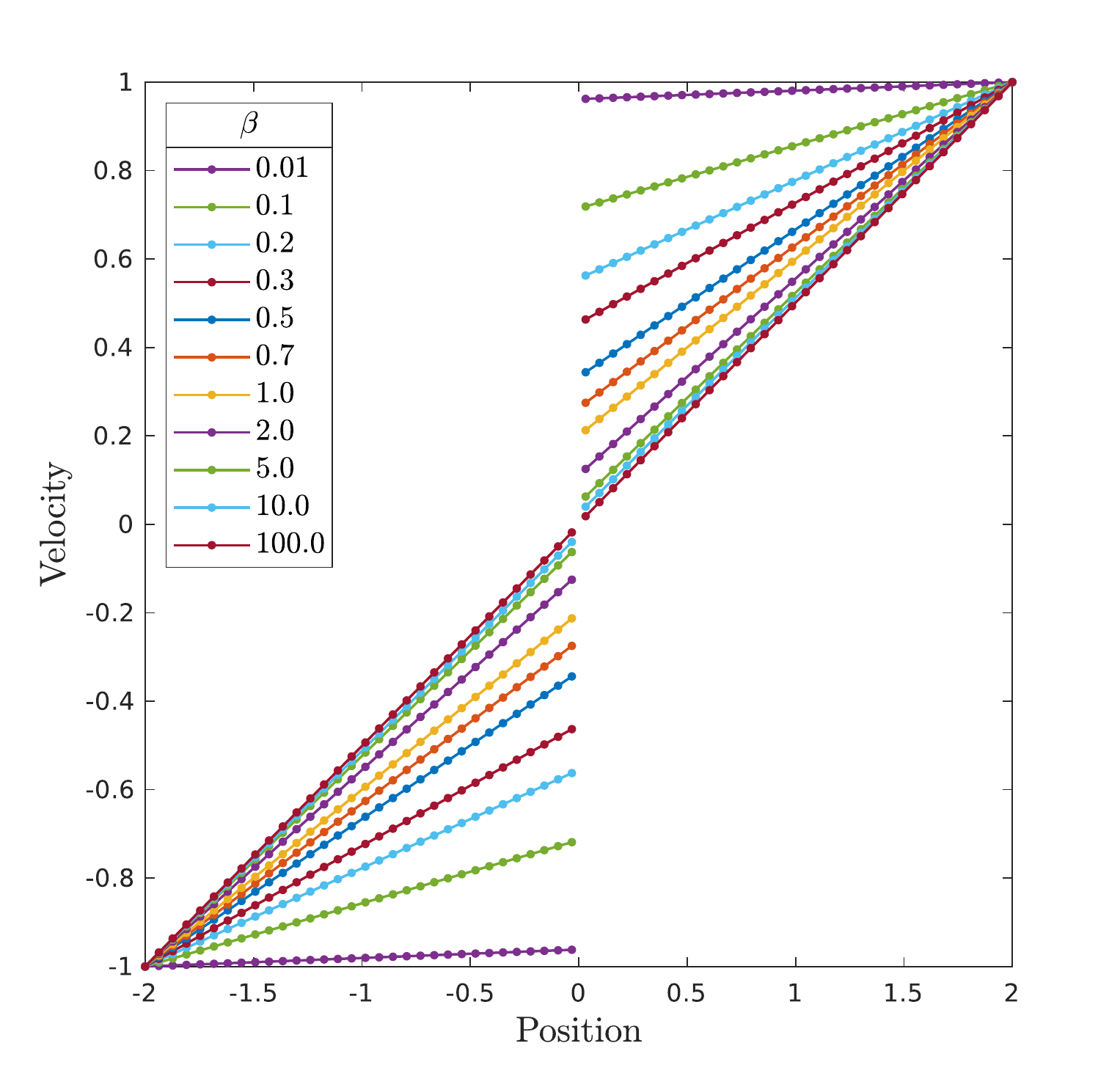}
    \caption{The velocity profile along the $x=0$ line as $\beta$ varies.}
  \end{subfigure}
  \hfill
  \begin{subfigure}[t]{0.495\textwidth}
    \centering
    \includegraphics[width=\textwidth]{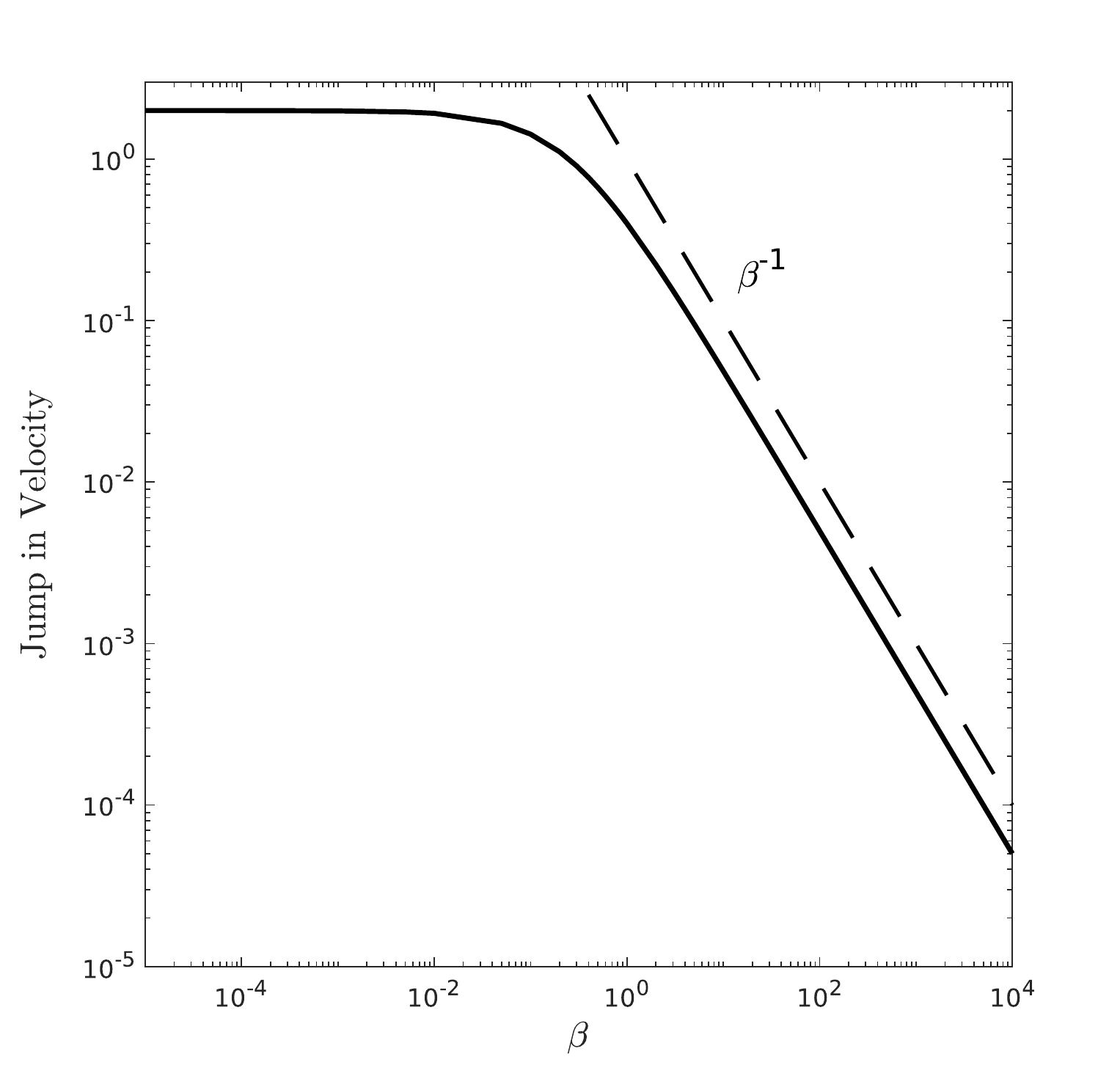}
    \caption{The magnitude of the velocity jump as a function of the friction coefficient. As expected the magnitude scales at $\beta^{-1}$.}
  \end{subfigure}
  \caption{The effect of the friction coefficient on the equilibrium velocity field at a liquid-liquid interface in simple Couette flow.}
  \label{fig:u_beta}
\end{figure}

The evolution of the velocity profile in the $y$-direction as a function of time for $\beta=1$ is presented in Fig.~\ref{fig:u_time} while the influence of the friction coefficient on the equilibrium velocity profile is shown in Fig.~\ref{fig:u_beta}. Due to the Reynolds number, initially the velocity discontinuity is small. As the wall-boundary effects influence the center of the domain the jump in the velocity grows until reaching a steady-state value. In this particular example the jump in the velocity grows quickly, reaching it's maximum value of 0.4. Due to the matched viscosity the effective shear rate in the two fluids is the same, unlike the example shown in Sec.~\ref{sec:res_PPI}.

The influence of the friction coefficient has two general regimes, as seen in Fig.~\ref{fig:u_beta}. For moderate to large values of $\beta$ the jump in the velocity scales as $\beta^{-1}$, which is to be expected due to it's form in Eq.~\eqref{eq:tangVelJumpSimp}. It is suspected that this will hold even for cases with a viscosity difference, but further investigations are needed to determine if this is true. Small values of the friction coefficient ($\lessapprox 10^{-2}$) approach the free-slip condition with the velocity jump approaching two.

\subsection{Droplets in extensional flow}\label{sec:dropExtensionalFlow}
To further examine the validity of the present model, a comparison with the numerical results of Ramachandran et al~\cite{Ramachandran2012} for droplets in uni-axial extensional flow under creeping-flow conditions is presented in this section. Specifically, the effect of interfacial slip as well as the Capillary number on the deformation of droplets submerged in a fluid is investigated. In this section only, in order to perform a better comparison with the axisymmetric simulations provided by Ramachandran the presented results by our model are obtained via performing 3D simulations. The computational domain is $[-8,8]^3$ with wall-boundary conditions in each direction, while the grid is $161^3$ and the time step is $\Delta t=10^{-2}$. The uni-axial flow is obtained via a boundary condition of $\vec{u}_{b}(x,y,z) = \dot{\epsilon}(-\frac{1}{2}x, -\frac{1}{2}y, z)$, where $\dot{\epsilon}$ is the flow strength which can be used to determine the characteristic time scale as $t_0 = 1/\dot{\epsilon}$. It is also assumed that the density and viscosity between the inner and outer fluids is matched. Therefore, the Capillary number can be written as $\Ca = (\dot{\epsilon}\mu^+a)/(\sigma)$, where $\sigma$ is the surface tension of the droplet. An increase in $\Ca$ for a given droplet corresponds to stronger flow strength. It must be mentioned that the non-dimensional parameter for surface tension that was presented earlier in this work as $\We$, is equivalent to Reynolds number multiplied by this Capillary number $\Ca$, in other words, $\We = \Re \times \Ca$.

\begin{figure}[H]
  \centering
  \includegraphics[width=0.8\textwidth]{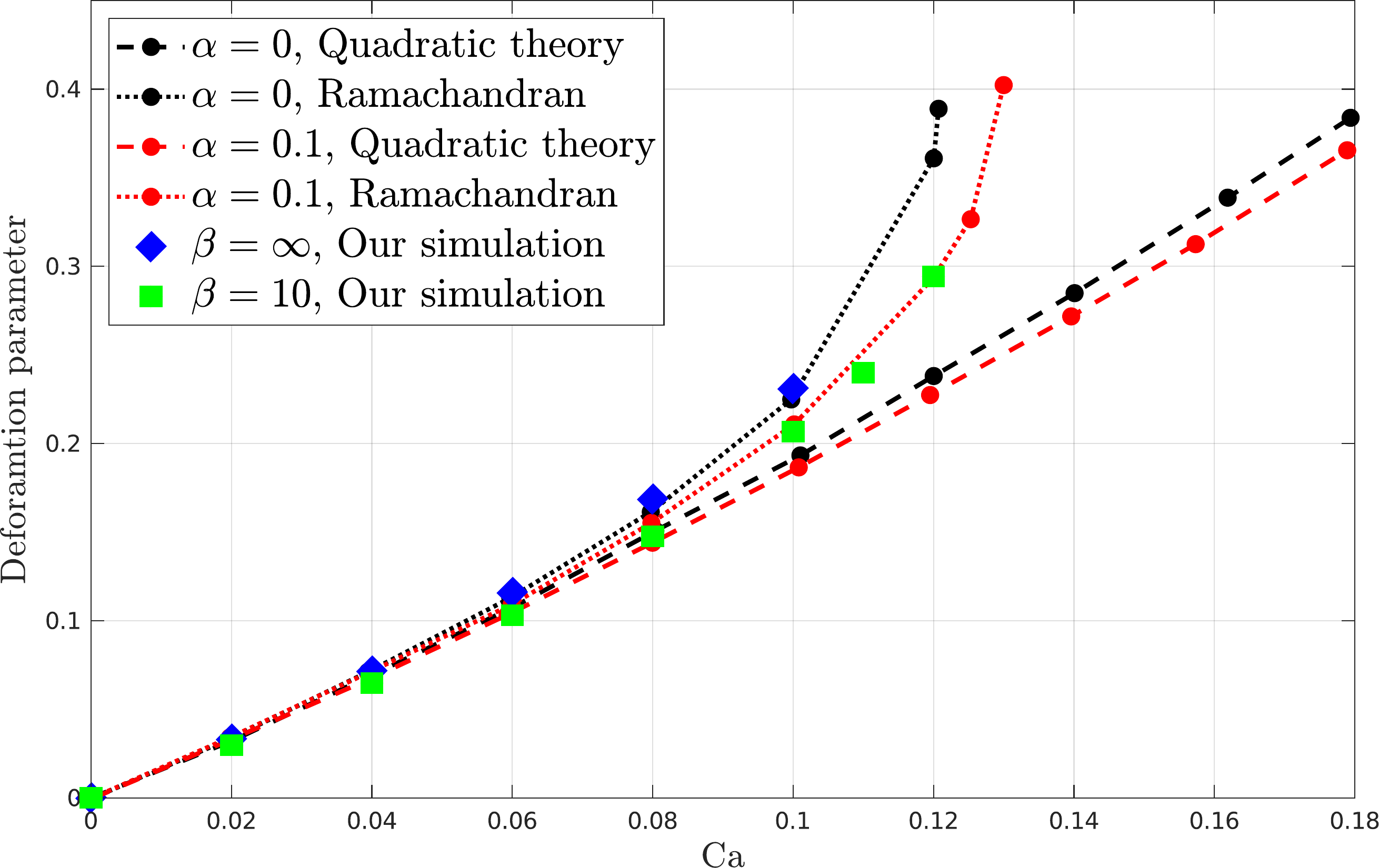}
  \caption{Deformation parameter vs capillary number ($\Ca$) for droplets in extensional flow, with and without slip and a viscosity ratio of $\eta=1$. The numerical results and results from linear theory by Ramachandran et al~\cite{Ramachandran2012} are shown as dotted and dashed lines, respectively. The black and red colors represent the friction coefficients of $\beta = \infty \ (\alpha = 0)$, and $\beta = 10 \ (\alpha = 0.1)$, respectively. Our numerical results are shown by blue diamonds and green squares for the same friction coefficients of $\beta = \infty$ and $\beta = 10$.}
  \label{fig:extensionDrop}
\end{figure}

An initially spherical droplet of radius equal to 1 is placed in the center of the domain and allowed to evolve until reaching a steady-state. Once at steady-state the Taylor deformation parameter as described in Sec.~\ref{sec:res_converg} is calculated. As shown in Fig.~\ref{fig:extensionDrop} and noting that Ramachandran defines a slip parameter rather than a friction coefficient that is given by $\alpha=\beta^{-1}$, our results are in excellent agreement with the numerical results presented by Ramachandran et al~\cite{Ramachandran2012}. As expected, the simulations predict that deformation decreases as the friction coefficient $\beta$ decreases, which results in stronger slip on the interface. We can also observe that as the Capillary number increases, meaning the applied flow becomes stronger compared to surface tension, the results diverge further away from the linear theory, as that theory is valid for smaller deformations.

\subsection{Droplet relaxation}\label{sec:relax}
\begin{figure}[H]
    \centering
    \begin{subfigure}[b]{0.475\textwidth}
        \centering
        \includegraphics[width=\textwidth]{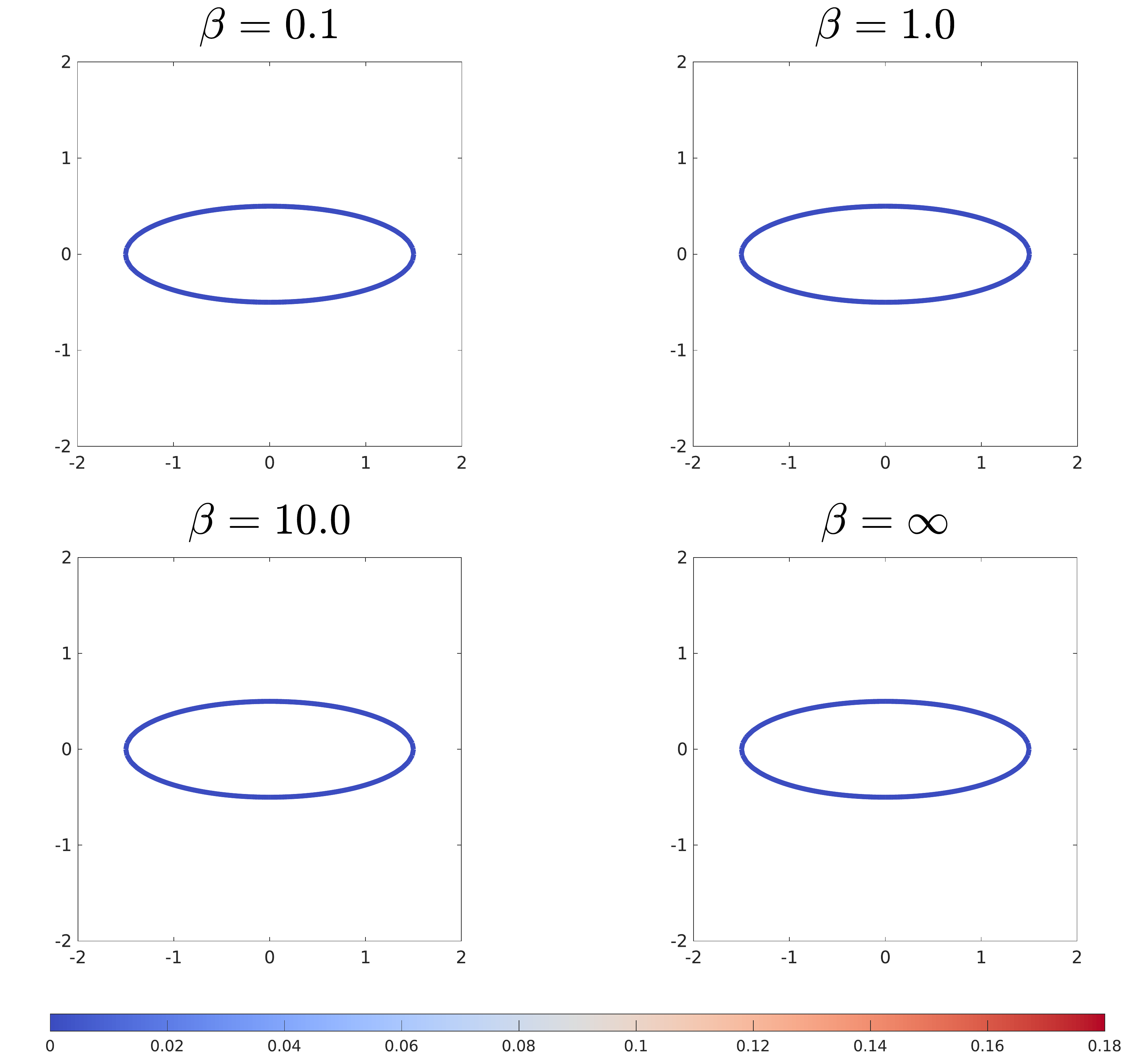}
        \caption[]%
        {{\small Time=0}}
        \label{fig:Relax0}
    \end{subfigure}
    \hfill
    \begin{subfigure}[b]{0.475\textwidth}
        \centering
        \includegraphics[width=\textwidth]{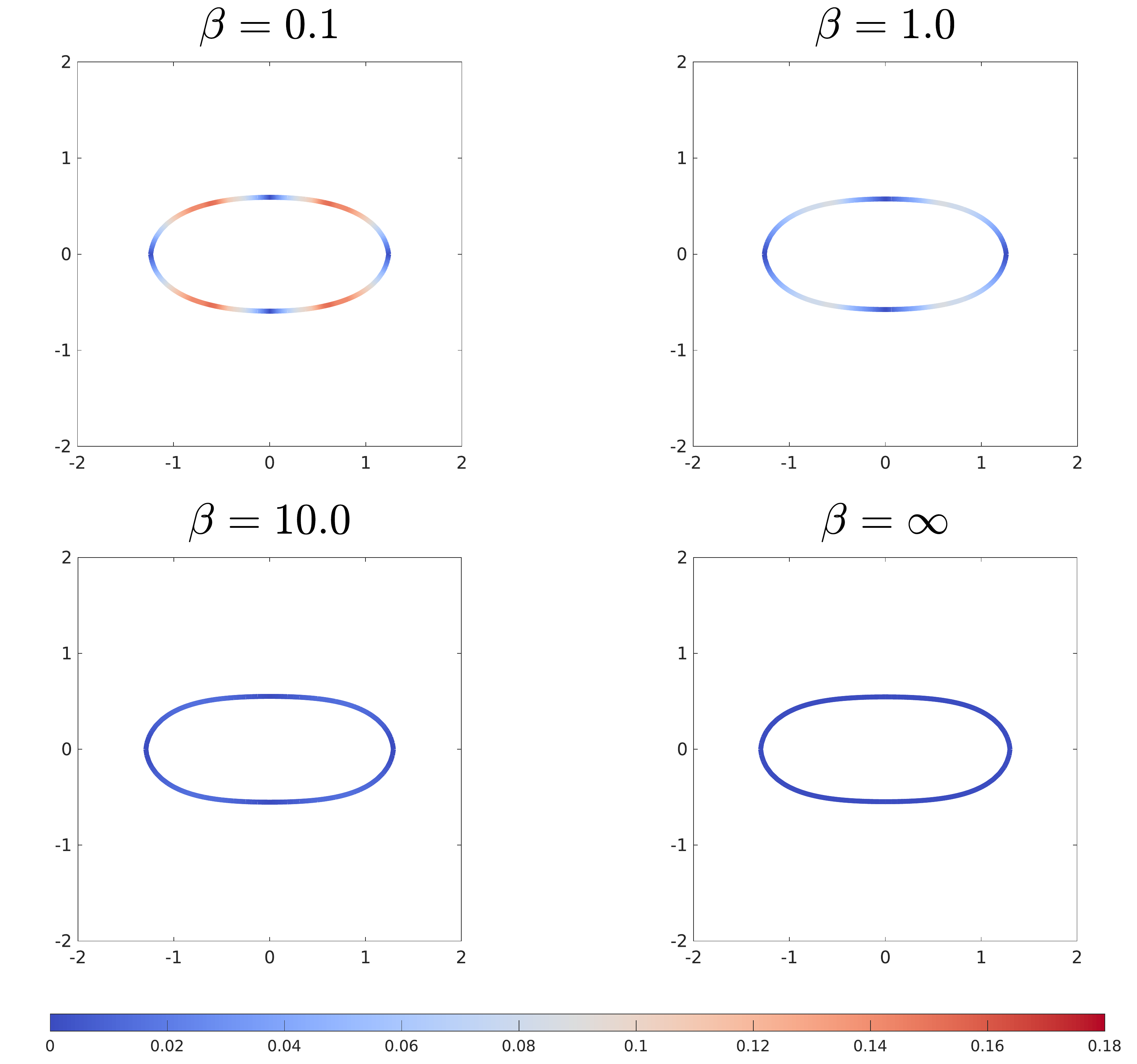}
        \caption[]%
        {{\small Time=2}}
        \label{fig:Relax20}
    \end{subfigure}
    \vskip\baselineskip
    \begin{subfigure}[b]{0.475\textwidth}
        \centering
        \includegraphics[width=\textwidth]{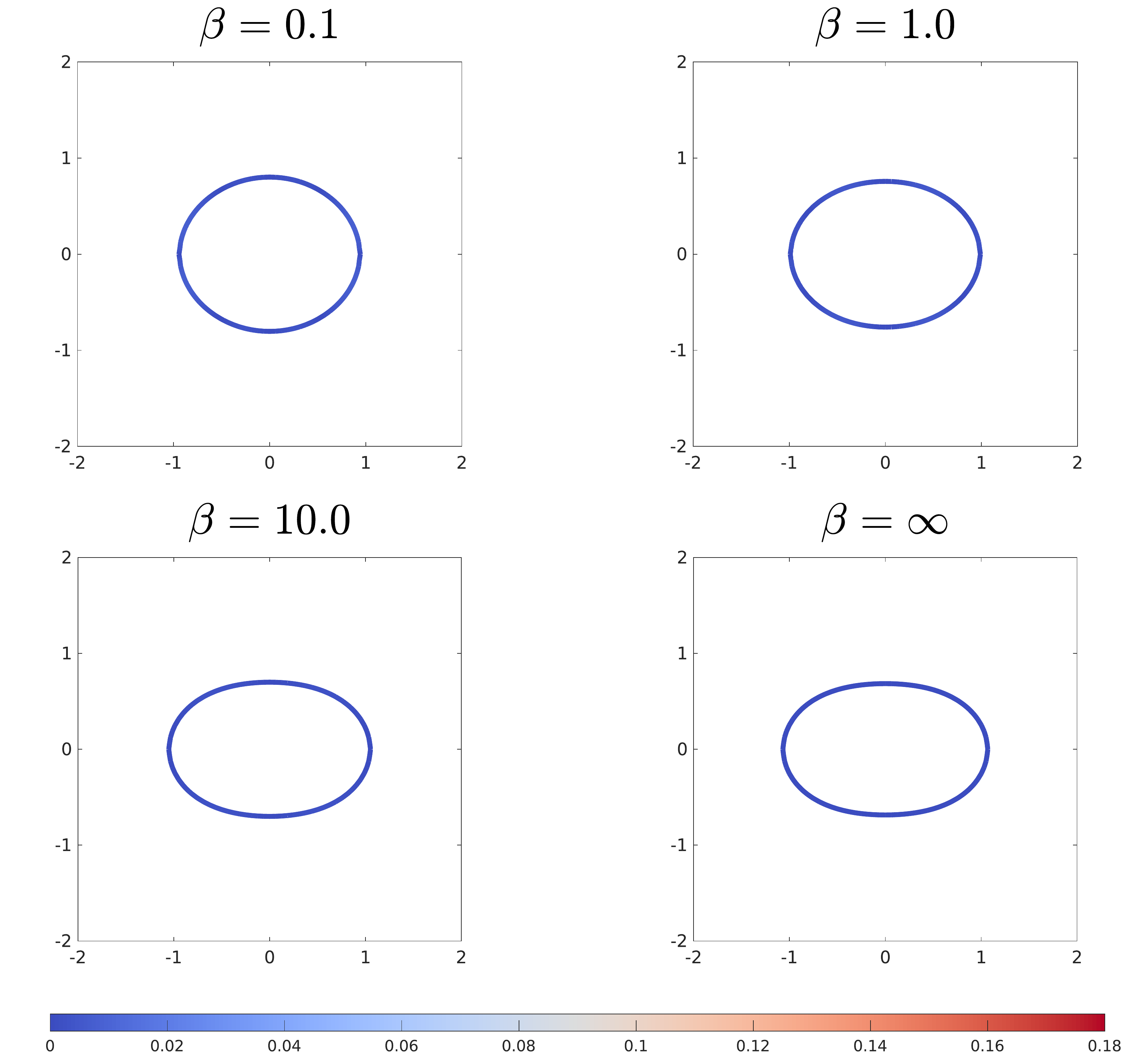}
        \caption[]%
        {{\small Time=4}}
        \label{fig:Relax40}
    \end{subfigure}
    \hfill
    \begin{subfigure}[b]{0.475\textwidth}
        \centering
        \includegraphics[width=\textwidth]{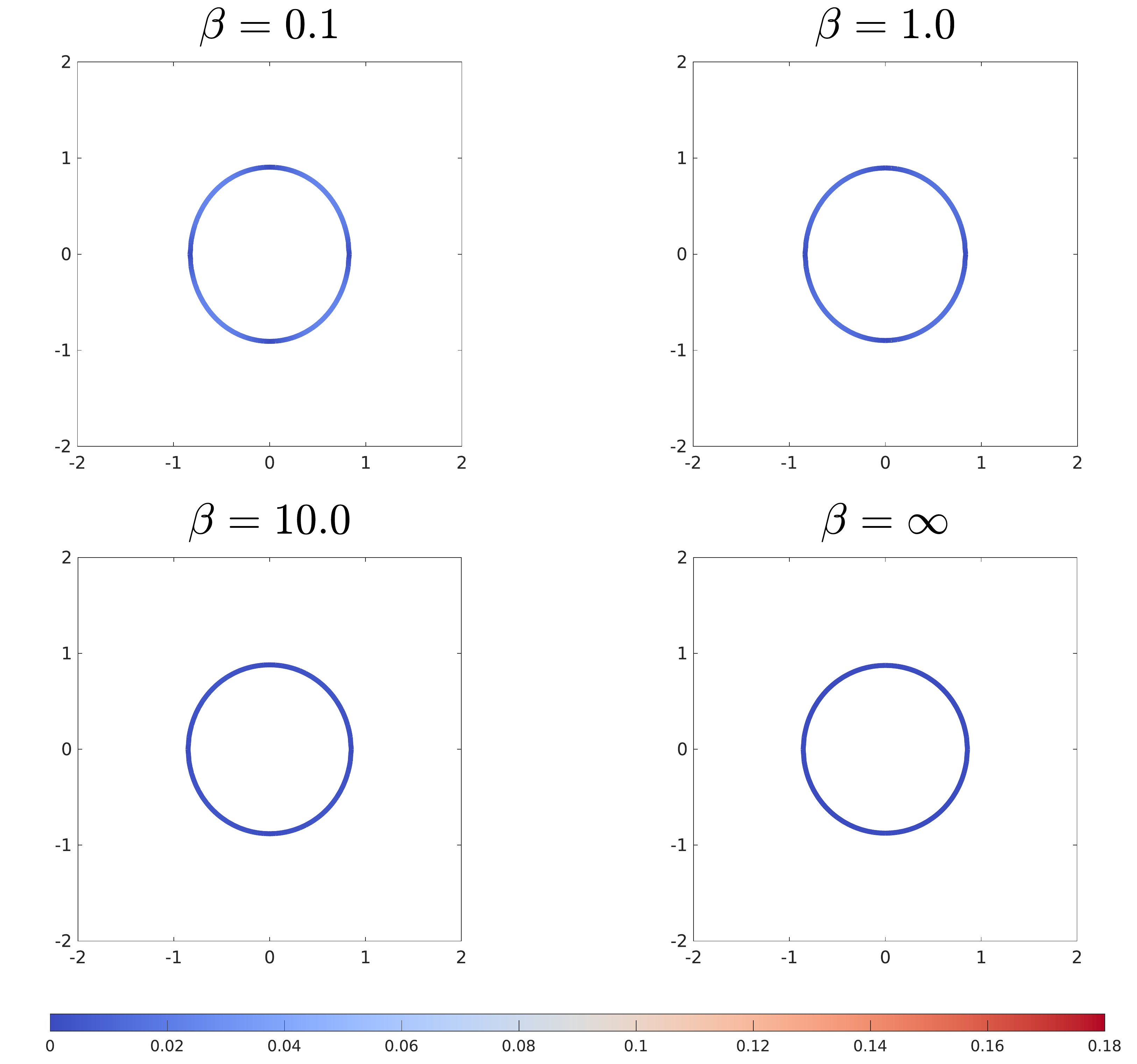}
        \caption[]%
        {{\small Time=7}}
        \label{fig:Relax70}
    \end{subfigure}
    \caption[  ]
    {Relaxation of initially elliptical droplets in absence of flow for different friction coefficients $\beta = 0.1, 1, 10, \infty$, at times of $0, 2, 4, 7$. The magnitude of the jump in velocity is indicated for each simulation.}
    \label{fig:relaxationFSF}
\end{figure}

\begin{figure}[H]
  \centering
  \includegraphics[width=0.6\textwidth]{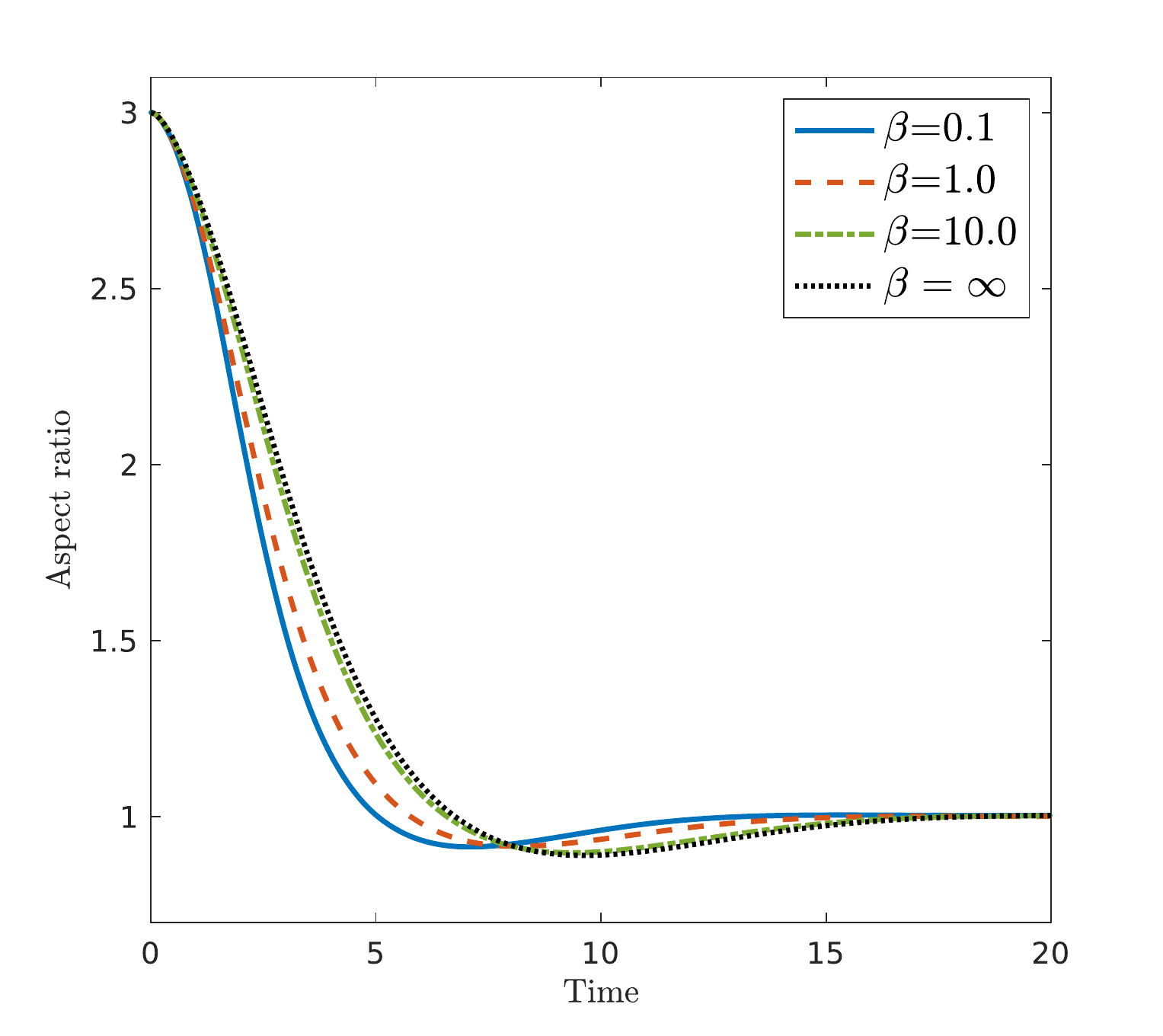}
  \caption{The effect of slip on relaxation of droplets for different friction coefficients of $\beta=0.1,1,10,\infty$ which are shown by solid, dashed, dotted-dashed, and dotted lines, respectively, where $\beta=\infty$ represents the no slip case. All of the droplets have the same initial aspect ratio of $3$, with $\Re=10$, $\We=10$, and matching density and viscosity for inner and outer fluids.}
  \label{fig:relaxationTime}
\end{figure}

We next consider the influence of slip on the relaxation of initially elliptical droplets and on long filaments. For the first case consider an initially elliptical droplet with axis lengths of 1.5 and 0.5 in a domain with a size of $[-4,4]^2$ with matched viscosity and density, Fig.~\ref{fig:Relax0}. We consider four friction coefficients: $\beta=0.1, 1, 10, \textnormal{ and } \infty$, a Reynolds number of $\Re=10$ and Weber Number of $\We=10$. The results for a $129\times 129$ grid and a time step of $\Delta t=10^{-2}$ are shown in Fig.~\ref{fig:relaxationFSF}, including snapshots of the jump in velocity on the interface. Droplets with smaller friction coefficients (larger slip) demonstrate faster relaxation dynamics, which can be confirmed by examining the aspect ratio, defined as the axis length in the $x$-direction divided by that in the $y$-direction, of the droplet over time, Fig.~\ref{fig:relaxationTime}. In all cases the droplets have an inversion of the shape, from having the long axis aligned with the $x$-axis to the $y$-axis, with this occurring earlier for the droplets with slip.

\begin{figure}[H]
  \centering
  \includegraphics[width=0.4\textwidth]{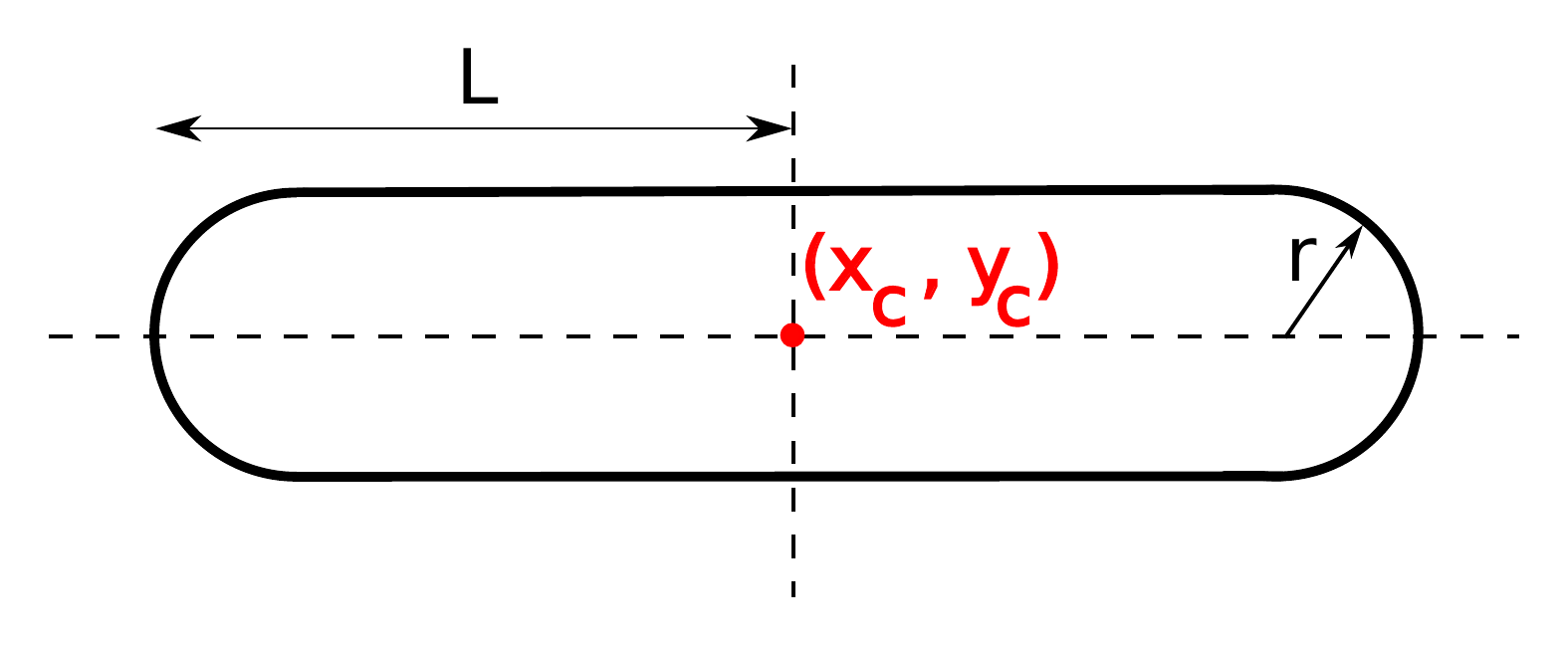}
  \caption{Schematic of a filament, where $L$ is half the total length, and $r$ is the radius of the half circles at both ends of the filament.}
  \label{fig:filament}
\end{figure}

A second set of simulations examine the relaxation of very long filaments, Fig.~\ref{fig:filament}. The filament is symmetric about both the $x-$ and $y$-axes. In a 2D domain the filament can be composed of a rectangular area centered at $(x_c, y_c)$ and length of $2L$, with caps at both ends in the shape of half circles with radius $r$. The aspect ratio for these filaments is defined as $L/r$.

\begin{figure}[H]
 \centering
 \includegraphics[height=0.88\textheight]{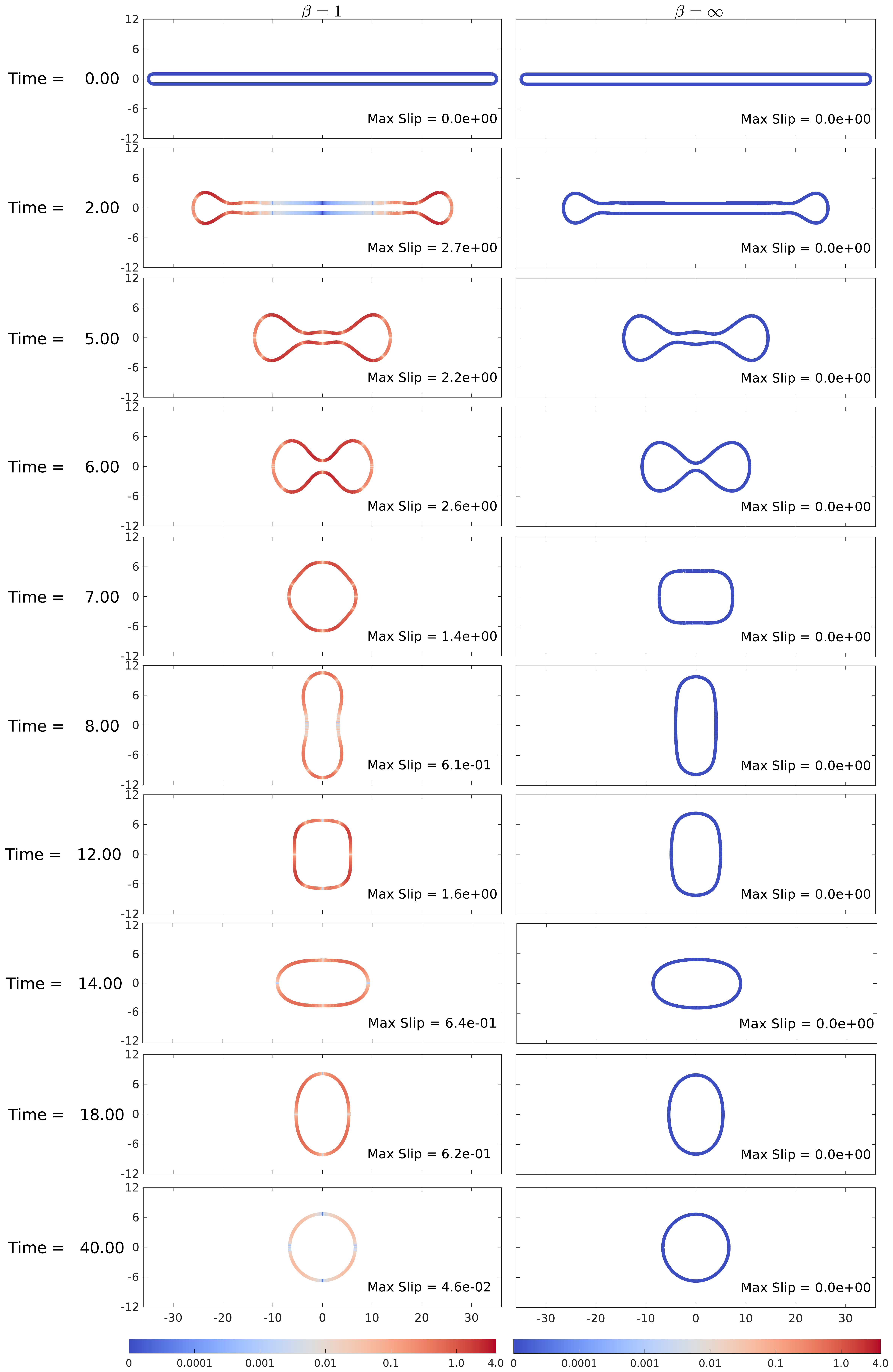}
 \caption{Relaxation of long filaments with $L=35$ and $r=1$ over time in both the presence $\beta = 1$, and absence $\beta=\infty$, of interfacial slip. The final time of simulation for both cases is $t=40$ and snapshots of the interface are shown at different times, $t= 0,2,5,6,7,8,12,14,18,40$. (In this figure only, in order to be able to see the variations of slip on the interface better, instead of plotting the values of jump in velocity across the interface, $log_{100}$ of those values is plotted. The color bar then shows the range in which these values change with color.)}
 \label{fig:relaxationL34}
\end{figure}

The relaxation of a filament with the an initial aspect ratio of $35$ and two different friction coefficients is studied. The filaments are centered at $x=0, y=0$ in a domain of $[-36, 36] \times [-15, 15]$, where a grid of $1440\times 600$ is used. The time-step for these simulations is $\Delta t = 0.01$ and we have $\Re=1.0$, $\We = 0.01$, and matching viscosity and density is used for the inner and outer fluids. Figure~\ref{fig:relaxationL34} shows snapshots of these filaments at different times until they reach equilibrium and fully relax into a spherical shape. Up to $t=5$ the dynamics are qualitatively similar despite a large amount of slip occurring at the interface. After this time the influence of slip become more pronounced, with a thicker center region at $t=6$ and squaring-off of the shape at $t=12$.

\begin{figure}[H]
        \centering
        \begin{subfigure}[b]{0.475\textwidth}
            \centering
            \includegraphics[width=\textwidth]{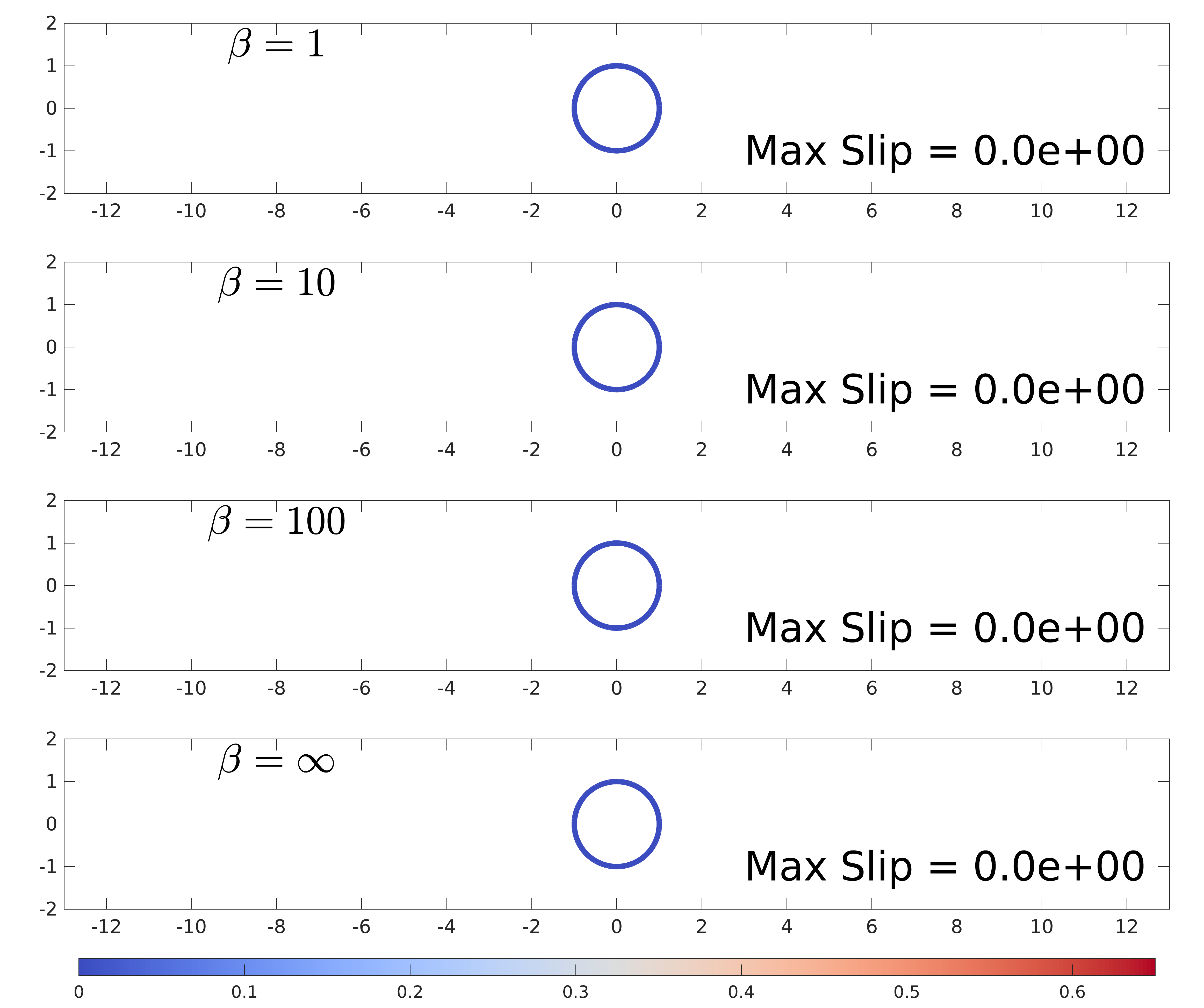}
            \caption[]%
            {{\small Time=0}}
            \label{fig:Shear0}
        \end{subfigure}
        \hfill
        \begin{subfigure}[b]{0.475\textwidth}
            \centering
            \includegraphics[width=\textwidth]{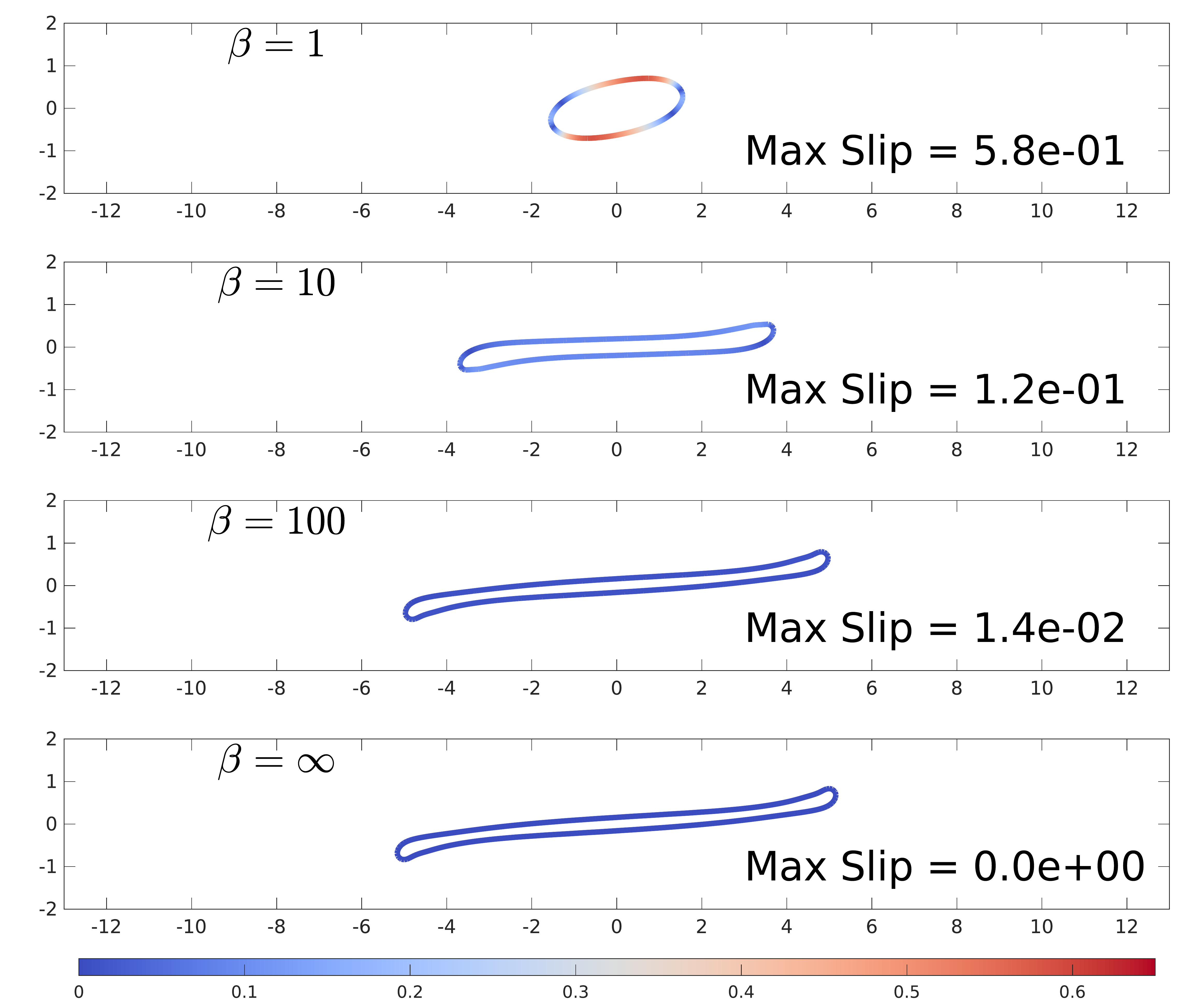}
            \caption[]%
            {{\small Time=12.5}}
            \label{fig:Shear50}
        \end{subfigure}
        \vskip\baselineskip
        \begin{subfigure}[b]{0.475\textwidth}
            \centering
            \includegraphics[width=\textwidth]{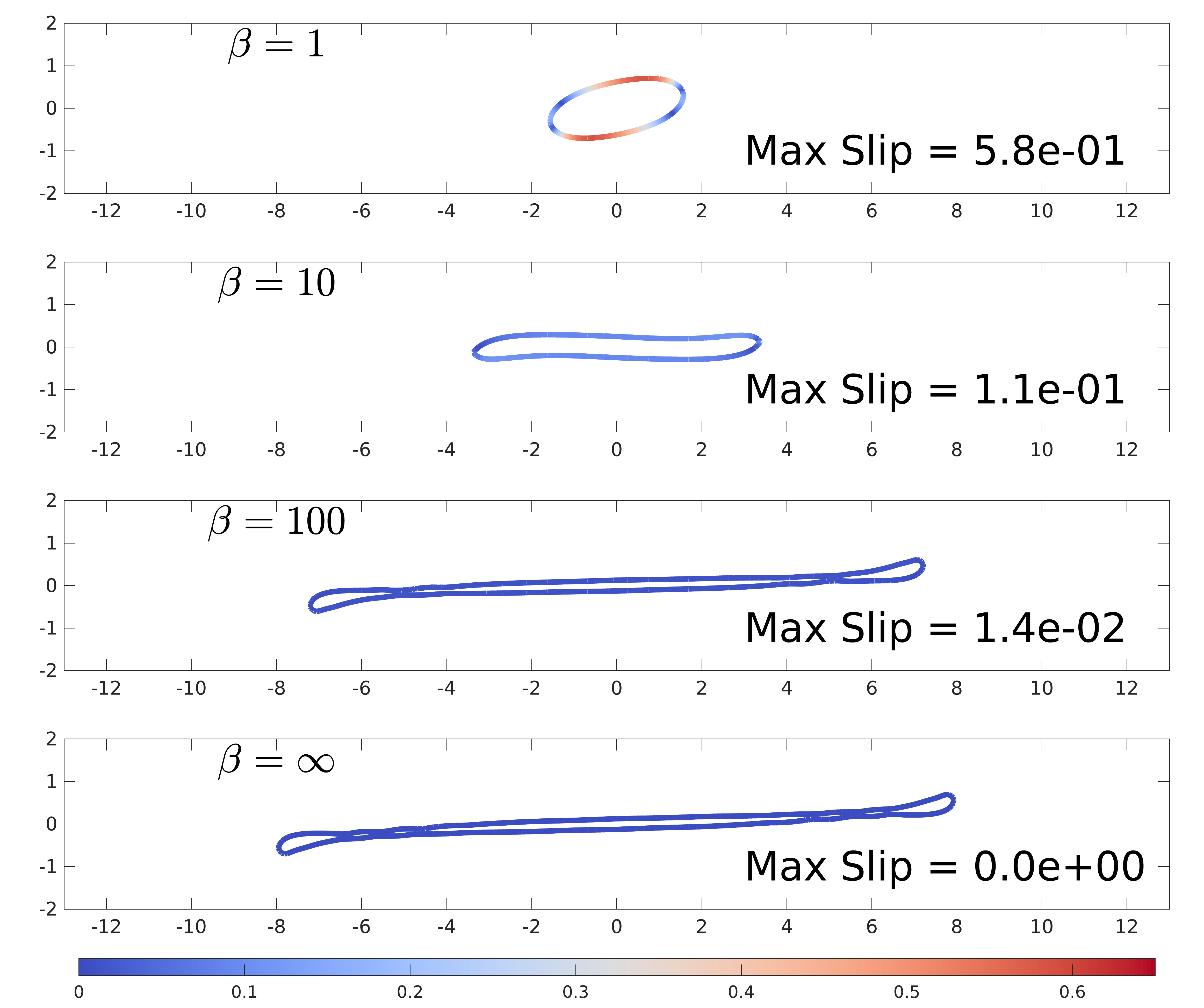}
            \caption[]%
            {{\small Time=25}}
            \label{fig:Shear100}
        \end{subfigure}
        \hfill
        \begin{subfigure}[b]{0.475\textwidth}
            \centering
            \includegraphics[width=\textwidth]{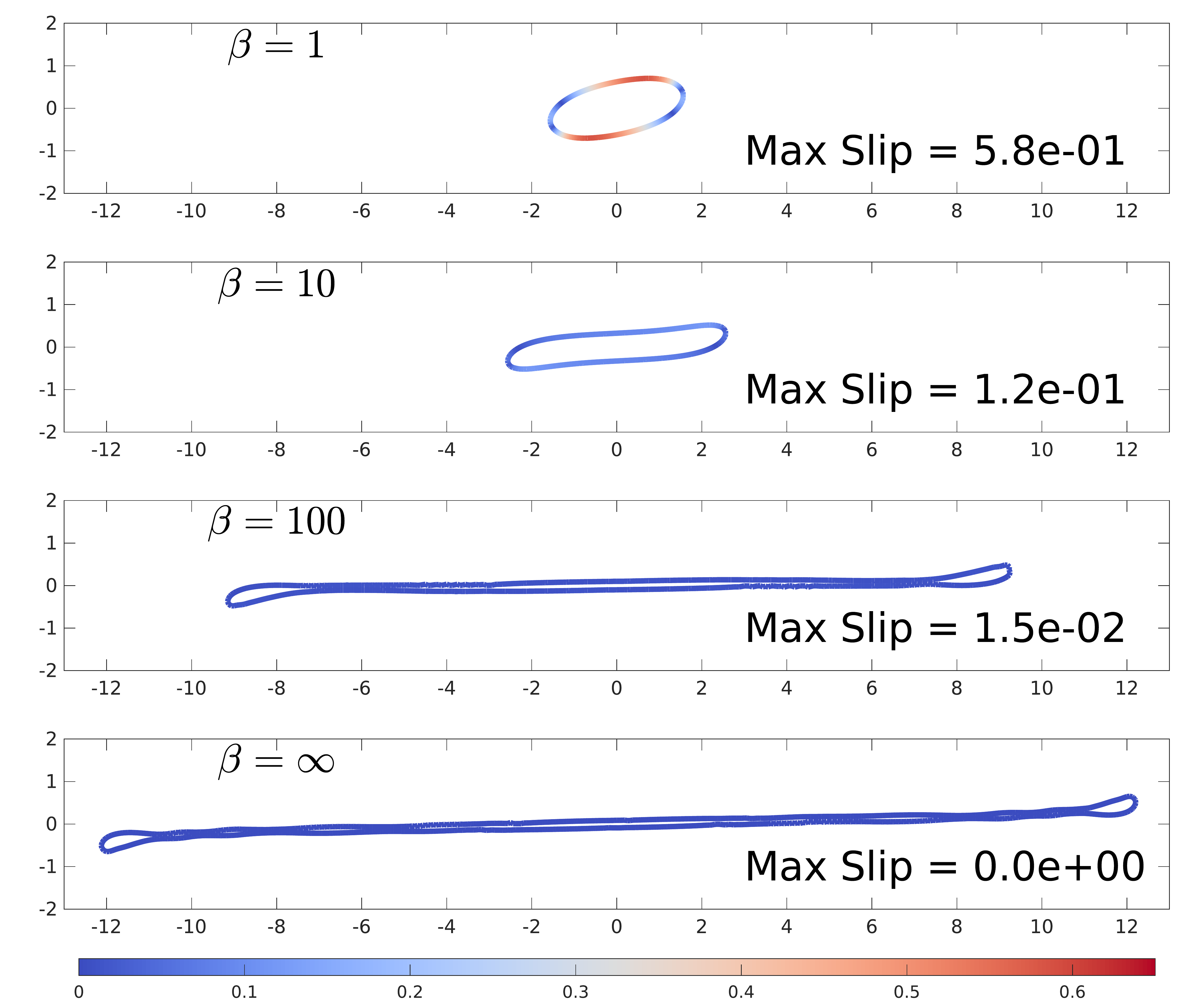}
            \caption[]%
            {{\small Time=50}}
            \label{fig:Shear200}
        \end{subfigure}
        \caption[  ]
        {Evolution of droplets in shear flow with different friction coefficients $\beta = 1, 10, 100, \infty$. Despite being above the critical no-slip shear rate necessary for droplet breakup interfacial slip results in either a steady-state shape or oscillatory behavior.}
        \label{fig:dropShearFSF}
    \end{figure}
\subsection{Droplet in Shear flow}\label{sec:res_dropShear}
In this section we study the influence of interfacial slip on shear-induced deformation of initially spherical droplets. In all the simulations in this section, a 2D sphere with radius of $r=1$ is placed in the center of a 2D domain of size $[-16, 16]\times[-2, 2]$ where the shear rate $\dot{\gamma}=1.0$ is applied at the boundaries at $y=\pm2$.  The grid size is $513\times 65$ and a time step of $\Delta t = 0.025$ is used. The density and viscosity ratios are $\lambda=1$ and $\eta=1$, and the other non-dimensional parameters used are $\Re=1.0$ and $\We=1.0$. Snapshots of the droplet shape at four different times is shown in Fig.~\ref{fig:dropShearFSF}. For the no-slip case, $\beta=\infty$, the droplet aligns with the flow and will extend indefinitely due to this system being above the critical shear rate~\cite{Vananroye2006,Grace1982,Stone1989,Taylor1934}. At large friction coefficients (moderate slip) the elongation is slowed but still substantial. For smaller friction coefficients there appear to be two different regimes. At $\beta=10$ the droplet elongates for a time before slowly retracting. It is suspected that this is due to the droplet fully aligning with the flow, which allows surface tension effects to reduce the interfacial length. At $\beta=1$ the droplet reaches a steady-state shape at a fixed inclination angle. Similar data regarding the time-evolution of this case was shown in Sec.~\ref{sec:res_converg}. Further investigations are needed to determine the critical friction coefficient necessary for elongation and eventual breakup of the droplet in shear flow.

\begin{figure}[H]
    \centering
    \begin{subfigure}[b]{0.485\textwidth}
        \centering
        \includegraphics[width=\textwidth]{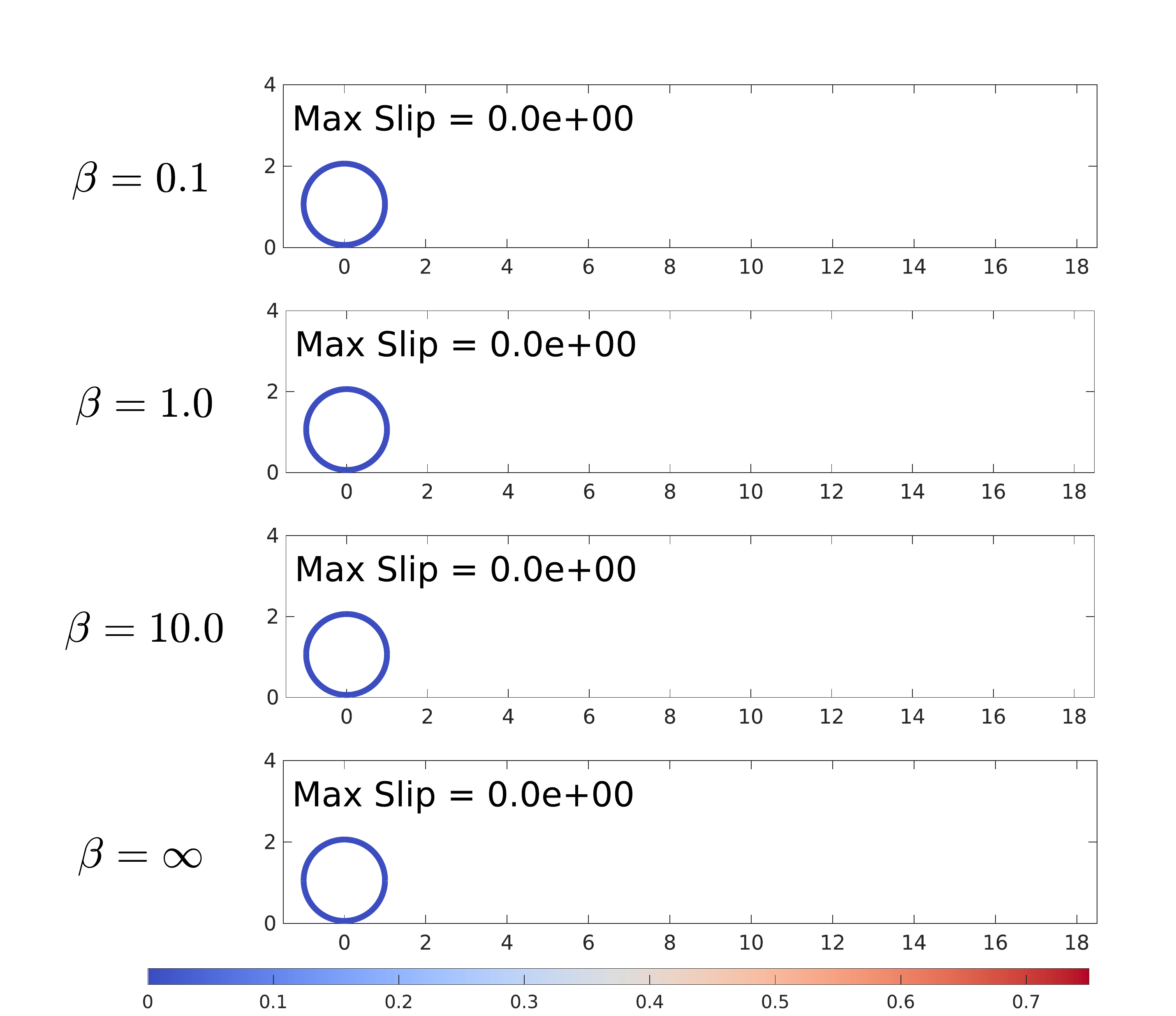}
        \caption[]%
        {{\small Time=0}}
        \label{fig:}
    \end{subfigure}
    \hfill
    \begin{subfigure}[b]{0.485\textwidth}
        \centering
        \includegraphics[width=\textwidth]{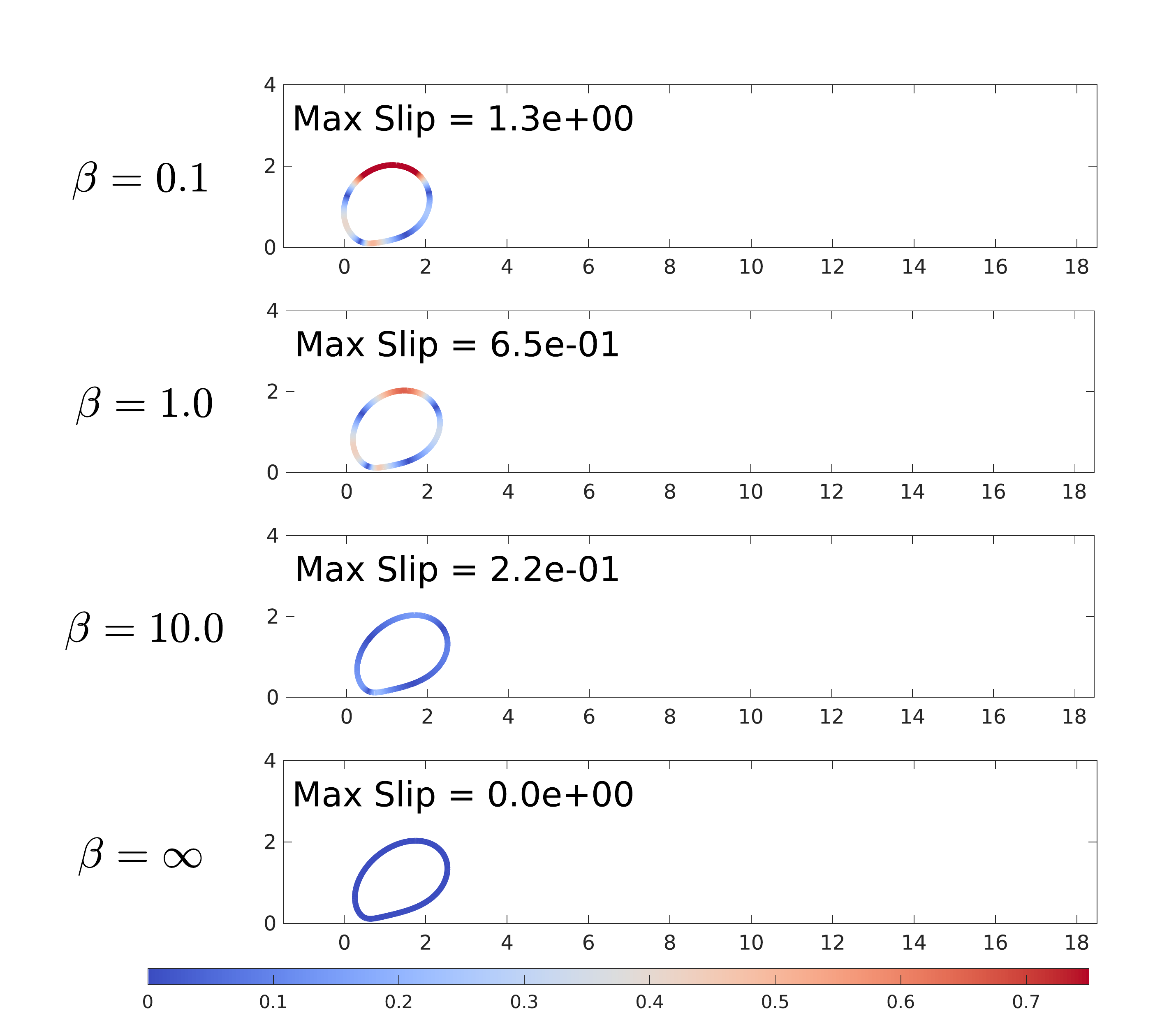}
        \caption[]%
        {{\small Time=2}}
        \label{fig:}
    \end{subfigure}
    \vskip\baselineskip
    \begin{subfigure}[b]{0.485\textwidth}
        \centering
        \includegraphics[width=\textwidth]{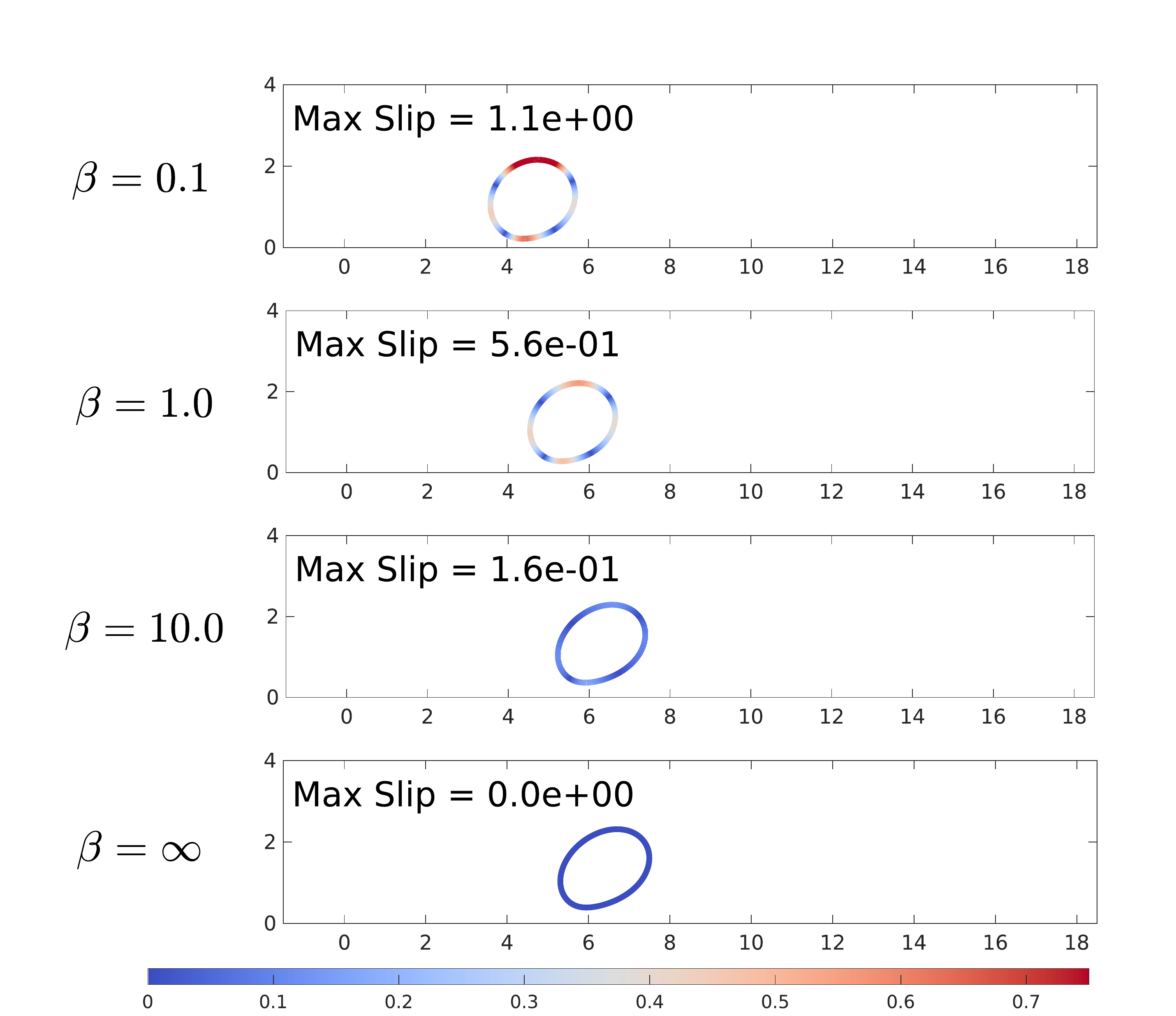}
        \caption[]%
        {{\small Time=7}}
        \label{fig:}
    \end{subfigure}
    \hfill
    \begin{subfigure}[b]{0.485\textwidth}
        \centering
        \includegraphics[width=\textwidth]{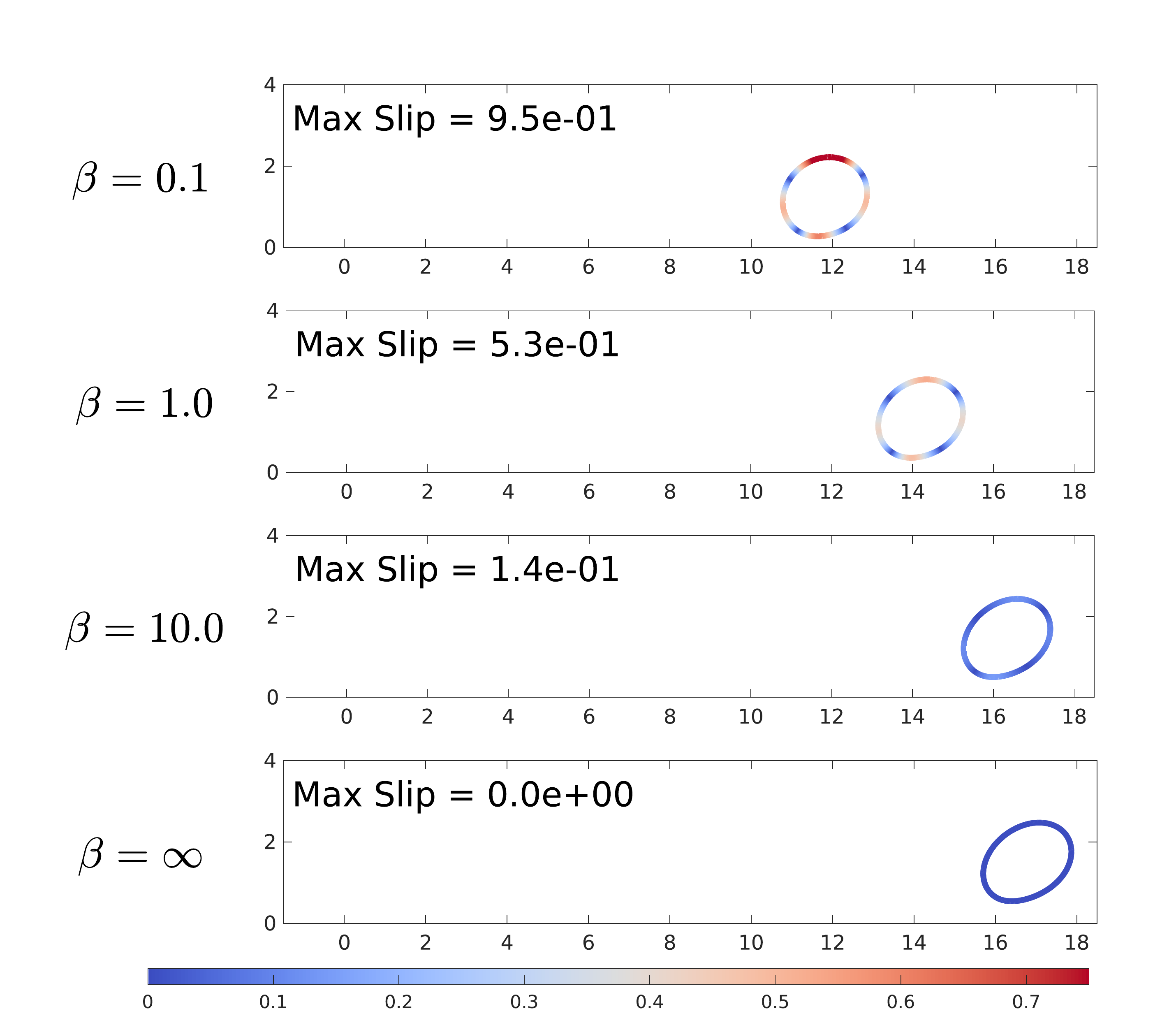}
        \caption[]%
        {{\small Time=15}}
        \label{fig:}
    \end{subfigure}
    \caption[  ]
    {Evolution of rising droplets in wall-bounded shear flow for different friction coefficients $\beta = 0.1, 1, 10, \infty$. The smaller friction coefficient (larger slip) results in droplets which do not experience as much of a lift force and migrate slower than those droplets with a larger friction coefficient.}
    \label{fig:risingDropFSF}
\end{figure}
\subsection{Wall-bounded droplet in Shear flow}\label{sec:res_wall}
\begin{figure}[H]
  \centering
  \includegraphics[width=0.8\textwidth]{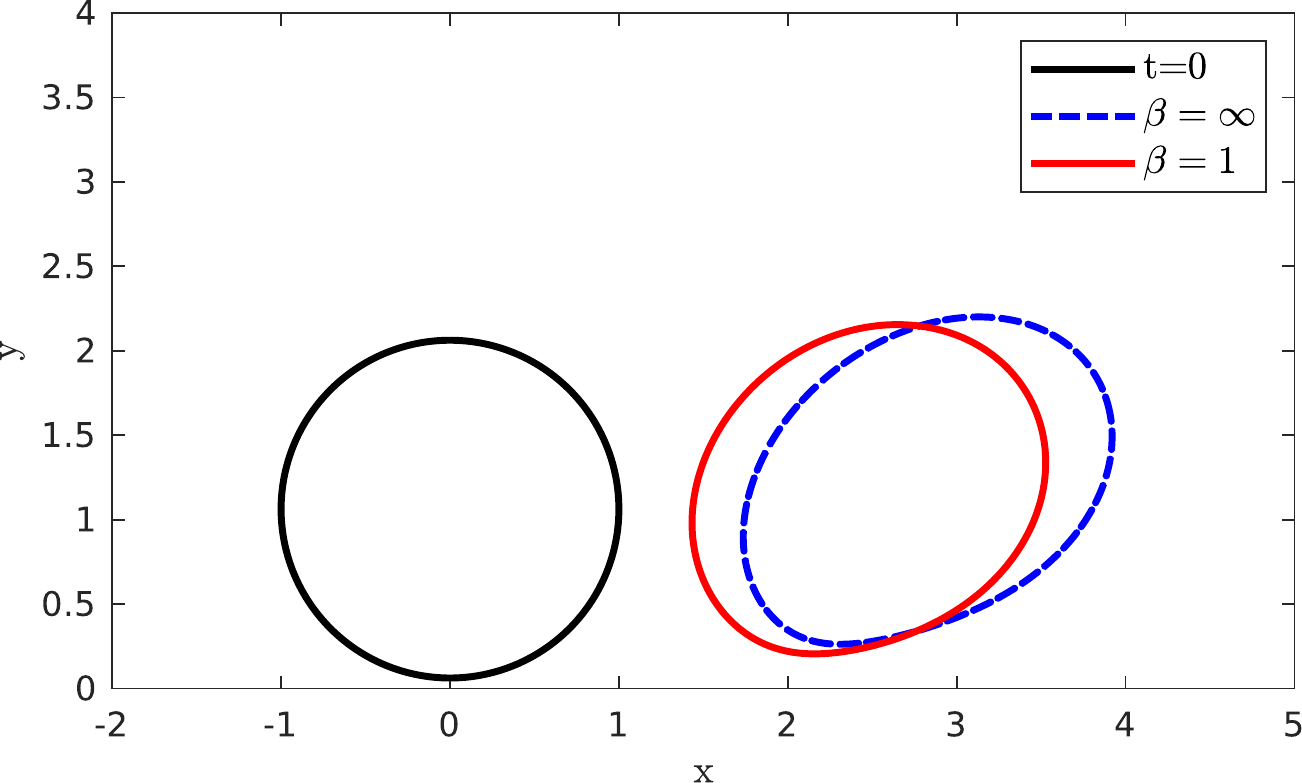}
  \caption{Comparison of the temporal evolution of droplets in absence ($\beta=\infty$) and presence of interfacial slip ($\beta=1.0$), at $t=4$ with the initial shape at $t=0$ shown in black centered at $x=0$.}
  \label{fig:risingDropCompare}
\end{figure}

As a final numerical example we will consider the effect of slip on droplets in wall-bounded shear flow in the presence of gravity. Consider a domain of $[-8,8] \times [0, 4]$ with wall boundary-conditions in the $y$-direction and periodic boundaries in the $x-$direction. A droplet of radius $r=1$ that is initially located at the bottom of the domain with its center at $(0, 1.0625)$ so that it does not touch the bottom wall. This droplet has matched viscosity, $\eta=1$, while the inner fluid density is five-times that of the outer fluid, $\lambda=5$. A shear rate of $\dot{\gamma}=1$ is then applied while the dimensionless parameters are $\Re=0.1$, $\We=0.01$, and $\Fr=1.0$.  For these results the numerical grid is of size $513 \times 257$ and the time step is $\Delta t = 10^{-3}$. It is well known that in such situations droplets~\cite{SMART1991, Karnis1967}, bubbles~\cite{Takemura2009}, and vesicles~\cite{Abkarian2005,Kaoui2009} will all experience a ``lift" force, driving the body away from the wall as they deform and move down the channel. This is demonstrated in Fig.~\ref{fig:risingDropFSF}, which shows the results over time for friction coefficients of $\beta=0.1, 1, 10, \textnormal{ and } \infty$. All of the droplets migrate away from the wall and down the channel, with the droplets experiencing slip staying closer to the wall and traveling less. This becomes more obvious when considering the location of the droplet at a time of $t=4$ for the no-slip ($\beta=\infty$) and moderate slip ($\beta=1$) case, Fig.~\ref{fig:risingDropCompare}. It is clear that the no-slip case has traveled farther than the slip case in the same amount of time.

\begin{figure}[H]
  \centering
  \includegraphics[width=0.8\textwidth]{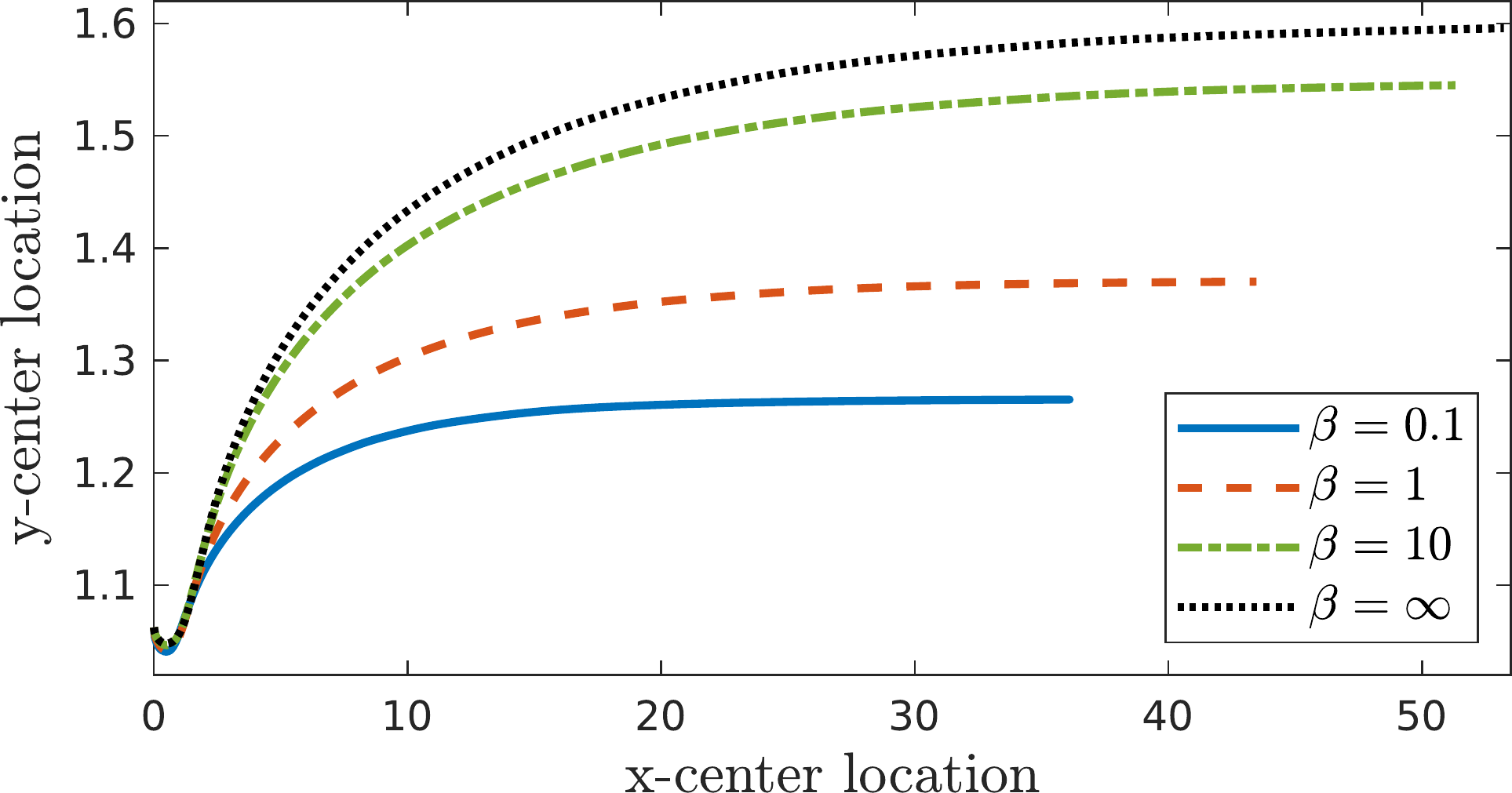}
  \caption{Location of a droplet's center over time for different friction coefficients $\beta=0.1, 1, 10, \textnormal{and } \infty$. All of these simulations have the same final run-time of $t=40$. The different final $x$-center locations for droplets with different friction coefficients shows how the interfacial slip affects the dynamics of the droplets, i.e., a lower friction coefficient results in less overall movement during the same amount of simulation time.}
  \label{fig:xcyc}
\end{figure}

The influence of slip can be further explored by tracking the location of the center-of-mass of each droplet over the time $t\in[0, 40]$. The results can be seen in Fig.~\ref{fig:xcyc}. Initially the $y$-location of the center-of-mass decreases for all cases due to the droplet undergoing deformation. The influence of slip then becomes apparent, as the no-slip case not only rose farther, 0.533 for no-slip versus 0.2 for $\beta=0.1$, but also travelled much farther, a distance of 53.2 for no-slip versus 36.1 for $\beta=0.1$. The height risen and lateral distance traveled are obviously coupled, as droplets which rise farther in to the flow field experience higher shear velocities.

\section{Conclusion}\label{sec:conc}
In this work a numerical model is presented of multiphase fluid systems which have interfacial slip. The work implements a hybrid projection method for the Navier-Stokes equations, whereby material properties and the pressure field are continuous across the interface while the velocity observes a jump. This velocity jump is accounted for via an Immersed Interface Method. This results in the tentative velocity step of the projection method requiring the solution of not only the velocity, but also the jump in a fully coupled manner.

The model is compared to experimental results demonstrating slip in polymer-polymer melts and to published numerical results of droplet elongation in extensional flow, showing excellent agreement with both. It is then used to explore the dynamics of several common multiphase systems, including the shearing of a planar interface, droplet and filament relaxation, and droplets in shear flow, both unbounded and wall-bound. The influence of slip is clearly demonstrated, with slip increasing the rate of droplet relaxation, inhibiting the continuous extension of droplet in shear flows, and decreasing the migration rate in wall-bounded shear flow.

This work is a step towards better models for multiphase fluid systems with complex phenomena, particularly those where interfacial effects can play a large role, such as microfluidics or rheological applications. Future work will investigate better numerical preconditioners to remove the dependence of solution time on the friction coefficient and begin to explore how interfacial slip can be used to gain a better understanding of novel engineering systems, along with how to design said systems for better performance.

\newpage
\appendix

\section{Immersed Interface Method}\label{app:IIM}
\setcounter{figure}{0}
The jump of any arbitrary variable $u$ can be written as $[u] = u^+ - u^-$, where $u^+$ is the value of $u$ approaching the interface from the outer domain and $u^-$ is its value approaching the interface from the inner domain. Therefore, given $[u]$ and either the outer and inner values of $u$ the inner and outer values can be found as
\begin{eqnarray}\label{eq:plusMinusValues}
  u^- &=& u^+ - [u], \nonumber \\
  u^+ &=& u^- + [u].
\end{eqnarray}
respectively.

In many situation the jump of a variable is only provided on an embedded interface, $\Gamma$, such as one described by the zero of a level-set function. To obtain the jump at a point away from the interface, specifically grid points $[u]_{gp}$, we can use $[u]$ and it's normal derivatives using a Taylor series expansion
\begin{equation}\label{eq:cpExtrapolate}
[u]_{gp} = [u]_{\Gamma} + d [\frac{\partial u}{\partial n}]_{\Gamma} + \frac{1}{2} d^2 [\frac{\partial^2 u}{{\partial n}^2}]_{\Gamma}
\end{equation}
where $[u]_{\Gamma}$ is the jump in $u$ at the closest point on $\Gamma$ to the grid point, while $[\partial u/\partial n]_{\Gamma}$ and $[\partial^2 u/\partial n^2]_{\Gamma}$ are the first and the second normal derivatives of $u$ at the closest point and $d$ is the signed distance from the grid point to the interface. In the case that the level set, $\phi$, describing the interface is a signed distance function we can determine the jump in $u$ at an arbitrary grid point located at $\vec{x}_{ijk}=({x}_i,{y}_j,{z}_k)$ in a three dimensional domain (or $\vec{x}_{ij}=({x}_i,{y}_j)$ in 2D), denoted as $[u]_{ijk}$ via
\begin{equation}\label{eq:cpExtrapolateijk}
[u]_{ijk} = [u]_{\Gamma} + \phi_{ijk} [\frac{\partial u}{\partial n}]_{\Gamma} + \frac{1}{2} \phi_{ijk}^2 [\frac{\partial^2 u}{{\partial n}^2}]_{\Gamma}
\end{equation}
where $\phi_{ijk}$ is the level set value at that grid point.

\begin{figure}[H]
  \centering
  \includegraphics[width=0.4\textwidth]{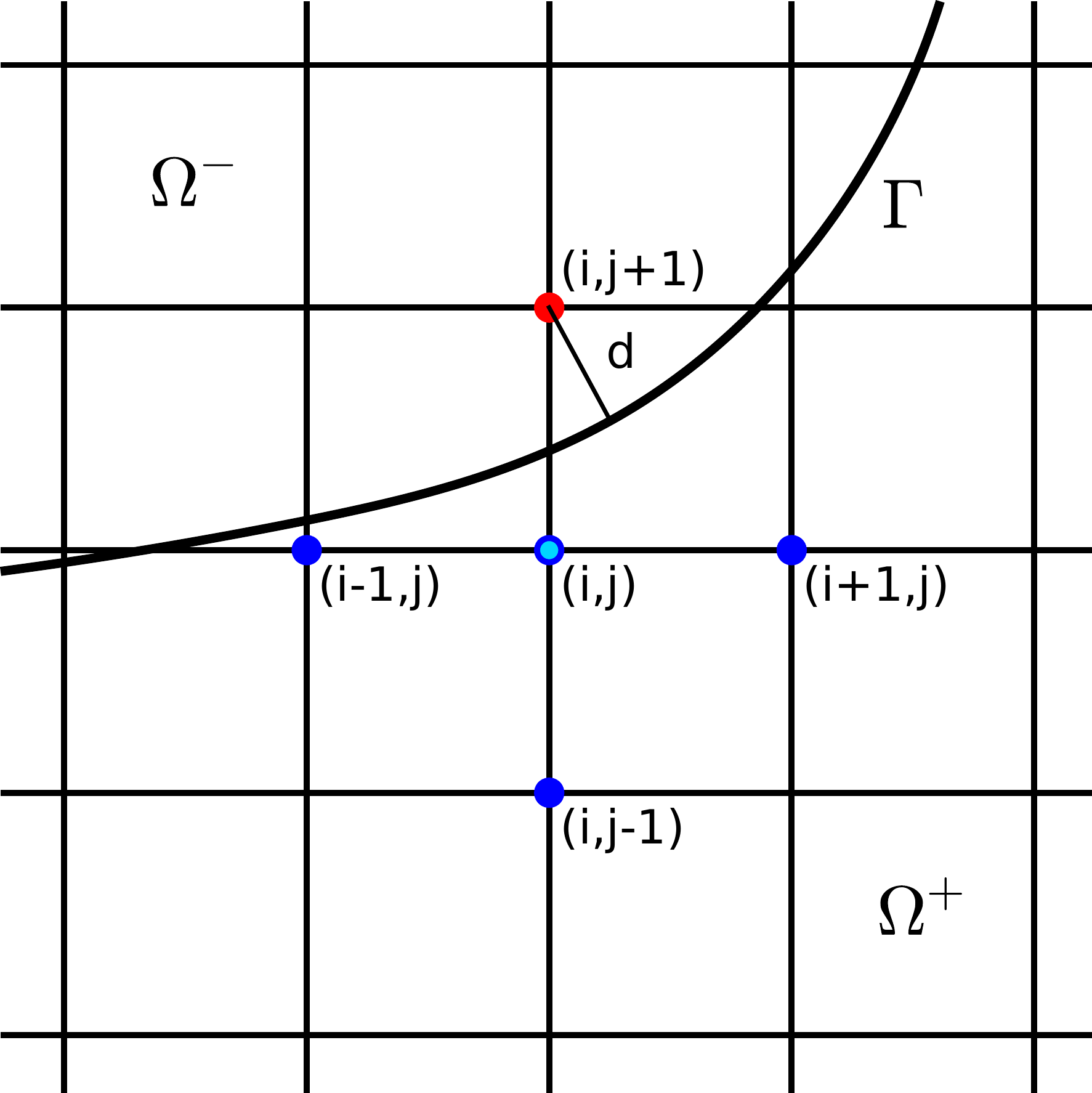}
  \caption{Schematic of a discretized 2D domain. The interface is shown as $\Gamma$ and the five points needed to discretize the Laplacian at point $(i,j)$ using a central difference method include: $(i-1,j), (i,j-1), (i,j), (i+1,j), (i,j+1)$, where point $(i,j+1)$ is located in inner domain, $\Omega^-$, and the rest of the points are in outer domain, $\Omega^+$. The symbol $d$ indicates the signed normal distance between a grid point and its closest point on the interface.}
  \label{fig:IIM}
\end{figure}

We demonstrate the use of the Immersed Interface Method via an example. Consider a Poisson equation in 2D: $\Delta u = f$, where $u$ and $f$ are both scalar variables.
According to Fig.~\ref{fig:IIM} the discretized form of the governing equation at point $(i,j)$, which is located in $\Omega^+$, using a uniform grid spacing $h$ both in $x$- and $y$-directions is given by
\begin{equation}\label{eq:discreteLap}
  \frac{u^+_{i,j-1} + u^+_{i-1,j} - 4u^+_{i,j} + u^+_{i+1,j} + u^+_{i,j+1}}{h^2} = f^+_{i,j}.
\end{equation}
As the grid point $(i,j+1)$ is located on the other side of the interface as point $(i,j)$, and hence the value $u^+_{i,j+1}$ is not available, corrections need to be added to the above equation for that particular grid point.
Note that all grid points where the discretization only contains points in the same domain require that no corrections to be added. According to Eq.~\ref{eq:plusMinusValues}, $u^+_{i,j+1}$ can be obtained by $u^+_{i,j+1} = u^-_{i,j+1} + [u]_{i,j+1}$, where $[u]_{i,j+1}$ can be calculated using the Taylor Series expansion shown in Eq.~\ref{eq:cpExtrapolateijk}:
\begin{equation}
  [u]_{i,j+1} = [u]_{\Gamma} + \phi_{i,j+1} [\frac{\partial u}{\partial n}]_{\Gamma} + \frac{1}{2} \phi_{i,j+1}^2 [\frac{\partial^2 u}{{\partial n}^2}]_{\Gamma},
\end{equation}
noting that it is assumed that the level set field is a signed distance function and thus $d_{i,j+1}=\phi_{i,j+1}$.
The unknown value $u^+_{i,j+1}$ can now be replaced with the known values $u^-_{i,j+1}+[u]_{i,j+1}$:
\begin{eqnarray}
  \frac{u^+_{i,j-1} + u^+_{i-1,j} - 4u^+_{i,j} + u^+_{i+1,j} + u^-_{i,j+1} + [u]_{i,j+1}}{h^2} &=& f^+_{i,j} \nonumber \\
  \frac{u^+_{i,j-1} + u^+_{i-1,j} - 4u^+_{i,j} + u^+_{i+1,j} + u^-_{i,j+1}}{h^2} &=& f^+_{i,j} -\frac{[u]_{i,j+1}}{h^2}. \\
  \label{eq:correctedPoissonP}
\end{eqnarray}
If the values of $[u]_\Gamma$, $[\partial u/\partial n]_\Gamma$, and $[\partial^2 u/\partial n^2]_\Gamma$ are known then this will result in a second-order accurate discretization of $\Delta u=f$ at grid point $(i,j)$.

On the other hand consider a situation where all points except for $(i,j+1)$ lie in $\Omega^-$. In this case the discretization of $\Delta u=f$ at $(i,j)$ is given by
\begin{eqnarray}
  \frac{u^-_{i,j-1} + u^-_{i-1,j} - 4u^-_{i,j} + u^-_{i+1,j} + u^-_{i,j+1}}{h^2} &=& f^-_{i,j} \nonumber \\
  \frac{u^-_{i,j-1} + u^-_{i-1,j} - 4u^-_{i,j} + u^-_{i+1,j} + u^+_{i,j+1} - [u]_{i,j+1}}{h^2} &=& f^-_{i,j} \nonumber \\
  \frac{u^-_{i,j-1} + u^-_{i-1,j} - 4u^-_{i,j} + u^-_{i+1,j} + u^+_{i,j+1}}{h^2} &=& f^-_{i,j} + \frac{[u]_{i,j+1}}{h^2}.
  \label{eq:correctedPoissonM}
\end{eqnarray}

This can be extended to account for \textit{any} linear operator evaluated at a grid location $\vec{x}_{ijk}$. Denote the corrections needed as $C_{ijk}$ such that $Lu_{ijk} + C_{ijk}$ is an accurate approximation of the continuous linear operator acting on $u$ at location $\vec{x}_{ijk}$. These corrections can be calculated via the introduction of an indicator function $\delta_{ijk,pqr}$ such that
\begin{equation}\label{eq:indicator}
  \delta_{ijk,pqr} = \begin{cases}
    1 & \phi_{pqr}\phi_{ijk} < 0 \\
    0 & \phi_{pqr}\phi_{ijk} \geq 0.
  \end{cases}
\end{equation}
The correction is then given by
\begin{equation}
  C_{ijk} = \sgn(\phi_{ijk})\sum_{pqr}l_{pqr}\delta_{ijk,pqr}[u]_{pqr}
\end{equation}
where the summation occurs over all grid-points associated with the linear discretization and $l_{pqr}$ are the corresponding weights.

Return to Eq.~\eqref{eq:discreteLap}. In this case $(pq)\in\{(i,j-1),(i-1,j),(i,j),(i+1,j),(i,j+1)\}$ where only $\delta_{(i,j),(i,j+1)}=1$, with all others zero. Therefore, the correction
to evaluate $\Delta u$ at $\vec{x}_{ij}$ is
\begin{equation}
  C_{ij}=\sgn(\phi_{(i,j)})l_{(i,j+1)}\delta_{(ij),(i,j+1)}[u]_{(i,j+1)}=(+1)\left(\frac{1}{h^2}\right)(1)[u]_{(i,j+1)}=\frac{[u]_{i,j+1}}{h^2}.
\end{equation}
To solve $\Delta u=f$ at $\vec{x}_{i,j}$ the corrected equation would be
\begin{align}
  Lu^+_{i,j} &= f^+_{i,j} - C_{i,j} \nonumber\\
  \frac{u^+_{i,j-1} + u^+_{i-1,j} - 4u^+_{i,j} + u^+_{i+1,j} + u^-_{i,j+1}}{h^2} &= f^+_{i,j} -\frac{[u]_{i,j+1}}{h^2},
\end{align}
which matches the result shown in Eq.~\eqref{eq:correctedPoissonP}.

Applying this procedure to the situation where all points except for $(i,j+1)$ lie in $\Omega^-$ results in a correction of
\begin{equation}
  C_{ij}=\sgn(\phi_{(i,j)})l_{(i,j+1)}\delta_{(ij),(i,j+1)}[u]_{(i,j+1)}=(-1)\left(\frac{1}{h^2}\right)(1)[u]_{(i,j+1)}=-\frac{[u]_{i,j+1}}{h^2},
\end{equation}
which when applied to $\Delta u=f$ results in the discretization of
\begin{equation}
  \frac{u^-_{i,j-1} + u^-_{i-1,j} - 4u^-_{i,j} + u^-_{i+1,j} + u^+_{i,j+1}}{h^2} = f^-_{i,j} +\frac{[u]_{i,j+1}}{h^2},
\end{equation}
matching the result shown in Eq.~\eqref{eq:correctedPoissonM}. The advantage of formulating the IIM using this method is that it can now be applied to any potential linear operator, including both finite difference approximations and Hermite interpolation.

\section{Derivation of jump conditions}\label{app:jump}
\setcounter{figure}{0}
In this appendix the jump conditions necessary to evaluate a discontinuous velocity field is presented, following the work of Xu et al~\cite{Xu2006}. Consider a fluid-fluid system in a Cartesian coordinate system $x_i\ (i=1,2,3)$ as shown in Fig.~\ref{fig:coord}, which are separated by an interface $\Gamma$ where its coordinates at time $t$ is shown by $\vec{X}$. In this figure $\alpha_i,\ (i=1,2,3)$ indicates the curvilinear coordinates near an arbitrary point on the interface, where $\alpha_3$ is in the same direction of the outward unit normal to the interface denoted by $\vec{n}$, and $\vec{\tau}$ and $\vec{b}$ are the two unit tangents to the interface given by
\begin{equation}\label{eq:unitVec}
  \vec{\tau} = \frac{\partial \vec{X}}{\partial \alpha_1} , \quad \vec{b} = \frac{\partial \vec{X}}{\partial \alpha_2} , \quad
  \vec{n} = \vec{\tau} \times \vec{b},
\end{equation}
where $\vec{X}=(X_1, X_2, X_3)$ is the arc-length parameterization of the interface $\Gamma$.

\begin{figure}[H]
  \centering
  \includegraphics[width=0.4\textwidth]{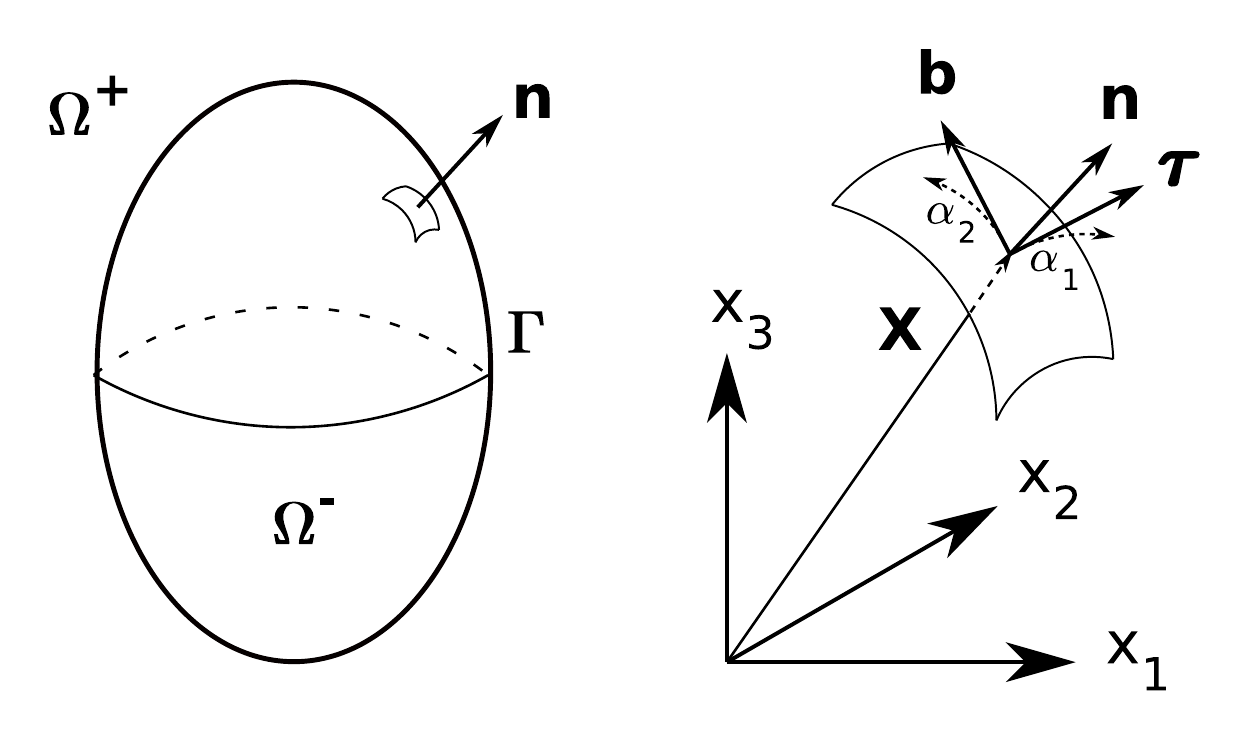}
  \caption{Schematics of a fluid-fluid system with interface $\Gamma$, where $x_i(i = 1,2,3)$ is the Cartesian coordinates, $\alpha_1$ and $\alpha_2$ are two Lagrangian parameters that parameterize the interface locally, $\vec{X}$ is the Cartesian coordinates of the interface, $\alpha_3$ would be aligned with the normal vector $\vec{n}$, and $\vec{\tau}$ and $\vec{b}$ are the two unit tangents at $\vec{X}$\cite{Xu2006,Xu2009}.}
  \label{fig:coord}
\end{figure}

According to Fig.~\ref{fig:coord} the jump of an arbitrary function $q(\vec{X})$ across $\Gamma$ at $\vec{X}$ is denoted by
\begin{equation}\label{eq:jump}
  [q] = \lim_{\varepsilon\to 0^+} q(\vec{X} + \varepsilon\vec{n}) - \lim_{\varepsilon\to 0^+} q(\vec{X} - \varepsilon\vec{n})
\end{equation}
where $\vec{n}$ is the unit normal to the interface pointing into $\Omega^+$ and can also defined as $\vec{n}= \nabla \phi/\|\phi\|$ where $\phi$ is a signed distance function.

In the following sections of this Appendix the derivation of the jump conditions of velocity and its normal derivatives are presented in which a few facts are being used that are worth mentioning in advance. First, the jump of the product of two arbitrary variables $q$ and $r$ can be written as
\begin{eqnarray}\label{eq:multjump}
[qr] &=& q^+r^+ - q^-r^- \nonumber \\
&=& q^+r^+ - q^+r^- +  q^+r^- - q^-r^-  = q^+(r^+ - r^-) + (q^+ - q^-)r^-\nonumber \\
&=& q^+[r] + r^-[q] =  q^-[r] + r^+[q]
\end{eqnarray}
Secondly, there is no jump in the unit vectors across the interface, therefore
\begin{equation}\label{eq:unitjumpzero}
  [\vec{n}] = 0 , \quad [\vec{\tau}] = 0 , \quad  [\vec{b}] = 0
\end{equation}
Lastly, the jump operator commutes with differentiation along the interface, in other words
\begin{equation}\label{eq:commute}
  \bigg[\frac{\partial q}{\partial \alpha_i}\bigg] = \frac{\partial[q]}{\partial \alpha_i} , \quad i = 1,2
\end{equation}

\subsection{Jump in continuity equation}
From continuity we have, $\mu \nabla \cdot \vec{u} = 0$. The expanded form of this relationship in terms of the unit vectors can be shown by
\begin{equation}\label{eq:cont}
  \mu \nabla \cdot \vec{u} =  \mu \frac{\partial \vec{u}}{\partial n} \cdot \vec{n} + \mu \frac{\partial \vec{u}}{\partial \tau} \cdot\vec{\tau} + \mu \frac{\partial \vec{u}}{\partial b} \cdot \vec{b}
\end{equation}
Therefore, the jump in the continuity equation would be
\begin{eqnarray}\label{eq:contJump}
  [\mu \nabla \cdot \vec{u}] &=&  [\mu \frac{\partial \vec{u}}{\partial n} \cdot \vec{n} + \mu \frac{\partial \vec{u}}{\partial \tau} \cdot\vec{\tau} + \mu \frac{\partial \vec{u}}{\partial b} \cdot \vec{b}] \nonumber \\
0 &=& [\mu \frac{\partial \vec{u}}{\partial n} \cdot \vec{n}] + [\mu \frac{\partial \vec{u}}{\partial \tau} \cdot\vec{\tau}] + [\mu \frac{\partial \vec{u}}{\partial b} \cdot \vec{b}]
\end{eqnarray}
By rearranging this equation we have
\begin{equation}\label{eq:dudnDnJump}
  [\mu \frac{\partial \vec{u}}{\partial n} \cdot \vec{n}] = -[\mu \frac{\partial \vec{u}}{\partial \tau} \cdot\vec{\tau}] - [\mu \frac{\partial \vec{u}}{\partial b} \cdot \vec{b}]
\end{equation}

According to Eq.~\ref{eq:unitjumpzero}, knowing there is no jump in the unit vectors, the equation above can be simplified to
\begin{equation}\label{eq:dudn.n}
[\mu \frac{\partial \vec{u}}{\partial n}] \cdot \vec{n} = -[\mu \frac{\partial \vec{u}}{\partial \tau}] \cdot\vec{\tau} - [\mu \frac{\partial \vec{u}}{\partial b}] \cdot \vec{b}
\end{equation}

\subsection{Jump in acceleration}
We can find the jump in acceleration by taking the total derivative of the jump in velocity, ($[u_i],$ $i=1,2,3$), as follows,
\begin{eqnarray}\label{eq:DuDtTensor}
\frac{D}{Dt}[u_i(\vec{X}, t)]
&=&  \frac{D}{Dt}u_i(\vec{X}^+, t) -  \frac{D}{Dt}u_i(\vec{X}^-, t)  \nonumber  \\
&=& \frac{\partial}{\partial t}u_i(\vec{X}^+, t) + \frac{\partial}{\partial x_j}u_i(\vec{X}^+, t)\frac{{dX_j}^+}{dt} \nonumber  \\
&&- \frac{\partial}{\partial t}u_i(\vec{X}^-, t) - \frac{\partial}{\partial x_j}u_i(\vec{X}^-, t)\frac{{dX_j}^-}{dt} \nonumber  \\
&=& [\frac{\partial u_i}{\partial t} +\frac{\partial u_i}{\partial x_j}\frac{dX_j}{dt}] = [\frac{\partial u_i}{\partial t} + u_j\frac{\partial u_i}{\partial x_j}]\nonumber  \\
&=& [\frac{D}{Dt}u_i(\vec{X}, t)]
\end{eqnarray}
In vector notation this becomes
\begin{equation}\label{eq:DuDtVector}
\frac{D}{Dt}[\vec{u}(\vec{X}, t)] = [\frac{D}{Dt}\vec{u}(\vec{X}, t)] = [\frac{\partial \vec{u}}{\partial t} + \vec{u} \cdot \nabla \vec{u}] = [\frac{\partial \vec{u}}{\partial t}] + [\vec{u} \cdot \nabla \vec{u}]
\end{equation}

If we use the Lagrangian method to find the acceleration term, the jump in acceleration term can be evaluated by
\begin{equation}\label{eq:DuDtLagrange}
\frac{D}{Dt}[\vec{u}](\vec{X}, t) = [\frac{D}{Dt}\vec{u}(\vec{X}, t)] = [\frac{\vec{u}^{n+1} - \vec{u}^n_d}{\Delta t}]
\end{equation}
where $\vec{u}_d$ is the departure velocity as mentioned in Section~\ref{sec:method}.

\subsection{Jump in first normal derivative of velocity}
The stress balance given by Eq.~\ref{eq:summary3}
can be written in a more general form as
\begin{eqnarray}
\ [\vec{T}_{hd}]\cdot\vec{n} &=& \vec{f} \nonumber \\ \
[-p \vec{I} + \mu(\nabla \vec{u} + \nabla^T \vec{u})]\cdot\vec{n} &=& \vec{f} \nonumber \\ \
-[p]\vec{n} + [\mu(\nabla \vec{u} + \nabla^T \vec{u})]\cdot\vec{n} &=& \vec{f}
\end{eqnarray}
where $\vec{f}$ can be any singular force on the interface such as tension. For example, for a droplet with a uniform surface tension on the interface $\vec{f}=\sigma \kappa \vec{n}$, as mentioned before in Section~\ref{sec:math}.

This equation can be written using index notation as
\begin{equation}\label{eq:stressBalanceTensor}
-[p]n_i+ [\mu(\frac{\partial u_i}{\partial x_j} +  \frac{\partial u_j}{\partial x_i})]n_j = f_i.
\end{equation}
By multiplying the above equation by $\tau_i$ and using the orthogonality of the unit vectors ($n_i\tau_i = 0$),
\begin{eqnarray}\label{eq:stressBal.tau}
-[p]n_i\tau_i + [\mu(\frac{\partial u_i}{\partial x_j} +  \frac{\partial u_j}{\partial x_i})]n_j\tau_i
&=& f_i\tau_i, \nonumber \\ \
[\mu(\frac{\partial u_i}{\partial x_j}n_j\tau_i +  \frac{\partial u_j}{\partial x_i}n_j\tau_i)]
&=& f_i\tau_i, \nonumber \\ \
[\mu( \frac{\partial u_i}{\partial x_j}n_j\tau_i +  \frac{\partial u_j}{\partial x_i}\frac{\partial X_i}{\partial \alpha_1} n_j)]
&=& f_i\tau_i, \nonumber \\ \
[\mu( \frac{\partial u_i}{\partial n}\tau_i +  \frac{\partial u_j}{\partial \tau}n_j)]
&=& f_i\tau_i, \nonumber \\ \
[\mu(\frac{\partial \vec{u}}{\partial n} \cdot \vec{\tau} + \frac{\partial \vec{u}}{\partial \tau} \cdot \vec{n})]
&=& \vec{f} \cdot \vec{\tau}.
\end{eqnarray}
Rearranging the results and using Eq.~\ref{eq:unitjumpzero} gives us
\begin{equation}\label{eq:dudn.tau}
  [\mu\frac{\partial \vec{u}}{\partial n}] \cdot \vec{\tau} = \vec{f} \cdot \vec{\tau} -[\mu \frac{\partial \vec{u}}{\partial \tau}] \cdot \vec{n}.
\end{equation}
In a similar way, by multiplying Eq.~\ref{eq:stressBalanceTensor} by $b_i$ results in
\begin{equation}\label{eq:dudn.b}
  [\mu\frac{\partial \vec{u}}{\partial n}] \cdot \vec{b} = \vec{f} \cdot \vec{b} - [\mu \frac{\partial \vec{u}}{\partial b}] \cdot \vec{n}.
\end{equation}

Using Eqs.~\ref{eq:dudn.n},~\ref{eq:dudn.tau},\ref{eq:dudn.b}, we can form a system of equations to find $[\mu\frac{\partial \vec{u}}{\partial n}]$
\begin{equation}\label{eq:dudnSystem}
\begin{bmatrix}
\tau_1 & \tau_2 & \tau_3\\
b_1 & b_2 & b_3 \\
n_1 & n_2 & n_3
\end{bmatrix}
\begin{bmatrix}
[\mu\frac{\partial u_1}{\partial n}] \\
[\mu\frac{\partial u_2}{\partial n}] \\
[\mu\frac{\partial u_3}{\partial n}]
\end{bmatrix} =
\begin{bmatrix}
\vec{f} \cdot \vec{\tau} -[\mu \frac{\partial \vec{u}}{\partial \tau}] \cdot \vec{n} \\
\vec{f} \cdot \vec{b} -[\mu \frac{\partial \vec{u}}{\partial b}] \cdot \vec{n} \\
-[\mu \frac{\partial \vec{u}}{\partial \tau}] \cdot \vec{\tau} - [\mu \frac{\partial \vec{u}}{\partial b}] \cdot \vec{b}
\end{bmatrix}.
\end{equation}

Assuming the matrix of coefficients is called $C$, its inverse is equal to its transpose due to orthogonality
\begin{equation}\label{eq:Cinverse}
C^{-1} = \begin{bmatrix}
\tau_1 & b_1 & n_1\\
\tau_2 & b_2 & n_2 \\
\tau_3 & b_3 & n_3
\end{bmatrix}.
\end{equation}
Therefore, the solution of this system of equations can be found by multiplying $ C^{-1}$ by the right-hand-side of Eq.~\ref{eq:dudnSystem},
\begin{equation}
\begin{bmatrix}
[\mu\frac{\partial u_1}{\partial n}] \\
[\mu\frac{\partial u_2}{\partial n}] \\
[\mu\frac{\partial u_3}{\partial n}]
\end{bmatrix} =
C^{-1}
\begin{bmatrix}
\vec{f} \cdot \vec{\tau} -[\mu \frac{\partial \vec{u}}{\partial\tau}] \cdot \vec{n} \\
\vec{f} \cdot \vec{b} -[\mu \frac{\partial \vec{u}}{\partial b}] \cdot \vec{n} \\
-[\mu \frac{\partial \vec{u}}{\partial \tau}] \cdot \vec{\tau} - [\mu \frac{\partial \vec{u}}{\partial b}] \cdot \vec{b}
\end{bmatrix}.
\end{equation}
This will finally result in
\begin{eqnarray}\label{eq:mududn}
[\mu\frac{\partial u_i}{\partial n}]
&=& \tau_i(\vec{f} \cdot \vec{\tau} -[\mu \frac{\partial \vec{u}}{\partial \tau}] \cdot \vec{n} ) + b_i(\vec{f} \cdot \vec{b} -[\mu \frac{\partial \vec{u}}{\partial b}] \cdot \vec{n}) - n_i([\mu \frac{\partial \vec{u}}{\partial \tau}] \cdot \vec{\tau} + [\mu \frac{\partial \vec{u}}{\partial b}] \cdot \vec{b}), \nonumber \\ \
[\mu\frac{\partial \vec{u}}{\partial n}]
&=& (\vec{f} \cdot \vec{\tau} -[\mu \frac{\partial \vec{u}}{\partial \tau}] \cdot \vec{n} )\vec{\tau} + (\vec{f} \cdot \vec{b} -[\mu \frac{\partial \vec{u}}{\partial b}] \cdot \vec{n})\vec{b} - ([\mu \frac{\partial \vec{u}}{\partial \tau}] \cdot \vec{\tau} + [\mu \frac{\partial \vec{u}}{\partial b}] \cdot \vec{b})\vec{n}, \nonumber \\
\end{eqnarray}
Knowing $(\vec{f} \cdot \vec{\tau})\vec{\tau} + (\vec{f} \cdot \vec{b})\vec{b} = \Vec{f} - (\vec{f} \cdot \vec{n})\vec{n} = \Vec{P}\Vec{f}$, where $\vec{P}$ is the projection operator, the above equation can be re-written as
\begin{equation}
[\mu\frac{\partial u_i}{\partial n}] = \Vec{P}\Vec{f} -([\mu \frac{\partial \vec{u}}{\partial \tau}] \cdot \vec{n} )\vec{\tau} - ([\mu \frac{\partial \vec{u}}{\partial b}] \cdot \vec{n})\vec{b} - ([\mu \frac{\partial \vec{u}}{\partial \tau}] \cdot \vec{\tau} + [\mu \frac{\partial \vec{u}}{\partial b}] \cdot \vec{b})\vec{n}.
\end{equation}

Assuming continuous fluid properties, i.e. $[\mu]=[\rho]=0$, the jump in normal derivative of velocity simplifies to
\begin{equation}\label{eq:dudnJumpCont}
[\frac{\partial \vec{u}}{\partial n}]
= \frac{1}{\mu_{avg}}\Vec{P} \vec{f} - ([ \frac{\partial \vec{u}}{\partial \tau}] \cdot \vec{n} )\vec{\tau} - ([\frac{\partial \vec{u}}{\partial b}] \cdot \vec{n})\vec{b} - ([\frac{\partial \vec{u}}{\partial \tau}] \cdot \vec{\tau} + [\frac{\partial \vec{u}}{\partial b}] \cdot \vec{b})\vec{n}
\end{equation}
where $\mu_{avg}=\mu^- + \left(\mu^+ - \mu^-\right)H_{\epsilon}(0) = 0.5\left(\mu^- + \mu^+\right)$ is the average of the inner and outer viscosities.

The non-dimensional form of this jump condition can be achieved by normalizing the velocity and viscosity by the characteristic velocity $u_0$ and the viscosity of the outer fluid $\mu^+$, and the directional derivatives $\frac{\partial}{\partial n}$, $\frac{\partial}{\partial \tau}$, and $\frac{\partial}{\partial b}$, using a characteristic length scale $a$. Assuming $\vec{f} = \sigma \kappa \vec{n}$, which can be normalized as $\hat{\vec{f}} = \frac{\vec{f}}{(\sigma/L_0)}$. Therefore,
\begin{eqnarray}
(\frac{u_0}{L_0})[\frac{\hat{\partial} \hat{\vec{u}}}{\partial n}]
&=& \frac{1}{\mu^+ \hat{\mu}_{avg}}\frac{\sigma}{L_0}\Vec{P}\hat{\vec{f}} - (\frac{u_0}{L_0})([ \frac{\hat{\partial} \hat{\vec{u}}}{\partial \tau}] \cdot \vec{n} )\vec{\tau} - (\frac{u_0}{L_0})([\frac{\hat{\partial} \hat{\vec{u}}}{\partial b}] \cdot \vec{n})\vec{b} \nonumber \\
 &-& (\frac{u_0}{L_0})([\frac{\hat{\partial} \hat{\vec{u}}}{\partial \tau}] \cdot \vec{\tau} + [\frac{\hat{\partial} \hat{\vec{u}}}{\partial b}] \cdot \vec{b})\vec{n} \nonumber \\ \
[\frac{\hat{\partial} \hat{\vec{u}}}{\partial n}]
&=& \frac{\sigma}{\mu^+ u_0}\frac{1}{\hat{\mu}_{avg}}\Vec{P}\hat{\vec{f}} - ([ \frac{\hat{\partial} \hat{\vec{u}}}{\partial \tau}] \cdot \vec{n} )\vec{\tau} - ([\frac{\hat{\partial} \hat{\vec{u}}}{\partial b}] \cdot \vec{n})\vec{b} \nonumber \\
 &-& ([\frac{\hat{\partial} \hat{\vec{u}}}{\partial \tau}] \cdot \vec{\tau} + [\frac{\hat{\partial} \hat{\vec{u}}}{\partial b}] \cdot \vec{b})\vec{n} \nonumber \\ \
[\frac{\hat{\partial} \hat{\vec{u}}}{\partial n}]
&=& \frac{1}{\Ca\hat{\mu}_{avg}}\Vec{P}\hat{\vec{f}} -([ \frac{\hat{\partial} \hat{\vec{u}}}{\partial \tau}] \cdot \vec{n} )\vec{\tau} - ([\frac{\hat{\partial} \hat{\vec{u}}}{\partial b}] \cdot \vec{n})\vec{b} \nonumber \\
 &-& ([\frac{\hat{\partial} \hat{\vec{u}}}{\partial \tau}] \cdot \vec{\tau} + [\frac{\hat{\partial} \hat{\vec{u}}}{\partial b}] \cdot \vec{b})\vec{n}
\end{eqnarray}
where $\Ca = (\mu^+ u_0)/\sigma$ and the hat notation can be dropped henceforth, for simplicity. Note that additional interfacial forces, such as a bending rigidity, will provide additional contributions.

\subsection{Jump in second normal derivative of velocity}
The jump in the momentum equations can be written as
\begin{eqnarray}\label{eq:momentumJump}
[\rho\frac{D\vec{u}}{Dt}] &=& -[\nabla p] + [\nabla \cdot(\mu(\nabla \vec{u} + \nabla^T \vec{u}))] + [\vec{g}] \nonumber \\
{[\rho \frac{D\vec{u}}{Dt} ]} &=& -[\nabla p] + [\mu \nabla^2 \vec{u}] + [(\nabla \vec{u} + \nabla^T\vec{u}) \nabla \mu] + [\vec{g}] \nonumber \\ {[\rho \frac{D\vec{u}}{Dt} ]} &=& -[\nabla p] + [\mu (\frac{\partial^2 \vec{u}}{\partial n^2} +  {\nabla_s}^2 \vec{u} + \frac{\partial \vec{u}}{\partial n}\kappa)] + [(\nabla \vec{u} + \nabla^T\vec{u}) \nabla \mu] + [\vec{g}] \nonumber \\ {[\rho \frac{D\vec{u}}{Dt} ]} &=& -[\nabla p] + [\mu \frac{\partial^2 \vec{u}}{\partial n^2}] +  [\mu{\nabla_s}^2 \vec{u}] + [\mu\frac{\partial \vec{u}}{\partial n}\kappa] + [(\nabla \vec{u} + \nabla^T\vec{u}) \nabla \mu]  \nonumber \\
&&+ [\vec{g}]
\end{eqnarray}

In deriving the above equation, the following relation is being incorporated: $\nabla^2 \vec{u} = \frac{\partial^2 \vec{u}}{\partial n^2} + \nabla^2_s \vec{u} +  \frac{\partial \vec{u}}{\partial n}\kappa$\cite{Xu2003}. Rearranging Eq.~\ref{eq:momentumJump} and assuming a continuous pressure field
\begin{equation}\label{eq:mududnfirst}
[\mu \frac{\partial^2 \vec{u}}{\partial n^2}] = [\rho\frac{D\vec{u}}{Dt}] -[(\nabla \vec{u} + \nabla\vec{u}^T)\nabla \mu] -[\mu {\nabla_s}^2 \vec{u}] - [\mu \frac{\partial \vec{u}}{\partial n}]\kappa - [\vec{g}]
\end{equation}
According to Eq.~\ref{eq:multjump}, the second term on the right hand side of this equation can be written as
\begin{equation}\label{eq:viscGrad}
[(\nabla \vec{u} + \nabla^T \vec{u})\nabla \mu] = [(\nabla \vec{u} + \nabla^T \vec{u})] \cdot \nabla \mu^{\pm} + (\nabla \vec{u}^{\mp} + \nabla^T \vec{u}^{\mp}) \cdot [\nabla \mu]
\end{equation}
Assuming a continuous viscosity, $\mu(\phi) = \mu^- + (\mu^+ - \mu^-)H_{\epsilon}(\phi(x)) = \mu^- + [\mu]H_{\epsilon}(\phi(x))$, where $H_{\epsilon}(\phi(x))$ is a smoothed Heaviside function. Therefore, we can find $\nabla \mu$ as follows:
\begin{eqnarray}
\nabla \mu &=& \nabla(\mu^- + [\mu]H_{\epsilon}(\phi(x))) = [\mu]\nabla H_{\epsilon}(\phi(x)) \nonumber \\
&=& [\mu] \frac{\partial H_{\epsilon}(\phi(x))}{\partial \phi} \nabla \phi = [\mu] \delta_{\epsilon}(\phi) \nabla \phi = [\mu] \delta_{\epsilon}(\phi) ||\nabla \phi|| \vec{n}
\end{eqnarray}
where $\delta_{\epsilon}$ is a smoothed Delta function. This relation also results in $\nabla \mu^+ = \nabla \mu^-$ and therefore $[\nabla \mu] = 0$. Equation~\ref{eq:viscGrad} will thus simplify to
\begin{eqnarray}\label{eq:viscGradfinal}
[(\nabla \vec{u} + \nabla^T \vec{u})\nabla \mu] &=& [(\nabla \vec{u} + \nabla^T \vec{u})] \cdot \nabla \mu = ([\mu] \delta_{\epsilon}(\phi) ||\nabla \phi|| )[(\nabla \vec{u} + \nabla^T \vec{u})] \cdot \vec{n}\nonumber \\
&=&([\mu] \delta_{\epsilon}(\phi) ||\nabla \phi||)\left([\nabla \vec{u}] \cdot \vec{n} + [\nabla^T \vec{u}] \cdot \vec{n} \right) \nonumber \\
&=&([\mu] \delta_{\epsilon}(\phi) ||\nabla \phi||)\left([\frac{\partial \vec{u}}{\partial n}] + [\nabla^T \vec{u}] \cdot \vec{n} \right)
\end{eqnarray}

By substituting Eq.~\ref{eq:viscGradfinal} into Eq.~\ref{eq:mududnfirst} and considering continuous viscosity, the jump in second normal derivative of velocity can be found by
\begin{eqnarray}
[\frac{\partial^2 \vec{u}}{\partial n^2}] &=&  \frac{\rho}{\mu}[\frac{D\vec{u}}{Dt}] - \frac{[\mu]}{\mu}( \delta_{\epsilon}(\phi) ||\nabla \phi||)\left([\frac{\partial \vec{u}}{\partial n}] + [\nabla^T \vec{u}] \cdot \vec{n} \right)- {\nabla_s}^2 [\vec{u}] - [\frac{\partial \vec{u}}{\partial n}]\kappa - \frac{[\vec{g}]}{\mu} \nonumber \\
&=&  \frac{\rho}{\mu}[\frac{D\vec{u}}{Dt}] - \frac{[\mu]}{\mu}\delta_{\epsilon}(\phi) ||\nabla \phi||[\nabla^T \vec{u}] \cdot \vec{n} - {\nabla_s}^2 [\vec{u}] - [\frac{\partial \vec{u}}{\partial n}]\left(\kappa + \frac{[\mu]}{\mu}\delta_{\epsilon}(\phi) ||\nabla \phi||\right)\nonumber \\ &-&\frac{[\vec{g}]}{\mu}
\end{eqnarray}

To normalize this jump condition, other than the characteristic parameters mentioned earlier, a characteristic time scale $t_0$ and also the density of outer fluid $\rho^+$ are being used to normalize time and density. As a result the non-dimensional jump in second normal derivative of the velocity is given by
\begin{eqnarray}
[\frac{\hat{\partial}^2 \hat{\vec{u}}}{\partial n^2}] &=&
 \Re \frac{\hat{\rho}_{avg}}{\hat{\mu}_{avg}}[\frac{D\hat{\vec{u}}}{Dt}] - \frac{[\mu]}{\hat{\mu}_{avg}}\delta_{\epsilon}(\phi) ||\hat{\nabla} \phi||[\hat{\nabla}^T \hat{\vec{u}}] \cdot \vec{n} - {\hat{\nabla}_s}^2 [\hat{\vec{u}}] - [\frac{\hat{\partial} \hat{\vec{u}}}{\partial n}]\left(\hat{\kappa} + \frac{[\mu]}{\hat{\mu}_{avg}}\delta_{\epsilon}(\phi) ||\hat{\nabla} \phi||\right)\nonumber \\ &-& \frac{\Re}{\Fr}\frac{(\hat{\rho}_{avg}-1)[\hat{\bar{\vec{g}}}]}{\hat{\mu}_{avg}}
\end{eqnarray}
where $\hat{\rho}_{avg}$ is the average of inner and outer densities, and the hat notation can be dropped for simplicity.

\section{Jump conditions at grid points}\label{app:jumpAtGrid}
\setcounter{figure}{0}
Combining the jump conditions derived in \ref{app:jump} with the Taylor series expansion shown in \ref{app:IIM}, we can find the jump in velocity at grid points. Starting from the Taylor Series
\begin{equation}
[\Vec{u}]_{gp} = [\Vec{u}]_\Gamma + \phi [\frac{\partial \vec{u}}{\partial n}]_\Gamma + \frac{1}{2} \phi^2 [\frac{\partial^2 \vec{u}}{{\partial n}^2}]_\Gamma
\end{equation}
where $[\vec{u}]_{gp}$ here is the velocity jump at a grid point and $[\vec{u}]_\Gamma$, $[\partial \vec{u}/\partial n]_\Gamma$, and $[\partial^2 \vec{u}/\partial n^2]_\Gamma$ are all computed at the closest point. By substituting the previously defined jump conditions we get
\begin{eqnarray}
  [\vec{u}_{gp}] &=& [\vec{u}] + \phi [\frac{\partial \vec{u}}{\partial n}] + \frac{1}{2} \phi^2 \bigg(\Re \frac{\rho_{avg}}{\mu_{avg}}[\frac{D\vec{u}}{Dt}] - \frac{[\mu]}{\mu_{avg}}\delta_{\epsilon}(\phi) ||\nabla \phi||[\nabla^T \vec{u}] \cdot \vec{n} - {\nabla_s}^2 [\vec{u}] \nonumber \\
                &-& [\frac{\partial \vec{u}}{\partial n}](\kappa + \frac{[\mu]}{\mu_{avg}}\delta_{\epsilon}(\phi) ||\nabla \phi||) - \frac{\Re}{\Fr}\frac{(\rho_{avg}-1)[\bar{\vec{g}]}}{\mu_{avg}}\bigg)
\end{eqnarray}
Further denoting contributions evaluated at previous time steps results in
\begin{eqnarray}
  [\vec{u}]_{gp} &=& [\vec{u}] + (\phi+\frac{1}{2} \phi^2 (\kappa - \frac{[\mu]}{\mu_{avg}}\delta_{\epsilon}(\phi) ||\nabla \phi||))[\frac{\partial \vec{u}}{\partial n}] +\frac{1}{2} \phi^2 (\Re \frac{\rho_{avg}}{\mu_{avg}}(\frac{[\vec{u}] -[\vec{u^n}]}{\Delta t}+[\vec{u^n}\cdot \nabla \vec{u^n}]) \nonumber \\
  &-& \frac{[\mu]}{\mu_{avg}}\delta_{\epsilon}(\phi) ||\nabla \phi||[\nabla^T \vec{u}^n] \cdot \vec{n} - {\nabla_s}^2 [\vec{u}] - \frac{\Re}{\Fr}\frac{(\rho_{avg}-1)[\bar{\vec{g}]}}{\mu_{avg}})\nonumber \\
  &=& (1+\frac{\phi^2 \Re\rho_{avg}}{2\Delta t\mu_{avg}})[\vec{u}] - \frac{\phi^2}{2} {\nabla_s}^2 [\vec{u}] - \frac{\phi^2 [\mu]}{2\mu_{avg}}\delta_{\epsilon}(\phi) ||\nabla \phi||[\nabla^T \vec{u}^n]\cdot \vec{n} \nonumber \\
  &+& (\phi+\frac{\kappa}{2} \phi^2 - \frac{ \phi^2[\mu]}{2\mu_{avg}}\delta_{\epsilon}(\phi) ||\nabla \phi||))[\frac{\partial \vec{u}}{\partial n}]   \nonumber \\
  &+& \frac{\phi^2 \Re}{2\mu_{avg}}(\frac{-\rho_{avg}}{\Delta t}[\vec{u^n}] + \rho_{avg}[\vec{u^n}\cdot \nabla \vec{u^n}]- \frac{(\rho_{avg}-1)}{\Fr}[\bar{\vec{g}]})\nonumber \\
  &=& (1+\frac{\phi^2 \Re\rho_{avg}}{2\Delta t\mu_{avg}})[\vec{u}] - \frac{\phi^2}{2} {\nabla_s}^2 [\vec{u}] - \frac{\phi^2 [\mu]}{2\mu_{avg}}\delta_{\epsilon}(\phi) ||\nabla \phi||[\nabla^T \vec{u}^n]\cdot \vec{n} \nonumber \\
  &-& (\phi+\frac{\kappa}{2} \phi^2 - \frac{[\mu]}{2\mu_{avg}}\phi^2 \delta_{\epsilon}(\phi) ||\nabla \phi||))(([ \frac{\partial \vec{u}}{\partial \tau}] \cdot \vec{n} )\vec{\tau} + ([\frac{\partial \vec{u}}{\partial b}] \cdot \vec{n})\vec{b} +([\frac{\partial \vec{u}}{\partial \tau}] \cdot \vec{\tau} + [\frac{\partial \vec{u}}{\partial b}] \cdot \vec{b})\vec{n})\nonumber \\
  &+& (\phi+\frac{\kappa}{2} \phi^2 - \frac{[\mu]}{2\mu_{avg}}\phi^2 \delta_{\epsilon}(\phi) ||\nabla \phi||))(\frac{1}{\Ca\mu_{avg}}\Vec{P}\vec{f})\nonumber \\
  &+& \frac{\phi^2 \Re}{2\mu_{avg}}(\frac{-\rho_{avg}}{\Delta t}[\vec{u^n}] + \rho_{avg}[\vec{u^n}\cdot \nabla \vec{u^n}]- \frac{(\rho_{avg}-1)}{\Fr}[\bar{\vec{g}]})
\end{eqnarray}
This expression can be split into a part which depends on the velocity jump at the interface, $\vec{j}_{[\vec{u}]}$, and a part which contains external forces and contributions from the bulk fluid velocity, $\vec{j}_{other}$,
where $[\vec{u}_{gp}]=\vec{j}_{[\vec{u}]} + \vec{j}_{other}$ and
\begin{eqnarray}
\Vec{j}_{[u]} &=& (1+\frac{\phi^2 \Re\rho_{avg}}{2\Delta t\mu_{avg}})[\vec{u}] - \frac{\phi^2}{2} {\nabla_s}^2 [\vec{u}] - (\phi+\frac{\kappa}{2} \phi^2 - \frac{[\mu]}{2\mu_{avg}}\phi^2 \delta_{\epsilon}(\phi) ||\nabla \phi||))(( \frac{\partial [\vec{u}]}{\partial \tau} \cdot \vec{n} )\vec{\tau}  \nonumber \\
  &+& (\frac{\partial [\vec{u}]}{\partial b} \cdot \vec{n})\vec{b} +(\frac{\partial [\vec{u}]}{\partial \tau} \cdot \vec{\tau} + \frac{\partial [\vec{u}]}{\partial b} \cdot \vec{b})\vec{n}) \nonumber \\
\Vec{j}_{other} &=& (\phi+\frac{\kappa}{2} \phi^2 - \frac{[\mu]}{2\mu_{avg}}\phi^2 \delta_{\epsilon}(\phi) ||\nabla \phi||))(\frac{1}{\Ca\mu_{avg}}\Vec{P}\vec{f})- \frac{\phi^2 [\mu]}{2\mu_{avg}}\delta_{\epsilon}(\phi) ||\nabla \phi||[\nabla^T \vec{u}^n]\cdot \vec{n}\nonumber \\
&+& \frac{\phi^2 \Re}{2\mu_{avg}}(\frac{-\rho_{avg}}{\Delta t}[\vec{u^n}] + \rho_{avg}[\vec{u^n}\cdot \nabla \vec{u^n}]- \frac{(\rho_{avg}-1)}{\Fr}[\bar{\vec{g}]}).
\end{eqnarray}
This formulation allows for the (linear) contributions of $[\vec{u}]$ to be computed implicitly via $\vec{j}_{[\vec{u}]}$ while the portions not dependent on the jump, $\vec{j}_{other}$ can be explicitly evaluated at the previous time step. This allows for the solution of a linear system, Eq.~\eqref{eq:MatVec}, rather than a set of non-linear equations.

\end{document}